\documentclass[10pt,aps,pra,reprint,amsmath,amsfonts,amssymb,showpacs,nofootinbib,floatfix]{revtex4-1}
\usepackage{epsfig,psfrag,graphicx,bm,color}
\usepackage{xspace}

\newcommand{\aap}{A{\&}A}

\newcommand{\aj}{AJ}
\newcommand{\apjl}{ApJL}

\newcommand{\mnras}{MNRAS}
\newcommand{\ssr}{SSRv}

\newcommand{\jcap}{JCAP}
\newcommand{\Fermi}{{\em Fermi}\xspace}

\newcommand{\rmn}{\mathrm}
\newcommand{\bx}{{\bf x}}
\newcommand{\fg}{{F_\gamma}}
\newcommand{\sg}{{S_\gamma}}
\newcommand{\psf}{\theta_\rmn{res}}
\newcommand{\ph}{\rmn{ph}}
\newcommand{\eph}{E_\ph}

\newcommand{\vir}{\rmn{vir}}
\newcommand{\gal}{\rmn{gal}}
\newcommand{\sd}{\rmn{SD}}
\newcommand{\clu}{\rmn{cl}}
\newcommand{\sfe}{\rmn{sfe}}
\newcommand{\sub}{\rmn{sub}}
\newcommand{\msun}{M_\odot}
\newcommand{\ee}{E_\rmn{e}}
\newcommand{\stars}{\rmn{stars}}
\newcommand{\dust}{\rmn{dust}}

\newcommand{\s}{\rmn{s}}
\newcommand{\Kp}{\rmn{K}^\prime}
\newcommand{\Ip}{\rmn{I}^\prime}
\newcommand{\Js}{\rmn{J}^*}
\newcommand{\Jp}{\rmn{J}^\prime}
\newcommand{\B}{\rmn{B}}
\newcommand{\bsub}{\B_\rmn{sub}}
\newcommand{\qCR}{q_{\rmn{CR}-\ensuremath{\pi^0}}}
\newcommand{\ds}{{\sc DarkSUSY}}
\newcommand{\mlim}{M_\rmn{lim}}
\newcommand{\sm}{\rmn{sm}}

\newcommand{\kpc}{\rmn{kpc}}
\newcommand{\mev}{\rmn{MeV}}
\newcommand{\cm}{\rmn{cm}}
\newcommand{\egt}{\tilde{E}_\gamma}
\newcommand{\eet}{\tilde{E}_\rmn{e}}
\newcommand{\bra}{\langle}
\newcommand{\ket}{\rangle}
\newcommand{\vst}{\vspace{-0.32mm}}
\newcommand{\vstt}{\vspace{-0.3mm}}
\newcommand{\el}{\rmn{el}}
\newcommand{\mytimes}{\hspace{-0.5mm}\times\hspace{-0.5mm}}
\newcommand{\hst}{\hspace{-1.0mm}}
\newcommand{\degs}{^\circ}
\newcommand{\colo}{}

\newcommand{\beq}{\begin{equation}}
\newcommand{\eeq}{\end{equation}}
\newcommand{\gev}{\rmn{GeV}}
\newcommand{\tev}{\rmn{TeV}}
\newcommand{\ev}{\rmn{eV}}
\newcommand{\dd}{\rmn{d}}
\newcommand{\mx}{\ensuremath{m_{\chi}}}

\newcommand{\ngamma}{\ensuremath{N_{\gamma}}}
\newcommand{\ngammaj}{\ensuremath{N_{\gamma,j}}}
\newcommand{\sigmaannv}{\ensuremath{\langle\sigma v\rangle}}
\newcommand{\sigv}{v_\rmn{cl}}

\newcommand{\egamma}{\ensuremath{E_{\gamma}}}

\newcommand{\CR}{\rmn{CR}}
\newcommand{\rhos}{\ensuremath{\rho_s}}
\newcommand{\rs}{\ensuremath{r_s}}
\newcommand{\rvir}{r_{200}}
\newcommand{\mvir}{M_{200}}
\newcommand{\rhoc}{\ensuremath{\rho_c}}
\newcommand{\e}{\rmn{e}}
\newcommand{\eg}{E_\gamma}

\begin{document}

\title{Prospects of detecting gamma-ray emission from galaxy clusters:
  cosmic rays and dark matter annihilations}

\author{Anders Pinzke$^{1}$} \email{apinzke@physics.ucsb.edu}
\author{Christoph Pfrommer$^{2}$}\email{christoph.pfrommer@h-its.org}
\author{Lars Bergstr\"om$^{3}$}\email{lbe@fysik.su.se}

\affiliation{$^{1}$University of California - Santa Barbara,
  Department of Physics, CA 93106-9530, USA}

\affiliation{$^{2}$Heidelberg Institute for Theoretical Studies
  (HITS), Schloss-Wolfsbrunnenweg 33, DE - 69118 Heidelberg, Germany}

\affiliation{$^{3}$The Oskar Klein Centre for Cosmoparticle Physics,
  Department of Physics, Stockholm University, AlbaNova University
  Center, SE - 106 91 Stockholm, Sweden}

\date{\today}


\begin{abstract}
We study the possibility for detecting gamma-ray emission from galaxy
clusters. We consider 1) leptophilic models of dark matter (DM)
annihilation that include a Sommerfeld enhancement (SFE), 2) different
representative benchmark models of supersymmetric DM, and 3) cosmic
ray (CR) induced pion decay. Among all clusters/groups of a
flux-limited X-ray sample, we predict Virgo, Fornax and M49 to be the
brightest DM sources and find a particularly low CR-induced background
for Fornax. For a minimum substructure mass given by the DM
free-streaming scale, cluster halos maximize the substructure boost
for which we find a factor of $\gtrsim 1000$. Since regions around the
virial radius dominate the annihilation flux of substructures, the
resulting surface brightness profiles are almost flat. This makes it
very challenging to detect this flux with imaging atmospheric
Cherenkov telescopes since their sensitivity drops approximately
linearly with radius and they typically have $5-10$ linear resolution
elements across a cluster. Assuming cold dark matter with a
substructure mass distribution down to an Earth mass and using
extended \Fermi upper limits, we rule out the leptophilic models in
their present form in 28 clusters, and limit the boost from SFE in M49
and Fornax to be $\lesssim 5$. This corresponds to a limit on SFE in
the Milky Way of $\lesssim 3$ which is too small to account for the
increasing positron fraction with energy as seen by PAMELA and
challenges the DM interpretation. Alternatively, if SFE is realized in
Nature, this would imply a limiting substructure mass of $\mlim >
10^4\,\msun$---a problem for structure formation in most particle
physics models. Using individual cluster observations, it will be
challenging for \Fermi to constrain our selection of DM benchmark
models without SFE. The \Fermi upper limits are, however, closing in
on our predictions for the CR flux using an analytic model based on
cosmological hydrodynamical cluster simulations. We limit the
CR-to-thermal pressure in nearby bright galaxy clusters of the \Fermi
sample to $\lesssim 10\%$ and in Norma and Coma to $\lesssim
3\%$. Thus we will soon start to constrain the underlying CR physics
such as shock acceleration efficiencies or CR transport properties.
\end{abstract}

\maketitle
\section{Introduction}
Dark matter (DM) has been searched for in direct detection experiments
\cite{Pato:2010zk}, at accelerators
\cite{Ellis:2001hv,Baer:2006ff,Khachatryan:2011tk} and also in
indirect detection experiments looking for signals in the cosmic-ray
(CR) spectra of antiprotons, positrons, neutrinos and all of the
electromagnetic spectrum from radio waves to gamma-rays
\cite{Bergstrom:2009ib}. So far, the improvements in direct detection
sensitivity have put this method into focus, but the situation may
change considerably the coming few years as the CERN LHC experiments
collect data, and new gamma-ray detectors are being planned, such as
the CTA \cite{Consortium:2010bc}. In fact, it has recently been
pointed out \cite{Bergstrom:2010gh} that a dedicated ground-based
gamma-ray detector would have potential that goes far beyond that of
the other methods, depending on presently unknown parameters in the
particle physics models for DM.

Among the astrophysical systems which will be very interesting to
detect, and study, with gamma-ray detectors (\Fermi, H.E.S.S., MAGIC,
VERITAS, and eventually large detectors like CTA)\footnote{{High
    Energy Stereoscopic System, Major Atmospheric Gamma Imaging
    Cerenkov Telescope, Very Energetic Radiation Imaging Telescope
    Array System, Cerenkov Telescope Array.}}  belong galaxy
clusters. The most promising directions in which to search for a
gamma-ray annihilation signal (from the annihilation process itself,
and also the accompanying bremsstrahlung and inverse Compton (IC)
components coming from charged particles produced in the
annihilations) are basically three:

1. The galactic center (g.c.). This is where all numerical simulations
of cold dark matter (CDM) predict the highest density. However, the
detailed DM density in the very central part is difficult to predict,
due a to a possibly very complicated interplay between baryons, DM,
and the central galactic black hole. Also, it is a very crowded region
with many gamma-ray sources like pulsars, CR-illuminated molecular
clouds, and other supernova remnants, which have to be subtracted from
the data to extract the DM induced signal. In fact, there is a recent
claim of an indication of a relatively light DM particle contribution
to the gamma-ray flux from the g.c. \cite{2011PhLB..697..412H}, but
other hypotheses seem to work at least as well
\cite{2010arXiv1012.5839B}.

2. The dwarf spheroidal galaxies orbiting the Milky Way (MW), like
Segue-1, Ursa Minor, Draco, Sagittarius, Sculptor, Carina or Willman-1
\cite{2009JCAP...01..016B,2010ApJ...720.1174A,2010JCAP...01..031S,2010JCAP...01..031S,2011arXiv1103.0477T,2011APh....34..608H}. The
problem here is that the nature of many of these small, dark
matter-dominated galaxies is not entirely clear, and the velocity
dispersion estimates are based on rather small numbers of stars.
Confusion with star clusters and tidal disruption are other
complications. Once a satellite dwarf galaxy is accreted by the MW,
the outer regions are severely affected by tidal stripping. The longer
a satellite has been part of our Galaxy, and the closer it comes to
the center during its pericentral passage, the more material is
removed \cite{2004MNRAS.355..819G}.  Thus the DM density profile is
very uncertain for most of them, especially for radii larger than
those probed by the stars. Nonetheless, by stacking the data together
from many dwarf spheroidals these uncertainties can be made less
severe, and preliminary results from \Fermi-LAT shows this method to
give quite promising results \cite{garde}.

3. Galaxy clusters. This possibility has been less studied
theoretically, however currently there is an ongoing observational
campaign to detect gamma-ray emission from galaxy clusters
\cite{2003ApJ...588..155R,2006ApJ...644..148P,2008AIPC.1085..569P,2009A&A...495...27A,2009arXiv0907.5000G,2009IJMPD..18.1627D,2009A&A...495...27A,2009A&A...502..437A,2009ApJ...704..240K,2009ApJ...706L.275A,2010ApJ...710..634A,2010JCAP...05..025A,2010ApJ...717L..71A}.
In fact, we noted in a previous Letter \cite{2009PhRvL.103r1302P} that
there are certain advantages that work in favour of this possible
target for gamma-ray detection of DM annihilation. Galaxy clusters
constitute the most massive objects in our Universe that are forming
today. This causes their DM subhalo mass function to be less affected
by tidal stripping compared to galaxy sized halos that formed long
ago. The annihilation luminosities of the DM halo component for
e.g. the Virgo cluster and the Draco dwarf scales in a way (see
\cite{2009PhRvL.103r1302P}) that the ratio of gamma-ray luminosities
from the smooth components is around 4, in favour of Virgo. In
addition, there may be a further enhancement due to substructure,
which to a large extent should be unaffected by tidal disruption, at
least in the outer regions. According to a recent estimate
\cite{2011MNRAS.410.2309G}, more massive haloes tend to have a larger
mass fraction in subhalos. For example, cluster size haloes typically
have 7.5 per cent of the mass within $r_{200}$ in substructures of
fractional mass larger than $10^{-5}$, which is 25 per cent higher
than for galactic haloes.\footnote{All halo masses and length scales
  in this paper are scaled to the currently favored value of Hubble's
  constant, $H_0 = 70\, \rmn{km~s}^{-1}\,\rmn{Mpc}^{-1}$. We define
  the virial mass $\mvir$ and virial radius $\rvir$ as the mass and
  radius of a sphere enclosing a mean density that is 200 times the
  critical density of the Universe $\rho_{\rmn{cr}}$.}

In this paper, we will investigate in detail the potential of several
of the most promising galaxy clusters to produce an annihilation
gamma-ray yield which could be observable with present and planned
gamma-ray detectors. Here, we sketch out the main arguments that need
to be considered for maximizing the expected signal-to-noise ratio of
a promising target cluster. First, we need to maximize the DM
annihilation flux of an unresolved cluster, $F=A_\rmn{dm} \int dV
\rho^2 / D^2 \propto \mvir^\alpha/D^2$, where $A_\rmn{dm}$ depends on
the particle physics model of DM, $\rho$ is the smooth DM density
profile, $D$ is the luminosity distance, and $\alpha=1$ if we assume
universality of the DM density profile. The dependence of $\rho$ on
the halo formation epoch breaks the universality and slightly modifies
the mass dependence, yielding $\alpha=0.83$
\cite{2009PhRvL.103r1302P}. Additionally, the presence of a hierarchy
of substructures down to small scales \cite{2008Natur.456...73S} and
the potential dependence of the particle physics cross section on the
relative DM particle velocities \cite{ArkaniHamed:2008qn} may
furthermore modify the scaling parameter $\alpha$ and shall be one of
the focus points of this work. Second, we need to minimize the
expected noise which is a sum of instrumental noise, galactic and
cluster-intrinsic foreground. While the galactic foreground varies
across the sky, it is typically lower for increasing latitude (away
from the galactic plane). It is thought that the gamma-ray foreground
from clusters is dominated by CRs which are accelerated at
cosmological formation shocks and transported over time although there
could be a substantial contribution of CRs from AGNs and
supernova-driven galactic winds. In this work we will also study (and
constrain) the CR-induced emission from clusters to identify those
objects that are expected to be especially dim which would imply a low
CR-induced background.  Since the CR-induced emission is expected to
scale with the thermal X-ray emission of clusters
\cite{2004A&A...413...17P}, we estimate gamma-ray fluxes of a nearby
X-ray flux complete sample of galaxy clusters from the Rosat all-sky
survey (extended HIghest X-ray FLUx Galaxy Cluster Sample, HIFLUGCS,
\cite{2002ApJ...567..716R,2007A&A...466..805C}).\footnote{Note that we
  have added the Virgo cluster to the sample that we nevertheless
  refer to as extended HIFLUGCS catalogue in the following.}  This
complements and extends previous work related to DM in clusters
\cite{2006A&A...455...21C,2009PhRvD..80b3005J,2011arXiv1104.3530S,2011ApJ...726L...6C,2011arXiv1110.1529H}
and to CRs in clusters
\cite{2010MNRAS.409..449P,2011MNRAS.410..127B,2008MNRAS.385.1211P,2009JCAP...08..002K,2010MNRAS.407.1565D}.

In Section 2 we discuss the theory of DM and gamma-rays. In
particular, we focus on the leptophilic (LP) and supersymmetric
benchmark (BM) DM models in this work, and outline the framework for
estimating the gamma-ray emission from various radiative processes. In
Section 3 we calculate the gamma-ray fluxes and spectral distributions
of four clusters that are identified as prime targets for DM
observations (Fornax and Virgo) as well as to have a high CR-induced
gamma-ray yield (Perseus and Coma). For the same clusters we derive
the gamma-ray surface brightness profiles in Section 4. To extend the
analysis to more clusters and increase the list of promising targets
we estimate the fluxes from DM and CRs of all clusters in the extended
HIFLUGCS sample in Section 5. In Section 6 we conclude and discuss our
results.

\section{Theory}
\label{sect:theory}
We start the section by discussing two different, but well motivated
DM models; LP and supersymmetric DM. We then present the framework
that is used to calculate the gamma-ray emission for these DM models
using an Einasto DM density profile and the expected enhancement from
DM substructures. This is followed by an outline of the framework of
IC emission where we take into account photon fields of the cosmic
microwave background (CMB), dust, and starlight. We end by summarizing
our formalism of calculating the CR-induced pion decay gamma-ray
emission which is thought to dominate the astrophysical gamma-ray
signal.

\subsection{Detecting Particle Dark Matter}
\label{sect:PF}
Besides the intrinsic interest in the gamma-ray flux from galaxy
clusters generated by conventional hadronic and electromagnetic
processes - which should be close to observability with the \Fermi-LAT
data
\cite{1997ApJ...487..529B,2007A&A...473...41E,2010MNRAS.409..449P} -
the possible contribution from WIMP DM is of greatest interest. A WIMP
(Weakly Interactive Massive Particle) which fulfils the WMAP bounds on
the relic density of CDM of \cite{Komatsu:2010fb}
$$\Omega_{CDM}h^2=0.112\pm 0.0056,$$ will in many cases naturally give
a gamma-yield which may be observable. In addition, there are possible
enhancement effects known, such as the astrophysical boost from dense
substructure of DM halos, or the particle physics boost from the
Sommerfeld effect, which may increase the chances of detection
further.

There are three methods for detecting DM candidates which are
presently employed: First, at particle physics accelerators such as
CERN's LHC, one is now entering into a new energy regime which may
allow the production of the heavy, electrically neutral and long-lived
particles which may constitute DM. From these experiments one may get
the first glimpse of the mass scale beyond the standard model where DM
may reside. However, to really show that any of the hypothetical, new
particles created at the LHC is actually the DM, one has to rely on
the two other methods available for the DM search, namely direct and
indirect detection. Second, direct detection methods, which are
presently evolving rapidly, use the feeble interaction between DM
particles of the galactic halo and nuclei such as Germanium, Sodium,
Iodine, Argon or Xenon to infer the scattering cross section and a
rough estimate of the mass of the DM particle (for the currently most
sensitive search, see \cite{Aprile:2010um,Aprile:2011hi}). A
characteristic of this method is that it basically only depends on the
local DM density (which is rather well determined to be around $0.4$
GeV/cm$^3$) and the basic cross section for DM - nucleus scattering. A
disadvantage is that it does not benefit from the two enhancement
mechanisms mentioned above, boost from substructure and/or the
Sommerfeld effect. Also, it cannot be excluded - although it seems at
present improbable - that the local halo structure is such that the
solar system happens to be in an underdense region.

Third, in indirect detection, in particular in the photon channel, one
searches for products of DM annihilation in the galactic halo and
beyond. Particularly interesting targets are the galactic center (for
a recent possible indication of a signal, see
\cite{2011PhLB..697..412H}, however, see also
\cite{2010arXiv1012.5839B}), the dwarf galaxies surrounding the MW
\cite{Strigari:2006rd,Essig:2009jx,2010JCAP...01..031S}, galaxy
clusters (the focus of this work)
\cite{Ghigna:1998vn,Lewis:2002mfa,Boyarsky:2006kc,2006A&A...455...21C,2009PhRvL.103r1302P},
and even the cosmological large scale structure
\cite{Bergstrom:2001jj,Ullio:2002pj,Taylor:2002zd,Elsaesser:2004ap,2011MNRAS.tmp..503C,Abazajian:2010sq,Abdo:2010dk,Zavala:2011tt}.

\subsubsection{Leptophilic models}
\label{sect:LP}
There is also a possibility to search for DM annihilation in the Milky
Way indirectly through annihilation to antimatter channels, which
however lacks the important directional signature of
gamma-rays. Recently, it has however been much in focus due to the
surprising findings of PAMELA \cite{Adriani:2008zr} and \Fermi
\cite{Abdo:2009zk}. Viable particle physics models are rather
constrained by other observations, however. For example, the
non-observation of an enhanced antiproton flux by PAMELA
\cite{Adriani:2010rc} means that quark-antiquark final states have to
be suppressed. This has led to the postulate of a whole class of {\em
  leptophilic} models with mainly or exclusively annihilation to
leptons and antileptons. In addition, one has to have a sizeable
fraction of final states containing muons or tau leptons as the shape
of the spectrum disfavors direct annihilation to electrons and
positrons only. A large enhancement, of the order of at least several
hundred, of the annihilation rate is also needed, something that may
be given by the {\em Sommerfeld effect} (for a pedagogical review of
viable scenarios, see \cite{ArkaniHamed:2008qn}). As still another
difficulty for these models, the lack of an IC signature from the
Galactic center means that the DM density distribution must be cored
rather than cuspy
\cite{Bertone:2008xr,Cirelli:2008pk,Bergstrom:2008ag}. (For a review
of the DM modeling of these effects, see \cite{Bergstrom:2009ib}. For
a recent treatment, showing still viable models, see
\cite{Finkbeiner:2010sm}.)

A more standard explanation of the PAMELA/\Fermi excess would be
pulsars or other supernova remnants (e.g.,
\cite{1995PhRvD..52.3265A,Hooper:2008kg,Ahlers:2009ae}). In that case,
DM would more likely be explained by the conventional scenario of a
WIMP, for example the thoroughly studied lightest supersymmetric
neutralino (for reviews, see
\cite{Jungman:1995df,Bergstrom:2000pn,Bertone:2004pz}). We note that
AMS-02 experiment on the International Space Station \cite{ams02} may
give interesting clues to the origin of the antimatter excess.

In the scenario with leptophilic DM, one has to rely on some
enhancement of the annihilation rate from the effects of DM halo
substructure, a possibility which was realized long ago
\cite{1993ApJ...411..439S,Bergstrom:1998zs,Moore:1999nt}. However, a
boost factor as large as several hundred is difficult to achieve in
the solar neighborhood, due to tidal stripping of subhalos in the
inner part of the Galaxy. Sommerfeld enhancement is thus important for
leptophilic models that claim to explain the PAMELA/\Fermi
result. This effect was computed for electromagnetism by Arnold
Sommerfeld many years ago \cite{sommerfeld} and recently rediscovered
in the quantum field theory of DM
\cite{2005PhRvD..71f3528H,2007NuPhB.787..152C,2009PhRvD..79a5014A}.
In quantum mechanics describing electron scattering and electron
positron annihilation, it is caused by the distortion of the plane
wave describing the relative motion of the annihilating particle pair,
due to the near formation of a bound state caused by photon
exchange. In the ladder approximation for QED, one reproduces the
Sommerfeld effect, and the square of the relative wave function at the
origin (which enters into the probability for the short-distance
process of annihilation) is increased by the factor
\cite{2009PhRvD..79a5014A}
\begin{equation}
S_{QED}=\frac{|\psi(0)|^2}{|\psi^{(0)}(0)|^2}=
\left|\frac{\frac{\pi\alpha}{v}}{1-e^{-\frac{\pi\alpha}{v}}}\right|,
\end{equation}
with $\alpha =e^2/\hbar c$ the fine-structure constant, and $v$ the relative
velocity. This amounts to $S_{QED}=\pi\alpha/v$ for small velocities. For a
Yukawa-like particle of mass $m_\phi$, mediating a weak attractive force with
coupling constant $\alpha_Y$ between DM particles $\chi$ with mass
$m_\chi$, the small-velocity limit of the enhancement becomes instead
\begin{equation}
S_Y\sim\frac{\alpha_Y m_\chi}{m_\phi}.
\label{eq:saturation}
\end{equation}
In the general case, with mediators that may also excite virtual
charged particles in the DM sector, one has to solve numerically a
coupled system of differential equations using appropriate boundary
conditions
\cite{2005PhRvD..71f3528H,2007NuPhB.787..152C,2009PhRvD..79a5014A}. In
some cases, the enhancement factor $S$ can be as high as several
hundred to a few thousand, depending however on the exact parameters
of the theory. The effect is usually strongly velocity-dependent,
depending on velocity as $1/v$ or, in the fine-tuning case of being
very near resonance, as $1/v^2$. This means that in a virialised
system (such as a galaxy cluster) with large velocity dispersion
$\sigv$ the SFE will be smaller than the one expected in a single
galaxy such as the MW, roughly by a factor $v_{\rm MW}/\sigv$. Note
that the $1/v$-scaling is valid only for $(v/c) \gtrsim
(m_\phi/m_\chi)$; at smaller velocities and outside resonances, the
$1/v$-enhancement saturates at $m_\phi/m_\chi$
\cite{2008PhRvL.101z1301K}.

As LP models, apart from being slightly contrived from the particle
physics point of view, are also rather limited by several sets of
astrophysical data. We take advantage of the recent reanalysis
\cite{Finkbeiner:2010sm} to define a benchmark LP model that is still
viable. It is found that the bounds from WMAP5 approximately imply a
Sommerfeld boost factor $S$ which has to satisfy $S(v\to
150\,\rmn{km}/\rmn{s})\lesssim 250/f\ (m_\phi/1$ GeV$)$, where $f$ is
the fraction of energy from DM annihilation that ionizes the
intergalactic medium, $f\sim 0.7$ for annihilation to electrons and a
factor of a few smaller for all other standard model final states.

Taking into account the whole cosmic history of LP models, also
utilizing limit on spectral and polarization distortions of the CMB,
the maximal value for the boost factor for a 1 to 2~TeV particle and
sub-GeV force carriers is found to be \cite{Finkbeiner:2010sm} between
400 and 800. Although lower than the first estimates (e.g.,
\cite{Bergstrom:2009fa,Meade:2009iu}) this is still enough to explain
the PAMELA and \Fermi excess, given other astrophysical uncertainties.

We adopt our benchmark LP model from \cite{Finkbeiner:2010sm} where we
use a DM mass $m_\chi=1.6$~TeV and a branching ratio of
($1/4:1/4:1/2$) into ($\mu^+\mu^-:\e^+\e^-:\pi^+\pi^-$). Furthermore,
since most of the gamma-ray flux is expected to come from dense
subhalos within clusters which have a velocity dispersion close to the
velocity limit $v_\rmn{sat}$ where the SFE saturates, we use
$S=\B_\sfe(v_\rmn{sat})\approx 530$ for the cluster
halos.\footnote{Note that $S$ is strictly speaking the Sommerfeld
  enhanced annihilation cross section, but here we have additionally
  included the order unity constraint that ensures that the relic
  density is not over-depleted at freeze-out.} However, for figures
where we neglect the substructure boost, we adopt a velocity dependent
SFE that is normalized to fit the electron and positron excess
observed at Earth,
\begin{equation}
\B_\sfe(\sigv) = \B_\rmn{sfe,MW} \frac{v_\rmn{MW}}{\sigv} =
70 \left(\frac{\mvir}{10^{15}\,\msun}\right)^{-1/3}\,,
\label{eq:B_sfe}
\end{equation}
where the local boost factor required to explain the data is
$\B_\rmn{sfe,MW}=300$. The velocity dispersion for the MW, $v_\rmn{MW}
\approx 220\,\rmn{km\,s}^{-1}$, and the mass dependent velocity
dispersion $\sigv$ for clusters is derived in
\cite{2005RvMP...77..207V}.

\subsubsection{Supersymmetric dark matter}
The most studied models for DM are supersymmetric ones, especially
models where the lightest supersymmetric particle, in most models the
lightest neutralino, is stable. The stability is assured in viable
supersymmetric models due to a discrete symmetry, $R$-parity. This
symmetry is needed from the particle physics point of view to avoid
fast proton decay, and automatically makes the lightest supersymmetric
particle stable, and therefore a good candidate for DM. In addition,
as the coupling to ordinary matter is given by gauge couplings, the
neutralino is automatically a WIMP candidate. Unfortunately, the
breaking of supersymmetry (which has to be there since otherwise,
e.g., there would exist a scalar electron with the same mass as the
electron) means that a large number of essentially free parameters
enter actual calculations. The neutralino is then a linear
combination of the supersymmetric partners of the U(1) gauge boson,
$\tilde B$, one of the SU(2) gauge bosons, $\tilde W_3$, and the two
Higgs doublets in the Minimal Supersymmetric extension of the Standard
Model (the MSSM), $\tilde H_1^0$ and $\tilde H_2^0$, with mixing
coefficients which depend on the supersymmetric breaking parameters
\cite{1984NuPhB.238..453E}. The parameter space is very large, too
large in general to scan efficiently even with the most powerful
computers. Therefore, often simplifying assumptions are made to limit
the number of free parameters. Even then, a large number of viable
models is found which give a relic density consistent with the WMAP
data. This means that they are indeed very good templates for WIMP DM,
something that explains their popularity besides the subjective
statement that few more well-motivated models for DM have been put forward
so far.

The wave function of the lightest neutralino can be written
\begin{equation}
\tilde\chi_1=a_1\tilde B+a_2\tilde W_3+a_3 \tilde H_1^0+a_4\tilde H_2^0\,,
\end{equation}
with
\begin{equation}
\sum_{i=1}^4 |a_i|^2=1\,,
\end{equation}
where the gaugino fraction of a given neutralino is $|a_1|^2+|a_2|^2$
whereas $|a_3|^2+|a_4|^2$ is the higgsino fraction.

We use the \ds package \cite{ds} to compute the mass, couplings and
relic density for a given set of parameters. Actually, we take
advantage of so-called benchmark (BM) models, which have been proposed
in different contexts (in particular in \cite{2009JCAP...01..016B}
giving predictions for imaging air Cherenkov telescopes, like MAGIC II
and CTA). They give a {\em good representation of models with high
  enough gamma-ray rates to become the first candidates for DM
  indirect detection} in imaging air Cherenkov telescopes. We use the
following supersymmetric BM models:

\begin{itemize}
\item
 $\Ip$: This model was introduced in \cite{2004EPJC...33..273B}, where
  its phenomenology at colliders was studied. Annihilation directly
  into lepton pairs is suppressed for neutralinos due to their
  Majorana nature and therefore helicity suppression for annihilation
  in the galactic halo \cite{1983PhRvL..50.1419G}. The higher order
  process (inner bremsstrahlung, IB) $\tilde\chi_1\tilde\chi_1\to
  \ell^+\ell^-\gamma$, which does not suffer from helicity suppression
  \cite{1989PhLB..225..372B,2008JHEP...01..049B}, gives a considerable
  contribution due to light sleptons in this model.

\item $\Jp$: This model, also from \cite{2004EPJC...33..273B} is in the
  so-called coannihilation tail. The sleptons are nearly degenerate
  with the neutralino, which causes the large IB from the leptonic
  final states to give a high enhancement of the flux.

\item $\Kp$: A representative model for the funnel region, where the
  annihilation dominantly occurs in the $s$-channel through exchange
  of the pseudoscalar Higgs boson \cite{2004EPJC...33..273B}. Here IB
  contributions are not important.

\item $\Js$: Annihilation in the so-called coannihilation region,
  introduced in \cite{2008JHEP...01..049B} as BM3, with a particularly
  large IB contribution.

\end{itemize}

The details for the DM BM models are summarized in Table~\ref{tab:BMpara}.

\begin{table}
\begin{tabular}{cccc}
\hline\hline
      BM & $m_{\chi}$ & $\Omega_{\chi} h^2$ & $\sigmaannv$\\
         & $[\rmn{GeV}]$ & & $[\rmn{cm}^3\,s^{-1}]$\\
\hline
$\Ip$ & 140 & 0.09 & $4.0\times 10^{-27}$ \\
$\Jp$ & 315 & 0.12 & $3.3\times 10^{-28}$ \\
$\Kp$ & 570 & 0.10 & $4.4\times 10^{-26}$ \\
$\Js$ & 234 & 0.09 & $8.9\times 10^{-29}$ \\
\hline\hline
\end{tabular}
 \caption{Relevant parameters for the four benchmark models. The mass
   of the DM particles are given denoted by $m_{\chi}$, the relic
   density by $\Omega_{\chi} h^2$ and the annihilation rate today by
   $\sigmaannv$.\label{tab:BMpara}}
\end{table}

\subsubsection{Final state radiation}
There are two types of radiative processes which are important for our
BM models, first the usual QED final state radiation for both the LP
model and for supersymmetric models with charged final states. In
addition, there may be direct emission from a virtual, charged,
exchanged particle (such as a spin-0 slepton or squark), internal
bremsstrahlung, IB. The latter is essentially only important when
there is a helicity suppression for the lowest order annihilation
process, which is the case for neutralinos, whose Majorana nature make
the annihilation rate in the $s$ wave proportional to $m_f^2$, for
final state fermion $f$. Interestingly, both the gamma-ray and fermion
energy spectra are peaked at the highest energy for this process,
which can in some models cause a rather spectacular bump in the
spectra for these models.

The photon spectrum resulting from final state radiation is universal
with only a weak dependence of the underlying particle physics
model. The photon yield from this process is given by (see
e.g. \cite{2008JHEP...01..049B})
\begin{equation}
\frac{\dd N_{X \bar{X}}}{\dd x} \approx \frac{\alpha Q_X^2}{\pi}
\mathcal{F}_X(x) \log\left[\frac{4 m_\chi^2\left(1-x\right)}{m_X^2}\right]\,.
\end{equation}
Here, the normalized photon energy $x=\eg/m_\chi c^2$,
$\alpha =e^2/\hbar c$ is the fine-structure constant,
$Q_X^2$ and $m_X$ the
charge and mass of the final state particle $X$, respectively. The function
$\mathcal{F}_X(x)$ depends on the spin of the final state and is given
by
\begin{equation}
\mathcal{F}_\rmn{fermion}(x) = \frac{1+\left(1-x\right)^2}{x}\,
\end{equation}
for fermions.

\begin{figure*}
\begin{minipage}{2.0\columnwidth}
 \includegraphics[width=0.49\columnwidth]{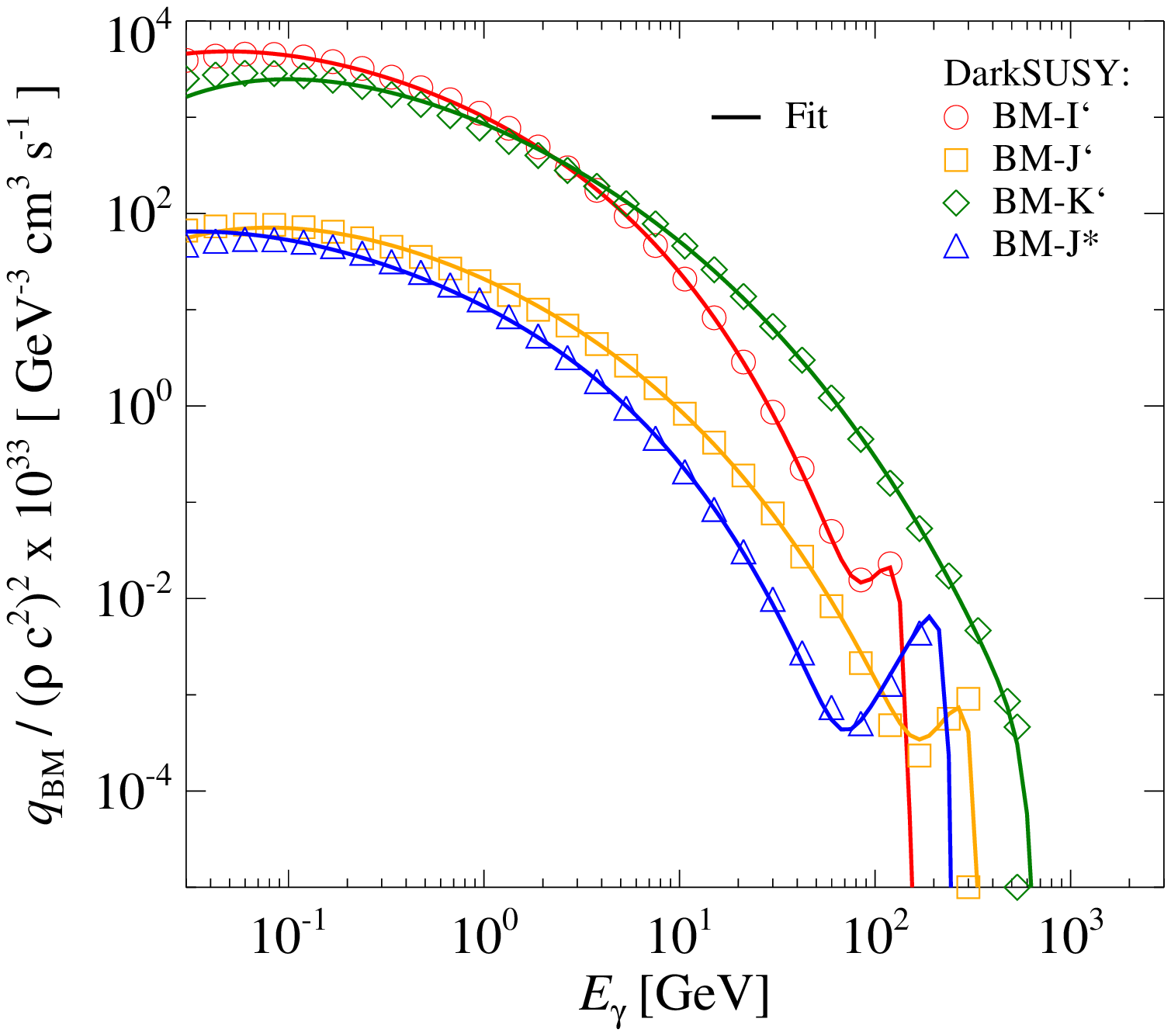}
 \includegraphics[width=0.49\columnwidth]{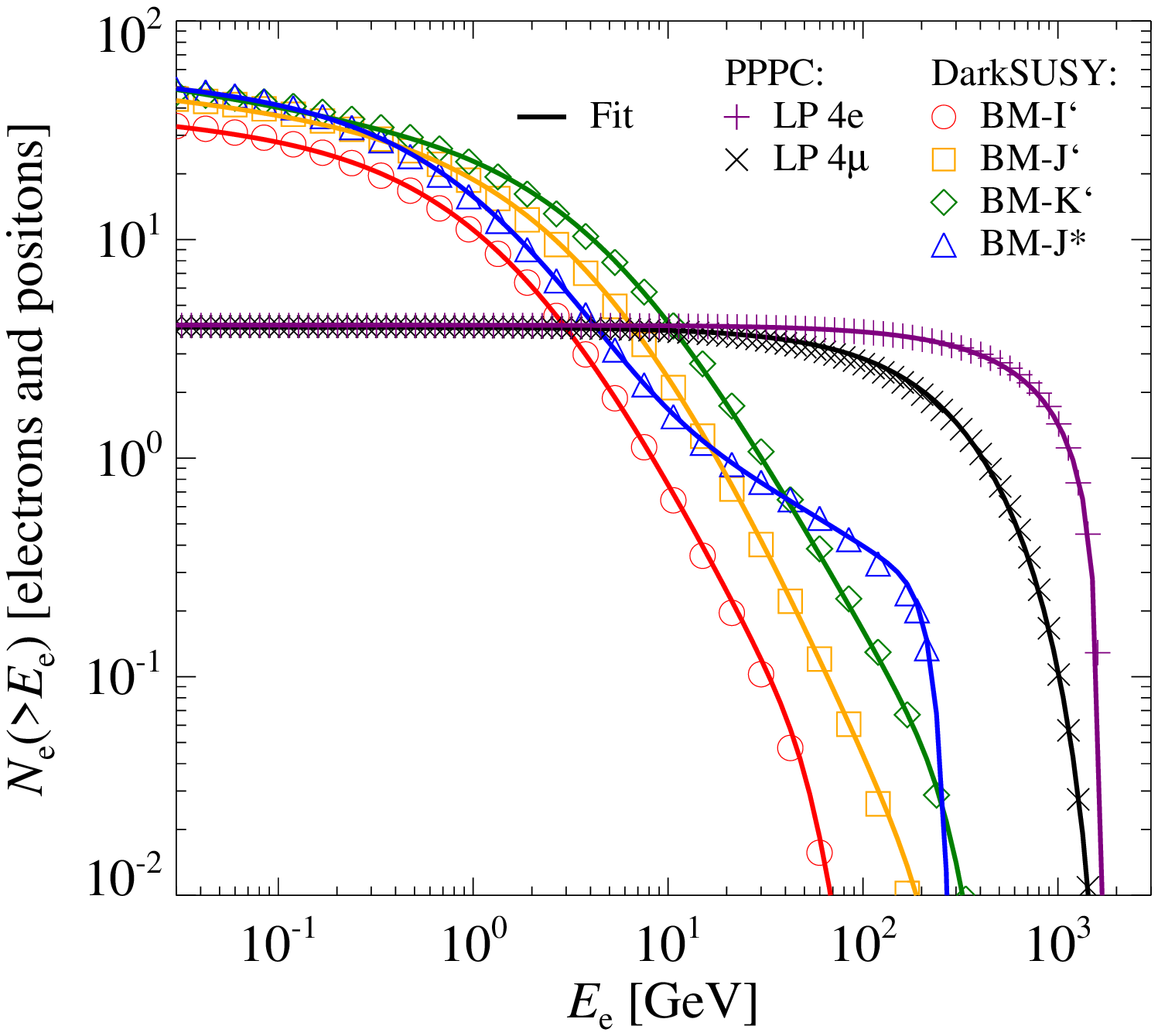}
 \caption{\colo Source functions for different dark
   matter (DM) models. We show the simulated data from {\sc DarkSUSY}
   \cite{ds}, and that generated by
   \protect\cite{2011JCAP...03..019C,2011JCAP...03..051C}; the solid
   lines show the fit to the data. Left panel: normalized differential
   continuum spectra for four different DM benchmark (BM) models;
   $\Ip$ model (red circles), $\Jp$ (orange squares), $\Kp$ (green
   diamonds), and $\Js$ (blue triangles). We use Eq.~(\ref{eq:bm_cont})
   to fit the continuum spectra. Right panel: number of electron and
   positron per DM annihilation above the electron energy $\ee$ for
   different DM models; $\Ip$ BM model (red circles), $\Jp$ BM model
   (orange squares), $\Kp$ BM model (green diamonds), and $\Js$ BM
   model (blue triangles), leptophilic (LP) DM annihilating indirectly
   into electrons and positrons (purple $+$) and into muons (black
   $\times$) without considering Sommerfeld boosts. We use
   Eq.~(\ref{eq:bm_elec}) and Eqs.~(\ref{eq:lp_mu}-\ref{eq:lp_ep}) to
   fit the spectra of electrons and positrons from BM and LP models,
   respectively.}
 \label{fig:q_DM}
\end{minipage}
\end{figure*}

The differential energy spectrum for the IB process is more
complicated (see \cite{1989PhLB..225..372B,2008JHEP...01..049B}), but
can be computed using {\sc DarkSUSY}. The differential photon energy
spectra for all our BM models are shown in Fig.~\ref{fig:q_DM}. The
peaking at high photon energy and the resulting bump caused by inner
bremsstrahlung is clearly seen for BM models $\Ip$, $\Jp$, and
$\Js$. Despite the different particle masses and cross sections, the
spectrum of the continuum emission shows a remarkably similar shape
which suggests that a simple rescaling of particle masses (responsible
for the exponential cutoff in the spectrum) and cross sections
(spectral amplitude) should yield a roughly scale-invariant continuum
emission spectrum except for the presence of the final state emission
feature which depends on the details of the specific decay
channels. The right panel of Fig.~\ref{fig:q_DM} shows the cumulative
number of leptons (electrons and positrons) above a given energy for
our different DM models which is of interested for the high-energy IC
emission of those leptons. Only LP models have energetic enough
electrons such that the IC emission is powerful enough to either be
constrained or detected at GeV energies and higher.

\subsection{Astrophysical modeling of DM induced emission}
\label{sect:AP}

We now turn to the detailed modeling of the surface brightness
profiles of DM annihilation emission and discuss the DM profiles for
the smooth distribution and substructures.

\subsubsection{General equations}

The differential photon flux within a given solid angle $\Delta
\Omega$ along a line-of-sight (los) is given by
\begin{equation}
\label{eq:dflux}
\frac{\dd \fg}{\dd \eg} \equiv \frac{\dd^3 \ngamma}{\dd A \,\dd t\, \dd
  \egamma} = \frac{1}{2}\int_{\Delta\Omega} \dd\psi \sin\psi \frac{\dd \sg}{\dd \eg}(\psi,\eg)\,,
\end{equation}
where
\begin{equation}
\label{eq:dtildeflux}
\frac{\dd \sg}{\dd \eg}(\psi,>\eg) = \int_{\Delta\Omega} \dd\Omega \int_{\rm los}
\dd l\, q_{\rm sum}\left(\eg,r\right)\Lambda(\theta)\,,
\end{equation}
and $\sg(\psi, >\eg)$ denotes the surface brightness above the photon
energy $\eg$. The integration along the line-of-sight $l$, in the
direction $\psi$ that the detector is pointing, is parametrized such
that the radius of the source $r=\sqrt{l^2+D_\clu^2-2 D_\clu
  l\cos\Psi}$, where $D_\clu$ is the distance from the Earth to the
center of the cluster halo and
$\cos\Psi\equiv\cos\theta\cos\psi-\cos\varphi\sin\theta\sin\psi$. The
angular integration $\dd \Omega= \sin\theta\dd \theta \,\dd \varphi$
is performed over a cone centered around $\psi$ and the opening angle
$\Delta \Omega$ is typically taken to be a few times the point spread
function (PSF) $\psf$. The limited angular resolution results in a
probability that a photon coming from a direction $\psi$' is instead
reconstructed to a direction $\psi$, where the underlying probability
distribution follow a Gaussian:
\begin{equation}
\Lambda(\theta)=\frac{1}{2\pi\psf^2}
\,\rmn{exp}\left[-\frac{\theta^2}{2\psf^2}\right]\,,
\quad \rmn{where}\quad \theta=\psi'-\psi \,.
\end{equation}
We denote the total differential source function by $q_{\rm
  sum}(\eg,r)$, where we include contributions from five main
processes; leptophilic DM annihilating through
$\chi\,\chi\to\phi\,\phi\to\{4e\mbox{~or~}4\mu\to4e
\mbox{~or~}4\pi\to4e\}$ (neglecting the produced neutrinos) where the
$\rmn{e}^+/\rmn{e}^-$ pairs IC upscatter background photons (LP-IC),
leptophilic DM emitting final state radiation (LP-FSR), supersymmetric
DM BM models where annihilating neutralinos generate
$\rmn{e}^+/\rmn{e}^-$ pairs that upscatter background photons (BM-IC)
and emit a continuum as well as final state radiation (BM-Cont), and
CR proton induced $\pi^0$ that decays into gamma-rays
(CR-$\pi^0$). The source function is given by
\begin{equation}
q_\rmn{sum} (\eg,r) = \qCR(\eg,r)+
\sum_i \,q_{\rmn{sm},i}(\eg,r)\,\B_{\rmn{tot},i}(\sigv,r)
\end{equation}
where the differential CR to gamma-ray source function is denoted
by $\qCR(\eg,r)$ (see Sec.~\ref{sect:CRs} and
\cite{2010MNRAS.409..449P} for further details). The subscript $i$
runs over the gamma-ray producing DM channels and the total
differential boost factor for DM is given by:
\begin{eqnarray}
\B_{\rmn{tot},i}(r,\sigv) = \left\{\begin{array}{cc}
\B_\sfe(\sigv)\,\B_{\sub,i}(r) &\rmn{for\,\,LP}\\
\B_{\sub,i}(r) &\rmn{for\,\,BM\,.}\end{array}\right.
\end{eqnarray}
It is the product of enhancement factors from SFE $\B_\sfe(\sigv)$
(see Eq.~\ref{eq:B_sfe} and Sec.~\ref{sect:LP}) and from substructure
enhancement over the smooth halo contribution $\B_{\sub,i}(r) =
1+\rho_{\sub,i}^2(r)/\rho^2(r)$ \footnote{Note that if the boost
    from substructures is $\gg 1$, then the Sommerfeld enhancement
    approaches the constant saturated boost of 530.} (see
Eqs.~\ref{eq:rho_sub}-\ref{eq:xvir} and Sec.~\ref{sect:subst}). The
DM source function from the smooth halo for each process is written in
the form:
\begin{equation}
\label{eq:q_sm}
q_{\rmn{sm},i} (\egamma,r) = \sum_j
\frac{\dd \ngammaj}{\dd E_\gamma} \Gamma_j(r)\,,
\end{equation}
where the annihilation rate density is given by
\begin{equation}
\label{eq:ann_rate}
\Gamma_j(r) = \frac{1}{2} \left[\frac{\rho(r)}{\mx}\right]^2
\, \sigmaannv_j\,.
\end{equation}
Here, the subscript $j$ runs over all kinematically allowed gamma-ray
producing channels, each with the spectrum $\dd
  \ngammaj /\dd\eg$ and annihilation cross section $\sigmaannv_j$.
We denote the DM mass with $\mx$ and the smooth DM density profile
with $\rho(r)$.

\subsubsection{The smooth DM density profile}
\label{sect:smooth}

 Typically the universal Navarro-Frenk-White (NFW) density
profile provides a good fit to both the observed and simulated
clusters. It can be considered as a special case of the more general
5-parameter profile:
\begin{equation}
\rho(r) = \frac{\rhos}{\left(r/r_\s\right)^\beta
  \left[1+\left(r/r_\s\right)^\alpha\right]^\delta}\,,\quad
\delta=\frac{\gamma - \beta}{\alpha}\,.
\label{eq:rho_nfw}
\end{equation}
Here, $\beta$ denotes the inner slope, $\gamma$ is the outer slope,
and $\alpha$ is the shape parameter that determines the profile shape
at the scaling radius $r_\s=\rvir/c$ that characterizes the transition
between the different power-law slopes. A cuspy NFW profile is given
by $(\alpha,\beta,\gamma)=(1,1,3)$. The characteristic overdensity for
an NFW profile is given by
\begin{equation}
\rhos(c)=\frac{200\,\rhoc}{3}\,\frac{c^3}
{\log\left(1+c\right)-c/(1+c)}\,,
\label{eq:rho_s}
\end{equation}
 where the halo mass dependent concentration parameter $c$ is derived
 from a power-law fit to cosmological simulations with $\mvir \gtrsim
 10^{10} \msun$ \cite{2008MNRAS.391.1940M},
\begin{equation}
\label{eq:cfit}
  c=3.56 \times \left(\frac{\mvir}{10^{15}\,\msun}\right)^{-0.098}\,.
\end{equation}
This mass scaling agrees well with \cite{2009ApJ...707..354Z} for
cluster-mass halos after converting the concentration definitions
according to \cite{2003ApJ...584..702H}. In this work we choose to
model the DM density by an Einasto density profile
\begin{equation}
\label{eq:dens_ein}
\rho_{\rm ein}(r) = \rho_{-2}\,\exp\left\{-\frac{2}{\alpha}
  \left[\left(\frac{r}{r_{-2}}\right)^\alpha -1 \right] \right\},\,
\alpha=0.17 \,,
\end{equation}
that is slightly shallower in the center than the conventional
Navarro-Frenk-White (NFW) profile, but provides a better fit to recent
simulated high resolution DM halos
\cite{2006AJ....132.2685M,2010MNRAS.402...21N}. It should also be
noted that recent observations of the Abell 383 galaxy cluster find a
density profile with a shallower inner slope of $\beta=0.6$ compared
to an NFW profile, and $\beta\gtrsim 1$ can be ruled out with $>95$\%
confidence \cite{2011ApJ...728L..39N}. These observations are based on
lensing and X-ray measurements as well as the stellar velocity
dispersion of the central galaxy. In Eq.~(\ref{eq:dens_ein}) we denote
the density where the profile has a slope of $-2$ by $\rho_{-2}$, and
the radius by $r_{-2}=\rvir/c$. We use that the density
$\rho_{-2}(c)=\rho_\s(c)/a(c)$ and determine $a(c)$ through $\mvir$:
\begin{equation}
\int_0^{\rvir}\dd V \rho_{\rm ein} = \mvir =
200\,\rho_{\rmn{cr}} \frac{4\pi\,\rvir^3}{3} \,.
\end{equation}
Here $a=(4.16-4.30)$ for $c=(3-10)$ and since $a$ is a slowly
increasing function with concentration $c$, we fix it for simplicity
to $a(c)\approx a \approx 4$.

In the recent dark matter simulation literature, it has become
standard to characterize halos by the value, $V_\rmn{max}$, where the
circular velocity $V_\rmn{circ}(r)=\sqrt{GM(<r)/r}$ attains its
maximum:
\begin{eqnarray}
V_\rmn{max} &=& V_\rmn{circ}(\rvir)
\left[\frac{0.216\,c}{\log(1+c)-c/(1+c)}\right]^{0.5}\propto \mvir^{0.32}.\nonumber\\
\end{eqnarray}
Especially for subhalos, this quantity seems to be more stable
compared to the virial mass that is subject to tidal stripping. For
comparison, we quote typical values for a large galaxy cluster and a
galaxy group:
\begin{eqnarray}
M_{200} = 1\times10^{15}\,\msun:\quad &&V_\rmn{max}=1480\,\rmn{km/s}, \nonumber\\
                                     &&R_\rmn{max}=0.61\,\rvir, \\
M_{200} = 4\times10^{13}\,\msun:\quad &&V_\rmn{max}=~520\,\rmn{km/s}, \nonumber\\
                                     &&R_\rmn{max}=0.44\,\rvir.
\end{eqnarray}

\subsubsection{Substructures}
\label{sect:subst}
High-resolution dissipationless DM simulations of MW type halos find
substantial amount of substructures in the periphery of DM halos,
while the substructures in the center suffer from dynamical friction
and tidal effects depleting their central number densities. Since the
DM annihilation rate depends on the density squared, the resulting
flux from substructures is boosted compared to the smooth density
distribution. While there is still a discrepancy in the literature of
the exact value of the predicted boost factor of the DM annihilation
luminosity from substructures in DM halos, this inconsistency starts
to become resolved, apparently converging toward predictions at the
high end \protect \cite{2008MNRAS.391.1685S, 2008Natur.456...73S,
  2010ApJ...718..899A}. Following recent high-resolution simulations
of Galaxy-sized halos, we adopt a boost factor due to substructures of
230 for such a halo \cite{2008MNRAS.391.1685S} (which needs to be
scaled to cluster halos as we will discuss below).

\begin{figure}
 \includegraphics[width=0.99\columnwidth]{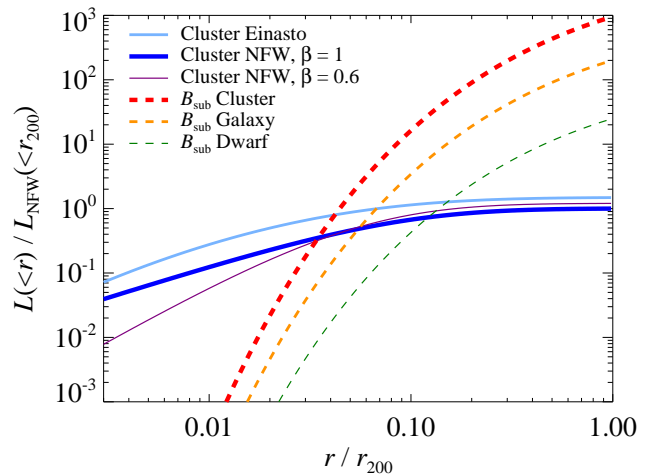}
 \caption{\colo Radial dependence of DM annihilation
   luminosity of smooth halo and substructures. The solid lines show
   the accumulative smooth luminosity from a cluster with the mass
   $\mvir=10^{15}\,\msun$ for three different density profiles; an
   Einasto profile with $\alpha=0.17$ (light blue), a cuspy NFW
   profile with $\beta=1.0$ (thick dark blue), and a core NFW profile
   with $\beta=0.6$ (thin purple). The dashed lines show the
   accumulative luminosity from substructures for three different mass
   scales; an $\mvir=10^{15}\,\msun$ galaxy cluster (thick red), an
   $\mvir=10^{12}\,\msun$ galaxy (orange), and an
   $\mvir=10^{8}\,\msun$ dwarf galaxy (thin green). All luminosities
   have been normalized with the luminosity within $\rvir$ from a
   cuspy NFW profile. We have assumed the standard value for the
   limiting substructure mass of $M_\rmn{lim}=10^{-6}\,\msun$. Note
   the large expected boost from substructures in clusters
   ($\sim1000$), and the relative small boost in dwarf galaxies
   ($\sim20$).}
 \label{fig:radial_lum}
\end{figure}

The initial suggestions of a small {\em total} boost ($<R_{200}$) of
order unity by the Via Lactea simulations \cite{2008JPhCS.125a2008K}
made assumptions of computing the boost assuming that the substructure
luminosity follows that of the smooth DM distribution. Physically,
this would imply substructure to follow the smooth DM distribution and
a radially independent concentration parameter of substructures. Both
assumptions are in conflict with tidal mass loss of satellites which
is at work in simulations and cause the radial number density to be
anti-biased with respect to the host's mass density profile yielding a
significantly flatter subhalo distribution compared to the smooth DM
distribution \cite{2010ApJ...718..899A}. This increases the
substructure boost preferentially in the outer parts of DM halos with
a factor that ranges between 20 and 1000 (for dwarfs and galaxy
clusters), hence depending sensitively on the host halo mass (see
Figure~\ref{fig:radial_lum}). Recent {\it Phoenix} simulations by Gao
et al. \cite{2011arXiv1107.1916G} confirm the large enhancement due to
substructures in clusters. In fact they find a boost of 1125 for a
Coma like cluster which agrees within 10\% with our estimated boost
(see e.g. Table~\ref{tab:flux_tab}).

We use a double power-law function to fit the luminosity from the
smooth component of substructures (i.e. substructures within
substructures are not included) inside radius $r$ which is determined
for the Aquarius simulations
\cite{2008MNRAS.391.1685S,2008Natur.456...73S}.\footnote{Our approach
  of fitting the scaling behavior of $L_\sub(<r)$ directly from
  numerical simulations self-consistently accounts for the radial
  dependence of the substructure concentration due to tidal mass
  losses \protect \cite{2008MNRAS.391.1685S}.} Our best fit is given
by
\begin{eqnarray}
  L_\sub(<x) &=& a_0\,C(\mvir)\,L_{200\sm}(\mvir)\,x^{f(x)}\,,\label{eq:Lsub}\\
   f(x)&=&a_1\,x^{a_2}\,, \nonumber\\
 a_0 &=& 0.76\,,\quad a_1 = 0.95\,,\quad a_2 = -0.27\,, \quad
 \nonumber\\
  C(\mvir) &=& \left(\frac{\mvir\,M_{\rm res,sim}}
{M_{\rm 200sim}\,M_{\rm lim}}\right)^{\alpha_C}
=0.023\,\left(\frac{\mvir}{M_{\rm lim}}\right)^{\alpha_C}.\nonumber\\
\label{eq:Csub}
\end{eqnarray}
Here, $\alpha_C=0.226$, $a_i$ denote our fit variables, $L_{200\sm}$
is the luminosity from the smooth halo without substructures within
$\rvir$ and $x= r/\rvir$. We derive the normalization function
$C(\mvir)$ in Eq.~(\ref{eq:Csub}) from the simulations in reference
\cite{2008Natur.456...73S} using a value of $M_{\rm
  200sim}=1.9\times10^{12}\msun$ for the mass of the MW halo in the
simulation and $M_{\rm res,sim}=10^5\msun$ for the mass of the
smallest resolved subhalos in the MW simulation. The smallest mass of
subhalos in reality, $M_{\rm lim}$, is determined from the free
streaming length of DM at decoupling---an effect that erases structure
on scales smaller than the free streaming length. In applying
Eq.~(\ref{eq:Csub}), we implicitly assume that the mass power-law
scaling relation is valid down to the free streaming mass of DM
halos. In the CDM universe, this is conventionally taken to be
$10^{-6}\msun$ \cite{2001PhRvD..64h3507H, 2005JCAP...08..003G} (see
\cite{2009NJPh...11j5027B} for a discussion of the range expected in
various DM models). Note that potentially the power-law could flatten
towards smaller mass scales although current simulations show no hints
of such a behavior and since Einstein's gravity is a scale-free
theory, we do not expect such a behavior on theoretical grounds
either. For DM halos more (less) massive than the MW we expect a
larger (smaller) boost from substructures, simply because of the
larger (smaller) mass range down to the minimum mass $M_\rmn{lim}$.

To motivate the scaling of the substructure luminosity boost with
limiting substructure mass in Eqs.~(\ref{eq:Lsub}-\ref{eq:Csub}), we
show how it derives from the substructure mass function, $\dd
N_\sub/\dd M_\rmn{sub} \propto M_\rmn{sub}^{-1.9}$
\cite{2008MNRAS.391.1685S,2008Natur.456...73S} and substructure
luminosity scaling with the limiting mass of satellites, $L_\sub
\propto M_\rmn{lim}^\delta$. The total luminosity of substructures scales
as
\begin{eqnarray}
  L_\rmn{sub,tot} &\simeq& L_\sub\,N_\sub \simeq L_\sub
\int_{M_\rmn{lim}} \dd M_\rmn{sub} \frac{\dd N_\sub}{\dd M_\rmn{sub}} \nonumber\\
 &\propto& M_\rmn{lim}^{\delta-0.9} \propto M_\rmn{lim}^{-0.226}\,,\quad
\rmn{for}\quad \delta=0.674\,. \nonumber\\
&&
\end{eqnarray}
Tidal truncation is responsible in shaping the substructure luminosity
scaling parameter $\delta$. Here we only sketch out qualitative
arguments and leave details and comparison to numerical simulations to
future work. First, tidal effects truncate the subhalo profile
primarily in the outer regions. As a result of this, the subhalo
acquires a steeper logarithmic slope than the canonical 3 from the
NFW-profile. Second, tidal stripping imposes a mass-dependency and
should be stronger for smaller substructure. The stripping efficiency
(for an effective cross section $\sigma$) only depends on the ambient
DM density of the host halo, $n_\rmn{dm}$, which implies a scaling of
the mean free path of a subhalo of $x_\rmn{mfp} \propto 1/(n_\rmn{dm}
\sigma)$. Since there is less mass at any given density for smaller
substructures, modest density inhomogeneities within the host halo
seen by the orbiting substructures cause a stronger mass loss of these
smaller satellites. Third, the tidal truncation radius should be a
function of host halo radius due to the increasing smooth density
profile for smaller radii. If we pick an effective host halo radius
which dominates the luminosity contribution of the substructures and
were to fit the tidally truncated subhalo density profile with an
NFW-profile, the resulting $\rs$ would be biased toward lower values
implying a higher concentration compared to an isolated halo. In order
to reproduce our value of $\delta$, we would need a concentration mass
relation of $c\propto M_\rmn{sub}^{-0.14}$ (employing the
$L(M_{200},c)$ formula derived by reference
\cite{2009PhRvL.103r1302P}). As expected, this is steeper than the
concentration-mass relation found for isolated halos on dwarf galaxy
scales, $c \propto M_\rmn{vir}^{-0.06}$ \cite{2011arXiv1101.2020I}, as
well as galaxy and cluster scales, $c \propto M_\rmn{vir}^{-0.098}$
\cite{2008MNRAS.391.1940M}.

We now derive the squared density profile for the substructures
using
\begin{eqnarray}
\rho_\rmn{sub}^2(r) &=& \frac{\dd L_\rmn{sub}}{\dd V} \frac{1}{A_{\rmn{dm}}}\,,\label{eq:rho_sub}\\
L_{200\sm} &=& A_{\rmn{dm}} \int_{\rvir} \dd r' 4\pi r'^2 \rho^2(r')\,,\label{eq:Lsm}\\
 \frac{\dd L_\rmn{sub}}{\dd V} &=& a_0\,a_1\,C(\mvir)\,x^{g(x)}\,
\left[\frac{1+a_2\log(x)}{4\pi\,r^3}\right]\,,\nonumber\\
\\
x &=& r/\rvir\,,\quad g(x) = a_2+a_1\,x^{a_2}\,.\label{eq:xvir}
\end{eqnarray}
Here, $A_{\rmn{dm}} = q_{\rmn{sm},i}(\eg,R) / \rho(R)^2$ represents
the particle physics factor, where $q_{\rmn{sm},i}(\eg,R)$ is defined
by Eq.~(\ref{eq:q_sm}). The different density profiles have some
impact on the luminosity from annihilating DM, although the details of
the density profile can be neglected compared to the dominating boost
from substructures (assuming DM to be cold). In
Fig.~\ref{fig:radial_lum} we compare the radial dependence of the
accumulative luminosity from different smooth cluster density profiles
to the boosted luminosity due to substructure for different mass
scales. We recalculate the overdensity, $\rhos$, for the shallower
density profile with $\beta=0.6$ and rescale the concentration
parameter in Eq.~(\ref{eq:cfit}) with $300/160$
\cite{2011ApJ...728L..39N} to account for the more centralized scale
radius in cluster with a shallow inner slope. The emission of this
profile with $\beta=0.6$ is about 30\% larger within $\rvir$ compared
to a cuspy NFW ($\beta=1.0$). This difference is built up within
$0.1\rvir$ (i.e. close to $r_\s$). Hence the slope of the central part
of a cluster has little influence for the DM luminosity within $\rvir$
as long as the degeneracy in the lensing mass measurements with $r_\s$
(which decreases for decreasing inner slope) has been taken into
account. The emission from an Einasto density profile is about 50\%
larger than the cuspy NFW profile in the periphery of the cluster,
where the difference is mainly built up at a few percent of
$\rvir$. The increase in luminosity due to substructures is negligible
in the center of halos, but integrated out to $\rvir$ the boost
relative to the smooth emission profile amounts to approximately 20
for dwarf galaxies, $200$ for galaxies, and $10^3$ for galaxy
clusters. We stress that these boost factors are only realized in the
region around the virial radius of each respective halo which is
mostly tidally stripped for dwarfs in the MW. Hence a more realistic
boost from substructures is probably much smaller in satellite dwarf
galaxies. In addition, these boost factors are only realized for
direct annihilation emission (continuum emission or final state
radiation) or IC scattering off homogeneous seed photon fields
(CMB). For IC scattering of SD photons, the overlap of final state
leptons and SD photons is smaller which causes the substructure boost
over the smooth emission to be reduced by roughly two orders of
magnitudes. It should also be noted that the substructure boosted
fluxes from clusters are much more extended than for dwarf galaxies,
hence more difficult to detect with Cherenkov telescopes. In fact, it
was shown in Sanchez-Conde et al. \cite{2011arXiv1104.3530S} that
inside $0.1\degs$, where the sensitivity of Cherenkov telescopes is
maximized, the expected DM flux including substructures from the
brightest dwarf galaxy is about an order magnitude higher than the
brightest cluster. However, that work assumed a substructure boost of
about 50 in the most massive cluster halos which is more than an order
magnitude smaller than what we use in this work, hence in projection
we expect this difference to be a factor few smaller.

\begin{figure}
  \includegraphics[width=0.99\columnwidth]{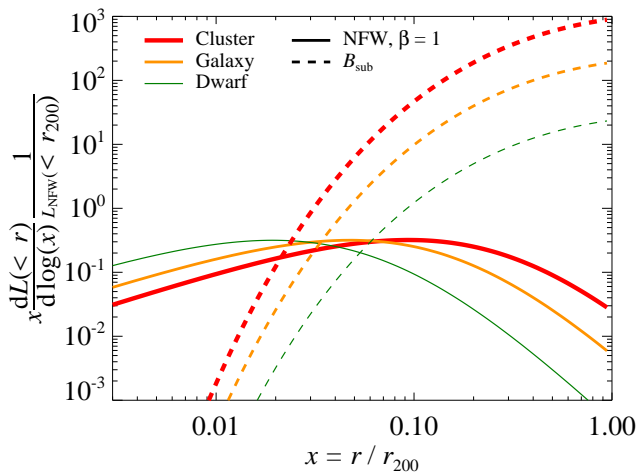}
  \caption{\colo Plot that shows where most of the DM
    annihilation luminosity originates. We show the differential
    contribution to the DM annihilation luminosity per logarithmic
    interval in radius which corresponds to $x^3\rho^2/\rho^2_{200}$
    where $\rho_{200}$ is the density at $\rvir$. The solid lines
    represent cuspy $\beta=1.0$ NFW density profiles for three
    different mass scales; an $\mvir=10^{15}\,\msun$ galaxy cluster
    (thick red), an $\mvir=10^{12}\,\msun$ galaxy (orange), and an
    $\mvir=10^{8}\,\msun$ dwarf galaxy (thin green). The dashed lines
    show the contribution from substructures for the same three mass
    scales. All luminosities have been normalized with the luminosity
    within $\rvir$ from a cuspy NFW profile. We have assumed the
    standard value for the limiting substructure mass of
    $M_\rmn{lim}=10^{-6}\,\msun$. For the smooth profile, the majority
    of the flux is delivered by a region around $r_\s/3$ as indicated
    by the maximum value of the curves. In contrast, for substructures
    the emission is dominated by regions around $r_{200}$.}
  \label{fig:radial_emis}
\end{figure}

In Fig.~\ref{fig:radial_emis} we show the radial regions that dominate
the DM annihilation luminosity, in particular we show the differential
contribution to the DM luminosity per logarithmic interval in radius
for three different mass scales. Solving for the maximum of this
curve, $\dd^2L_{200\sm}/\dd (\log x)^2=0$ in combination with
Eq.~(\ref{eq:Lsm}), we find that the luminosity from the smooth NFW
profile peaks at $r \simeq \rvir/3c$. Despite the cuspy nature of the
density profile, the luminosity is not dominated by the central region
but by the transition region where the profile steepens because of the
larger volume available there. For large clusters with typical
concentrations of $c = \rvir/r_\s \simeq 4$, the luminosity from the
smooth profile is focused to the regime around 10\% of $\rvir$. In
contrast, emission from substructures is mainly contributed by the
outer parts of DM halos. As shown in Fig.~\ref{fig:radial_emis}, the
product of annihilation emissivity and emission volume increases
towards $\rvir$ and only starts to drop outside this radius. Note
that, even though most of the substructure mass density has been
erased in the central regions of DM halos, a cluster in projection has
a significant enhancement due to substructures at a radius of just a
few percent of $\rvir$.

\subsection{Inverse Compton emission}
\label{sect:IC}

In this section we outline the basics of inverse Compton (IC)
emission. As target radiation fields we consider cosmic microwave
background (CMB) photons, and the light from stars and dust (SD). We
derive an analytic model from which we can estimate the spectral
and spatial distributions of SD as a simple function of cluster mass.

The standard IC source function is given by
\cite{1979rpa..book.....R}:
\begin{eqnarray}
  q_{\rmn IC}(\eg, r) &=& \frac{\dd^3 N_\gamma}{\dd V\,\dd t\,\dd \eg} =
 \frac{3}{4}c\,\sigma_\rmn{T}
\int\dd \eph \frac{n_\rmn{ph}(\eph)}{\eph}\nonumber\\
&\times& \int \dd \ee\,\frac{\dd n_\e}{\dd \ee}(\ee,r)\,
 \frac{\left(m_\e c^2\right)^2}{\ee^2}G(\Gamma_\e,q)\,,\nonumber\\
  \label{eq:ICemiss}
\end{eqnarray}
where $\ee$ is the energy of the upscattering electrons and $\eph$ is
the energy of the background photon field. We represent the Thomson
cross section with $\sigma_\rmn{T}$ and the full differential
Klein-Nishina (KN) cross section is captured by
\cite{1970RvMP...42..237B}:
\begin{equation}
\label{eq:KN_spec}
G(\Gamma_\e,q) = 2q\ln{q}+(1+2q)(1-q)+
\frac{1}{2}\frac{\left(\Gamma_\e q\right)^2\left(1-q\right)}
     {1+\Gamma_\e q}\,,
\end{equation}
where
\begin{equation}
\Gamma_\e=\frac{4\eph \ee}{\left(m_\e c^2\right)^2}\,,\qquad \rmn{and} \qquad
q=\frac{\eg}{\Gamma_\e\left(\ee-\eg\right)}\,.
\end{equation}
The full KN cross section accounts for the less efficient energy
transfer between the photon and electron once the energy of the
Lorentz-boosted photon in the electron rest frame comes close to $m_\e
c^2$ such that the scattering electron experiences a significant
recoil. This results in a steepening of the IC gamma-ray spectrum. In
the low energy Thomson regime the IC spectrum $F_\gamma\sim
E_\gamma^{-(\alpha_\e-1)/2}$, however when $\Gamma_\e \gg 1$ the IC
spectrum steepens due to the KN suppression to
$\eg^{-\alpha_\e}\log(\eg)$. Here we denote the steady state electron
spectrum by $\alpha_\e$.

We account for two major contributions to the number density of
radiative background fields $n_\ph$; the CMB photons and the infra-red
(IR) to ultra-violet (UV) light emitted by SD. The number density for
the SD is given by $n_\ph\equiv\dd^2 N_\ph/ (\dd V \,\dd \eph)
(\eph,r)= u_\sd(\eph,r) /\eph^2$ where the specific SD energy
density $u_\sd(\eph,r)$ is given in the Appendix,
Eq.~(\ref{eq:u_SD_er}). We model the CMB photon spectrum as a photon
gas that is isotropically distributed and follows a black body
spectrum with the temperature $T=2.73\,$~K:
\begin{equation}
\label{eq:photon_gas}
  n_\rmn{ph}(\eph) = \frac{\dd^2 N_\ph}{\dd V \,\dd \eph} =
  \frac{1}{\pi^2(\hbar c)^3}\frac{\eph^2}{\exp(\eph/k_\B T)-1}\,.
\end{equation}
Note that the typical energy of a black body photon before scattering
is given by $\langle\eph\rangle=\epsilon_\ph/\tilde{n}_\ph\approx\,2.7\, k_\B T$,
where $\tilde{n}_\ph$ and $\epsilon_\ph$ are the number- and
energy-density derived by integrating $n_\ph(\eph)$ and $\eph
n_\ph(\eph)$ over the photon energy $\eph$, respectively.

The electrons injected from annihilating DM also suffer from diffusive
and radiative losses. Hence we have to calculate the equilibrium
spectrum of the electrons plus positrons denoted by $\dd n_\e/\dd \ee$
in Eq.~(\ref{eq:ICemiss}). We derive this stationary solution using
the cosmic ray transport equation (neglecting convection and
re-acceleration effects):
\begin{eqnarray}
\label{eq:CRdiffusion}
\frac{\partial}{\partial t}\left(\frac{\dd n_\e}{\dd \ee}\right) =
&\nabla& \left[D\left(\ee,\bx\right)\nabla\frac{\dd n_\e}{\dd \ee}\right] +
\frac{\partial}{\partial \ee}
\left[b\left(\ee,\bx\right) \frac{\dd n_\e}{\dd \ee}\right]
 \nonumber \\
&+& q_\e(\ee,\bx)\,,
\end{eqnarray}
where $D(\ee,\bx)$ denotes the diffusion coefficient and $b(\ee,\bx)$
the energy loss term. The source function $q_\e(\ee,\bx)$ yields the
number of electrons and positrons produced per unit time, energy and
volume element at the position $\bx$:
\begin{equation}
q_\e(\ee,r)=\sum_f\frac{\dd N_\e^f}{\dd \ee}(\ee) B_f \Gamma_f(r) \,,
\end{equation}
where the annihilation rate density $\Gamma_f(r)$ is defined in
Eq.~(\ref{eq:ann_rate}). The sum runs over the kinematically allowed
annihilation final states $f$, each with a branching ratio $B_f$ and a
differential spectrum $\dd N_\e^f/\dd \ee$ that represents the number
of electrons plus positrons resulting from an annihilation event. We
use the differential spectra derived from high-statistics simulations
in \cite{2011JCAP...03..019C,2011JCAP...03..051C} to compute the
cumulative number of electrons and positrons resulting from
neutralinos annihilate indirectly into $e^+/e^-$ pairs as well as
$\mu^+/\mu^-$ pairs in the LP model. We use \ds to compute the
$e^+/e^-$ spectra from our four BM models where only a fraction of the
annihilating neutralinos is converted into electrons and positrons
(see Sec.~\ref{sect:PF} and Fig.~\ref{fig:q_DM} for further details).

The electrons and positrons loose their energy on a timescale that is
shorter than the diffusive timescale in the ICM of galaxy clusters for
cosmic ray electrons which is larger than the Hubble time in our
energy range \cite{1997ApJ...487..529B,2011A&A...527A..99E}. Hence, we
neglect the first term of the r.h.s. in Eq.~(\ref{eq:CRdiffusion}),
and derive an expression for the equilibrium number density:
\begin{eqnarray}
{\frac{\dd n_\e}{\dd \ee}}\left(\ee, r \right) &=&
 \frac{1}{b(\ee, r)} \int_{\ee}^{\mx c^2} \dd \ee' \,
  q_\e(\ee', r),
\label{eq:nds}\\
b(\ee,r) &=& \tilde{b}
\left[\frac{B^2_{\rm CMB}}{8\pi}+\frac{B^2(r)}{8\pi}+u_\sd(r)\right] \ee^2\,,
\\\tilde{b}&=&\frac{4\sigma_{\rm T}c}{3(m_{\rm e}c^2)^2}\,.
\end{eqnarray}
Here we include the three main radiative loss processes for the cosmic
ray electrons and positrons: (1) IC losses on CMB photons with the
equivalent field strength of the CMB of $B_{\rm
  CMB}=3.24\mu\rmn{G}\,(1+z)^2$, where $z$ is the cosmological
redshift. (2) Synchrotron losses on ambient magnetic fields where we
parametrize the magnetic field in the galaxy cluster by $B(r) =
3\mu\rmn{G} \,[n_{\rm e}(r)/n_{\rm e}(0)]^{\alpha_\B}$. We adopt a
magnetic decline of $\alpha_\B=0.7$ in this work which follows from
flux frozen magnetic fields. (3) IC losses on starlight and dust with
an energy density $u_\sd(r)$ given by Eq.~(\ref{eq:u_SD_r}), where
  we outline the derivation in the following.

The emission of galaxy clusters at infra-red (IR) and ultra-violet
(UV) wavelengths emerges from dust and starlight in both the galaxies
and the ICM (e.g. \cite{2006ApJ...648L..29P} and
\cite{2009MNRAS.399.1694G}). Three distinctive components dominate
these wavelengths: a central galaxy, the intra-cluster light (ICL),
and individual cluster galaxies. We decided to use the accurately
measured spectral shape of dust and starlight in the interstellar
medium of our Galaxy to model the emission from clusters. We then
normalize the two spectral components -- far IR dust and starlight at
wavelengths ranging from the near IR to UV -- individually by using
stacked cluster data and employ a measured mass-to-starlight
luminosity scaling relation derived from observations of the brightest
cluster galaxy (BCG) \cite{2010ApJ...713.1037H}.

\begin{figure}
 \includegraphics[width=0.99\columnwidth]{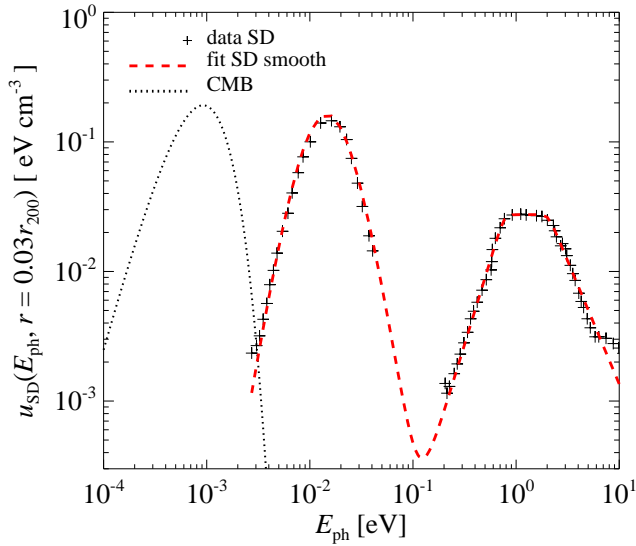}
 \caption{\colo Spectral dependence of radiation fields
   in a cluster of galaxies. The black dotted line of the left peak
   show the spectrum of CMB photons using a black body with a
   temperature of $2.73$~K. The crosses of the middle and right peaks
   represent the measured spectra from stars extending from the near
   IR to UV and dust at far IR wavelengths (SD), respectively, and are
   derived in \protect \cite{2006ApJ...648L..29P} for a galaxy. We
   normalize the individual SD spectrum separately using the observed
   luminosity from SD in clusters. The SD luminosity is related to the
   cluster mass through Eqs.~(\ref{eq:E_SD}-\ref{eq:N_dust}), where we
   use have used the mass $\mvir=6.0\times10^{14}\msun$ in this
   figure. We renormalize the SD spectra to the radius $r=0.03\rvir$,
   where the smooth energy density of the SD light (see
   Fig.~\ref{fig:SD_Edens}) equals the energy density of the CMB. The
   red dashed lines show the fitted SD spectral model.}
 \label{fig:SD_spectra}
\end{figure}

\begin{figure}
 \includegraphics[width=0.99\columnwidth]{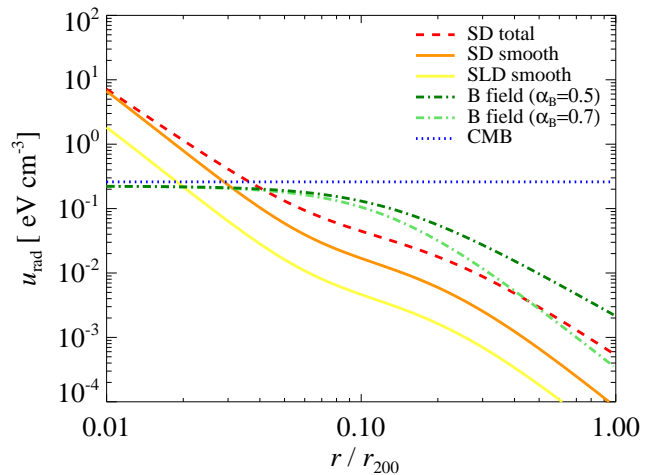}
 \caption{\colo Spatial dependence of the energy density
   of radiation fields in a cluster of galaxies. The energy density of
   the CMB (blue dotted line) is isotropic with
   $u_\rmn{cmb}=0.26\,\rmn{eV}\,\rmn{cm}^{-3}$. The energy density of
   the light from stars and dust (SD) is denoted by the red dashed
   line and the solid orange line for the total SD light and the
   smooth SD light, respectively. For comparison we show the energy
   density of the stars and a low dust model (SLD) with the solid
   yellow line. The SD light has been renormalized to a cluster with
   the mass $\mvir=6.0\times10^{14}\msun$. Finally we show the energy
   density of two magnetic field models with a central magnetic field
   of 3~$\mu G$. The magnetic field scales with the gas density to the
   power $\alpha_\rmn{B}$; dark green dash-dotted line
   ($\alpha_\rmn{B}=0.5$) and light green dash-dotted line
   ($\alpha_\rmn{B}=0.7$). Note that the SD radiation is dominating
   the energy density inside $\sim0.03\,\rvir$.}
 \label{fig:SD_Edens}
\end{figure}

In Fig.~\ref{fig:SD_spectra} we characterize the spectral shape
through a fit to the galactic spectra presented in
\cite{2006ApJ...648L..29P}. The figure is showing the spectra at
$r=0.03\rvir$, which is the radius there the SD energy density of a
galaxy cluster equals the energy density of the CMB black-body
distribution. Inside this central radius the SD component is
dominating, which is shown in Fig.~\ref{fig:SD_Edens} where we compare
the energy densities from different radiation fields in a galaxy
cluster with the mass $\mvir=6.0\times10^{14}\msun$. For this figure we
use two different profiles for the SD energy density, where the total
profile includes the contribution from the ICL, the BCG and all the
galaxies, while the galaxies are excluded in the smooth profile. To
compute the IC emission from SD, we require a non-negligible overlap
of the relativistic lepton distribution resulting from DM annihilation
and SD. In fact, one can show with a simple order of magnitude
calculation that the overlap, $f_\rmn{IC-ol}$, of the photon field of
individual galaxies (starlight and dust emission) and the smooth DM
density is very small so that we can neglect the starlight
contribution from galaxies to the IC emission for the reminder of this
work,
\begin{eqnarray}
\label{eq:SD_overlap}
f_\rmn{IC-ol} = \frac{N_\rmn{gal} V_\rmn{gal} f_\rmn{light}}{V_\rmn{clu}}
\lesssim \frac{N_\rmn{gal} M_\rmn{gal}}{M_\rmn{clu}\,c_\rmn{gal}^{3}}=10^{-4}.
\end{eqnarray}
Here we assume that the exponential scale height of the stellar light
is less than the scale radius of the galaxy halo, $f_\rmn{light}
\lesssim (r_\rmn{s} / r_{200,~\rmn{gal}})^3 \sim c_\rmn{gal}^{-3} \sim
10^{-3}$ and $N_\rmn{gal} M_\rmn{gal} \sim 0.1 M_\rmn{clu}$. We denote
the number of galaxies within $\rvir$ with $N_\rmn{gal}$ and the
concentration of a galaxy DM halo with $c_\rmn{gal}$.

We find that even for a cluster with a relative small mass the energy
density of the SD components dominates over the CMB and the magnetic
fields (with a central B field of 3~$\mu G$) within about 10\% of
$\rvir$. Outside this radius the CMB is dominating the energy density
of the cluster. Note that while the magnetic field is always
subdominant in the cluster for our assumptions, we keep its
contribution to the total energy density for consistency. Also note
that we extract the spatial distribution of the SD light in clusters
from a stacked emission analysis of Sloan Digital Sky Survey (SDSS)
data at the redshift $\sim 0.25$ \cite{2005MNRAS.358..949Z} and do not
attempt to correct for a potential evolution in this component
(c.f. Fig.~\ref{fig:SD_spatial}).

\begin{figure}
 \includegraphics[width=0.99\columnwidth]{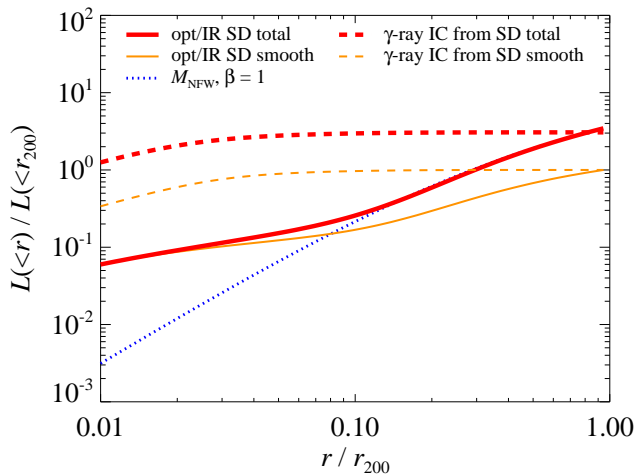}
 \caption{\colo Comparing the spatial dependence of emission from
   stars and dust (SD) to IC upscattered SD light.  We show the
   luminosity without substructures inside radius $r$. The solid lines
   show the optical/IR emission from SD and the IC upscattered SD
   light is shown by the dashed lines. The thick red lines show the
   total emission including the brightest cluster galaxy, the
   intra-cluster light and the additional intra-cluster galaxies,
   while these galaxies have been cut out in the smooth component
   shown by the thin orange lines. We normalize the SD light and the
   IC upscattered SD with the smooth luminosity within $\rvir$ for
   each component. We find that the gamma-ray IC luminosities are
   dominated by the central regime, while the SD light is mainly build
   up in the outer parts of the cluster. For comparison, we show that
   the SD light traces the NFW mass profile (dotted blue) outside
   $0.1\rvir$.}
 \label{fig:SD_lum}
\end{figure}

The spatial distribution of far-IR to UV light emitted by SD is quite
different from what is expected for the IC upscattered SD photons. We
illustrate this in Fig.~\ref{fig:SD_lum}, where the accumulative
luminosity from SD is compared to the IC SD photons for a
$\mvir=6.0\times10^{14}\,\msun$ cluster where the boost from
substructures is excluded. The gamma-ray luminosity from IC
upscattered SD photons is dominated from the inner parts where there
is the largest overlap of the cuspy DM profile and the peaked SD
distribution. This is in marked contrast to the accumulative SD
luminosity in the optical that is dominated by the outer parts of the
cluster. Interestingly, it rises as a function of radius and follows
the distribution of an NFW mass profile outside $0.1\rvir$. In the
center, the electrons and positrons mainly cool by IC upscattering SD
photons, hence the spatial dependence of the cooling cancels the
distribution of the SD source function, which results in a
distribution that approaches a density square profile in the center.
We also find that the luminosity within $\rvir$ from the total SD
model is a factor three larger compared to the more realistic smooth
SD model.


\subsection{Cosmic-ray induced gamma-ray emission}
\label{sect:CRs}

In supernovae remnants and on scales of galaxies, there are convincing evidences
of non-thermal populations. Especially, in the MW, the cosmic rays are observed
directly as well as indirectly through radio, X-ray, and gamma-ray emission. On
larger scales of the order of few 100~kpc up to Mpc, there are currently a vast
number of observations of radio emission coming from radio mini halos in the
centers of cooling flow clusters, radio relics in the periphery of clusters
\cite{2004rcfg.proc..335K}, as well as giant radio halos amounting to a total
number of more than 50 diffuse radio sources in clusters
\cite{2003ASPC..301..143F,2008SSRv..134...93F}. This type of emission is
expected in clusters since the formation process of a galaxy cluster is a very
energetic processes that induces both turbulence as well as frequently occurring
merging and accretion shocks which both are thought to accelerate relativistic
non-thermal protons and radio emitting electrons to high energies. The precise
origin of these electrons, in especially relics and giant radio halos, is still
not settled. One possible scenario for the production of these electrons are
hadronic CR interactions with ambient gas protons which results in charged and
neutral pions that decay into electrons, neutrinos, and gamma-rays (see
\cite{1980ApJ...239L..93D,1982AJ.....87.1266V,1999APh....12..169B,2000A&A...362..151D,2001ApJ...559...59M,2003MNRAS.342.1009M,2004A&A...413...17P,2004MNRAS.352...76P,2008MNRAS.385.1211P,2008MNRAS.385.1242P,2009JCAP...09..024K,2010MNRAS.401...47D,2010MNRAS.407.1565D,2010ApJ...722..737K,2011A&A...527A..99E}
).\footnote{An alternative scenario is the second order turbulent
  re-acceleration through the interaction of a previously injected relativistic
  population of electrons by supernova driven winds or active galactic nuclei
  with plasma waves and magneto-turbulence. \protect
  \cite{1987A&A...182...21S,1993ApJ...406..399G,2004MNRAS.350.1174B,2005MNRAS.363.1173B,2007MNRAS.378..245B,2011MNRAS.410..127B,2009A&A...507..661B}}
Supporting evidence comes from the smoothness of the extended radio emission
that often resembles that observed in thermal X-rays. This can be easily
explained in the hadronic model since the long cooling time of cosmic ray
protons (CRs) of order the Hubble time allows for a cluster-filling population
of CRs to build up over the formation history \cite{1996SSRv...75..279V,
  1997ApJ...477..560E, 1997ApJ...487..529B}. The production of these secondaries
depend both on the gas and CR densities in the cluster, where the CR density
roughly traces the gas outside the core regime and is slightly enhanced in the
center. This density scaling implies that clusters are great targets for
Cherenkov telescopes with a high sensitivity for the central parts of nearby
clusters. Detecting the cluster gamma-ray emission is crucial in this respect as
it potentially provides the unique and unambiguous evidence of CR populations in
clusters through observing the $\pi^0$ bump at about 100~MeV in the spectra.

We adopt the universal spectral and spatial gamma-ray model developed by Pinzke
\& Pfrommer \cite{2010MNRAS.409..449P} to estimate the emission from decaying
$\pi^0$:s that dominates over the IC emission from primary and secondary
electrons above 100~MeV in clusters. The gamma-ray formalism was derived from
high-resolution simulations of clusters of galaxies that included radiative
hydrodynamics, star formation, supernova feedback, and followed CR physics using
a novel formulation that trace the most important injection and loss processes
self-consistently while accounting for the CR pressure in the equation of motion
\cite{2008A&A...481...33J,2007A&A...473...41E,2006MNRAS.367..113P}.  We
highlight two main uncertainties of the models, namely the acceleration
efficiency at formation shocks and CR transport parameters. First, we note that
the overall normalization of the CR and gamma-ray distribution scales with the
maximum acceleration efficiency at structure formation shock waves. Following
recent observations at supernova remnants \cite{2009Sci...325..719H} as well as
theoretical studies \cite{2005ApJ...620...44K}, we assume the maximally allowed
acceleration efficiency for {\em strong} shocks that transfers 50\% of the
shock-dissipated energy (kinetic energy corrected for adiabatic compressional
heating at the shock) to CRs. These efficiencies drop quickly for weaker shocks
\cite{2007A&A...473...41E} which dominate the gravitational energy dissipation
inside galaxy clusters \cite{2006MNRAS.367..113P}. These efficiencies will have
to be corrected down if this value is not realized at strong structure formation
shocks. Second, these simulations (and by extension the analytic model) neglect
active CR transport such as streaming and diffusion relative to the gas, i.e.,
we assume that advective transport dominates and CRs are tightly coupled to the
gas via magnetic fields tangled on sufficiently small scales which produces
centrally enhanced profiles. However, CR diffusion and streaming tends to drive
the CR radial profiles towards being flat, with equal CR number density
everywhere. While the CR streaming velocity is usually larger than typical
advection velocities and becomes comparable or lower than this only for periods
with trans- and supersonic cluster turbulence during a cluster merger. As a
consequence a bimodality of the CR spatial distribution may result with merging
(relaxed) clusters showing a centrally concentrated (flat) CR energy density
profile \cite{2011A&A...527A..99E}. This translates into a bimodality of the
expected diffuse radio and gamma-ray emission of clusters, since more centrally
concentrated CR will find higher target densities for hadronic CR proton
interactions \cite{2011A&A...527A..99E}. As a result of this, relaxed clusters
could have a reduced gamma-ray luminosity by up to a factor of five.

\section{Gamma-ray spectra}
\label{sect:spectral}
Spectrally resolved indirect DM searches have the advantage of probing
different DM models through their characteristic spectral
distributions. To make current and future DM searches more effective
it is important to know in which energy band to focus the efforts in
order to maximize potential DM signals over the expected background.

We focus in this section on the spectral distribution of gamma-rays
from clusters. In the LP model, DM annihilation radiation includes
final state radiation and gives rise to substantial amounts of
electrons and positrons that IC upscatter background radiation fields
to high energies. We also consider four supersymmetric DM models with
a high gamma-ray yield in the form of continuum emission and IC
induced emission. In addition to the annihilating DM, we estimate the
gamma-ray flux induced by shock-accelerated CRs. Note that all fluxes
in this section are derived within an angle corresponding to $\rvir$
and are convolved with a PSF of $0.1\degs$.

We compare the calculated fluxes to gamma-ray upper limits (95\% c.l.)
set by \Fermi-LAT after 18 months of observations
\cite{2010ApJ...717L..71A}. In particular, to achieve a more reliable
comparison we adopt, if nothing else is stated, the maximal spatially
extended limits since the gamma-ray flux from our brightest clusters
all have an angular extent $>1^\circ$ on the sky when the boost from
substructures is included. In fact, published \Fermi upper limits are
not derived for the kind of extended emission that we find with our
treatment of substructures, hence these \Fermi limits may not be
adequate for some of these clusters. This and other recent work
highlights that the improved substructure models imply that \Fermi
limits will have to be re-calculated to accommodate for the large
source extensions. Note, however, that for most clusters the
assumption of a point source is well justified. The flux upper
limits, that we compare to, are a function of spectral index
$\alpha$. However, in the relevant energy range, $0.1-100$~GeV, the
spectral index varies within $1.5 < \alpha < 3.0$. This changes the
photon flux upper limits by $<50\%$, with \Fermi-LAT being more
sensitive to a hard spectrum \cite{2010ApJ...717L..71A}.

\begin{figure*}
\begin{minipage}{2.0\columnwidth}
 \includegraphics[width=0.49\columnwidth]{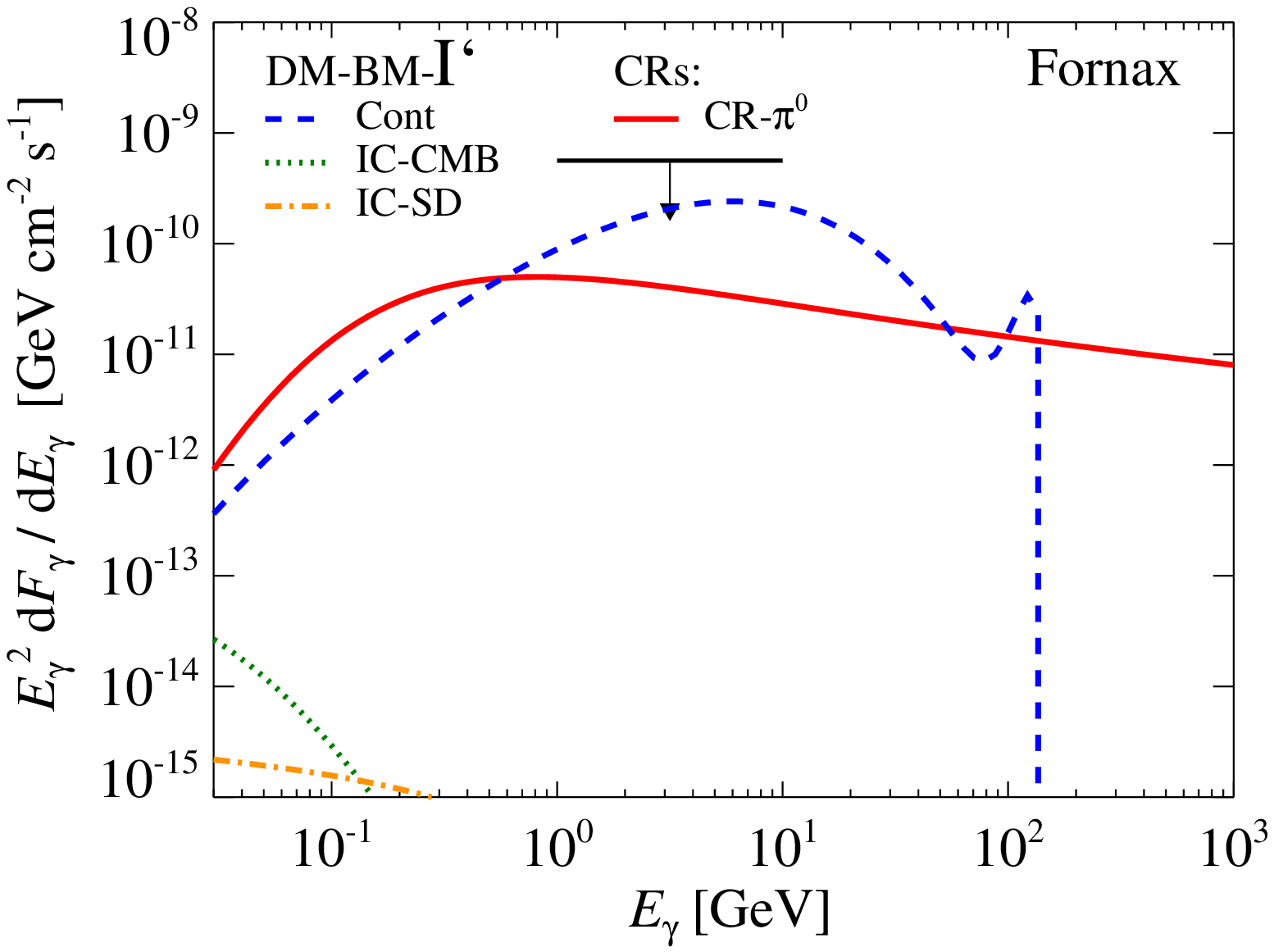}
\includegraphics[width=0.49\columnwidth]{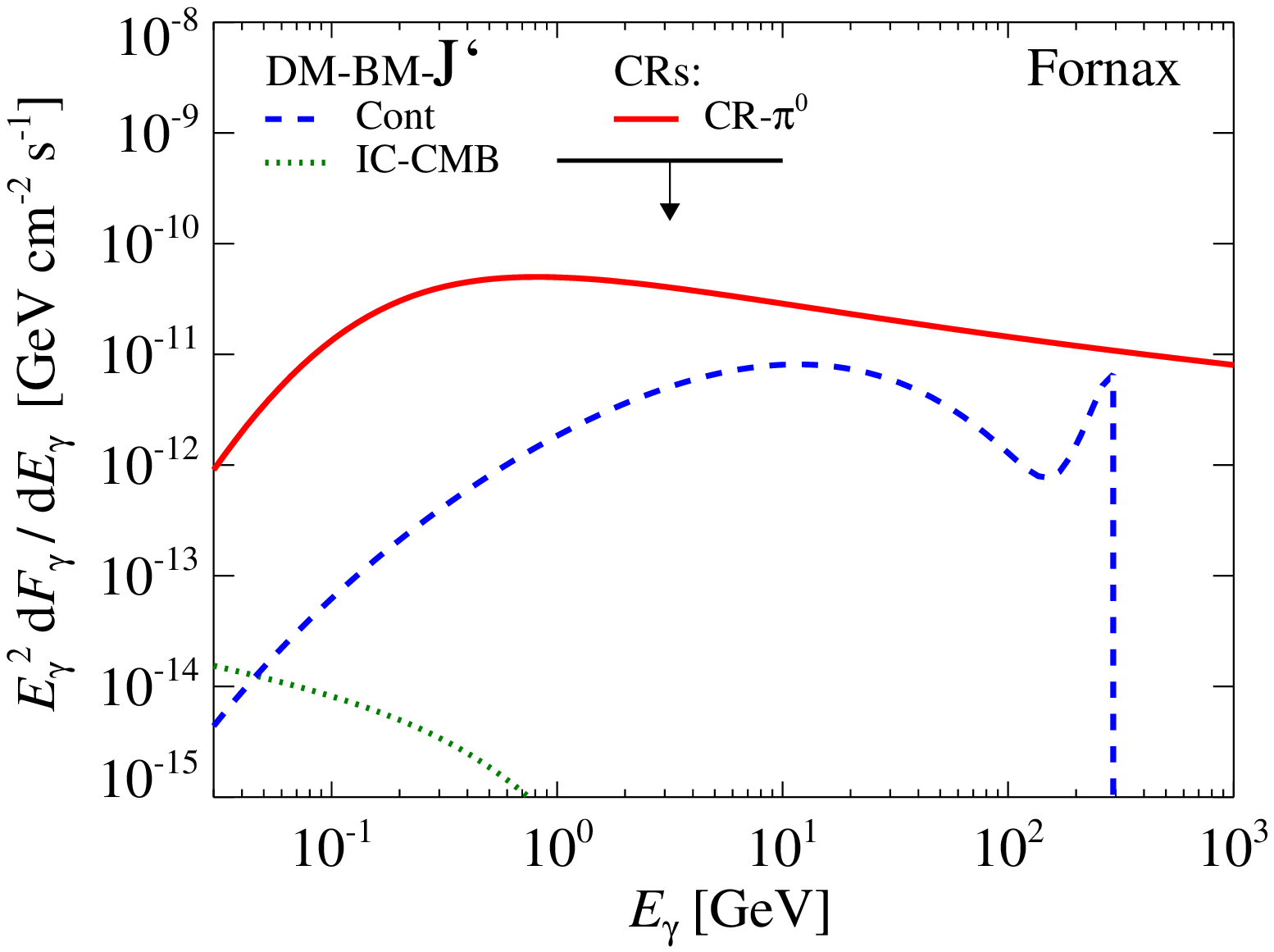}
\includegraphics[width=0.49\columnwidth]{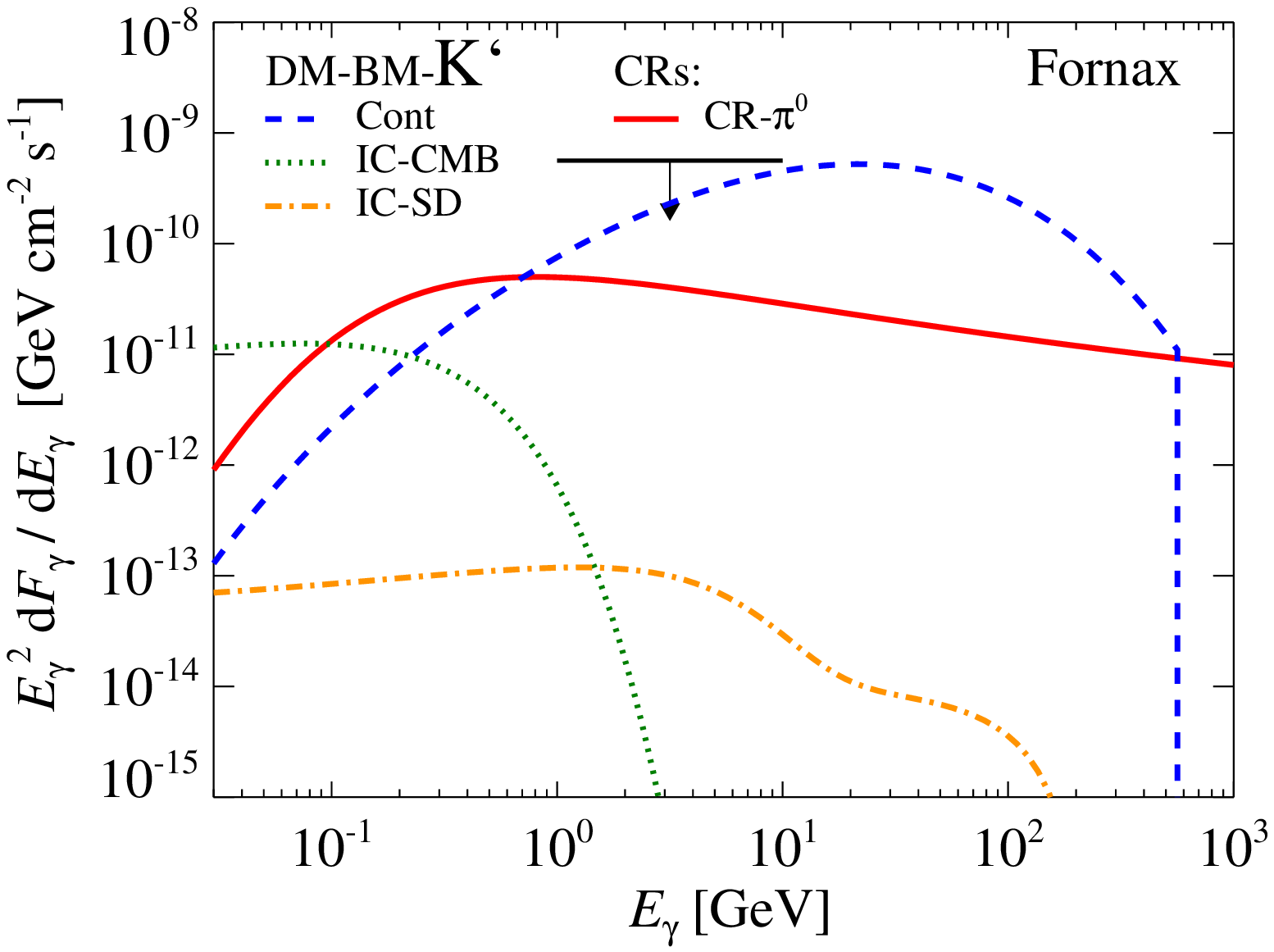}
\includegraphics[width=0.49\columnwidth]{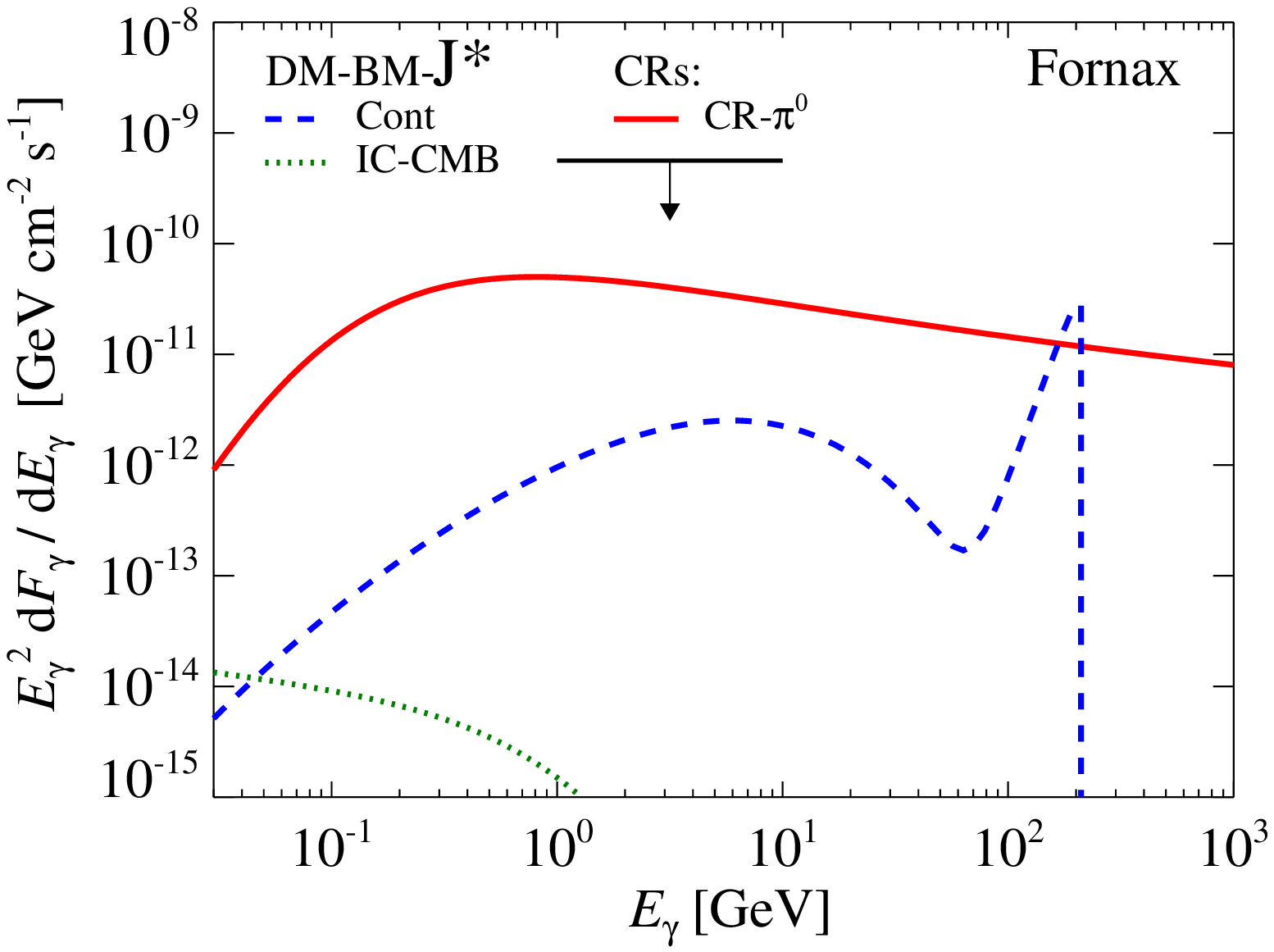}
\caption{\colo Comparing the differential flux from
  different models: we show the continuum emission from DM benchmark
  (BM) models (blue dashed), electrons and positrons from DM BM models
  that Compton-upscatter CMB photons (green dotted) as well as dust
  and star photons (orange dash-dotted), and CR induced gamma-ray
  emission (red solid). Each panel is associated with an individual DM
  BM model; upper left $\Ip$, lower left $\Kp$, upper right $\Jp$, and
  lower right $\Js$. The emission is calculated for the Fornax cluster
  using a point spread function of $0.1\degs$. The substructures boost
  the gamma-ray flux from IC upscattered CMB and continuum emission by
  a factor of 890 while the IC upscattered SD photons are only boosted
  by 20. Extended upper limits ($<1^\circ$) from \Fermi-LAT after 18
  months \protect \cite{2010ApJ...717L..71A} are also shown. In the
  near future we find it hard to detect even the brightest BM models,
  $\Ip$ and $\Kp$, where the continuum emission is dominating the
  total emission in the GeV energy range.}
 \label{fig:diff_BM}
\end{minipage}
\end{figure*}

In Fig.~\ref{fig:diff_BM} we show the differential flux from the
Fornax cluster which is one of the best clusters for indirect DM
searches due to its high DM annihilation fluxes and low CR induced
fluxes. We show the emission of four different supersymmetric BM
models and contrast it to the emission induced by CRs. Comparing this
emission to the differential gamma-ray upper limits set by \Fermi-LAT,
we find that the upper limits are not violated. The predicted DM flux
that is dominated by the continuum emission from $\Kp$ and $\Ip$
models (shown in the left panels) are about a factor few below the
upper limits, making it hard for \Fermi to probe these kind of DM
models in the near future without a significant improvement in the
analysis from e.g. stacking of clusters and improved methods for
analyzing extended sources. Furthermore, the gamma-ray signal induced
by CRs is expected to be about a factor 10 below the DM continuum flux
from the $\Kp$ and $\Ip$ BM models at 10~GeV. However, the IC emission
from upscattered CMB and SD photons is at least a factor of a few
lower for $\Kp$ and a factor 1000 lower for $\Ip,\Jp,\Js$, than the
expected flux from CRs above 100~MeV, making it very hard to
distinguish such a signal from the foreground due to the similar
spectral index. For energies below 100~MeV we expect IC emission by
primary shock accelerated electrons to be dominating
\cite{2010MNRAS.409..449P} over the supersymmetric DM induced
leptons. Hence in clusters, the IC emission from supersymmetric DM BM
models can be neglected compared to the CR pion and DM continuum
emission.

For the LP DM models, however, the main contributions
to the expected extended gamma-ray flux is coming from the IC
upscattered photons on various photon background fields. Depending on
the spatial and spectral distribution of electrons and positrons as
well as the photon background fields, the resulting spectral
distributions of gamma-rays can differ greatly. Hence it is interesting
to understand which spectral features and energy regimes are
dominating.

If the enhancement due to substructures in clusters is significant,
then the distribution of electrons and positrons follows the radial
profile of substructures outside the cluster center where the smooth
DM density profile is dominating. In addition, the cooling of the
steady state electron and positron distribution is dominated by SD
photons in the center. Especially since the SD distribution is
centrally peaked and the substructure distribution peaks in the
outskirts around $\rvir$, the overlap between electrons/positrons and
SD photons is small. This results in a suppression of the IC
upscattered SD photons relative to the IC upscattered CMB photons, the
final state radiation, and the continuum emission. However, if
substructures are only marginally dominating over the smooth
distribution in a cluster, the SD component becomes relatively more
important.

\begin{figure*}
\begin{minipage}{2.0\columnwidth}
\includegraphics[width=0.49\columnwidth]{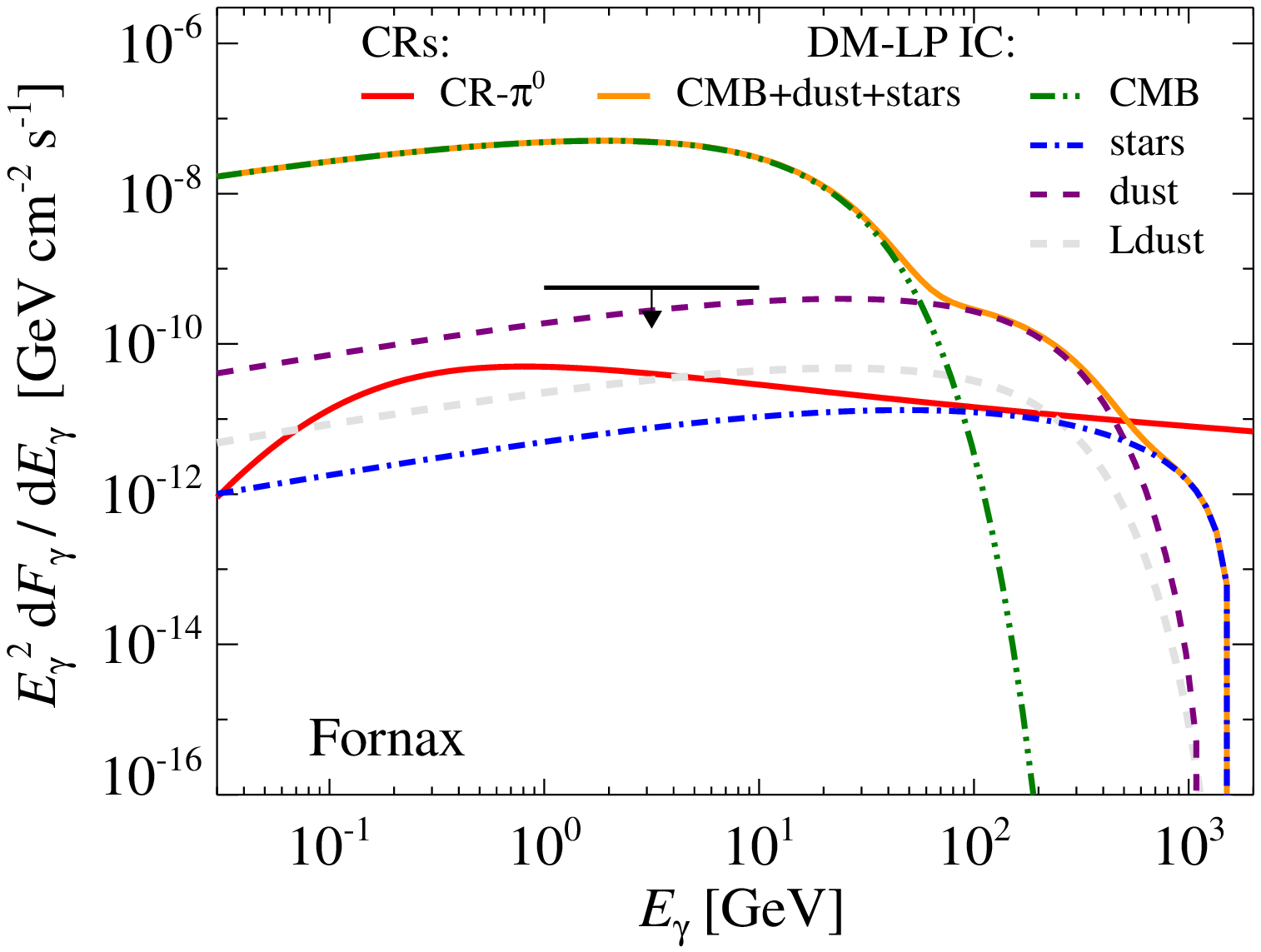}
\includegraphics[width=0.49\columnwidth]{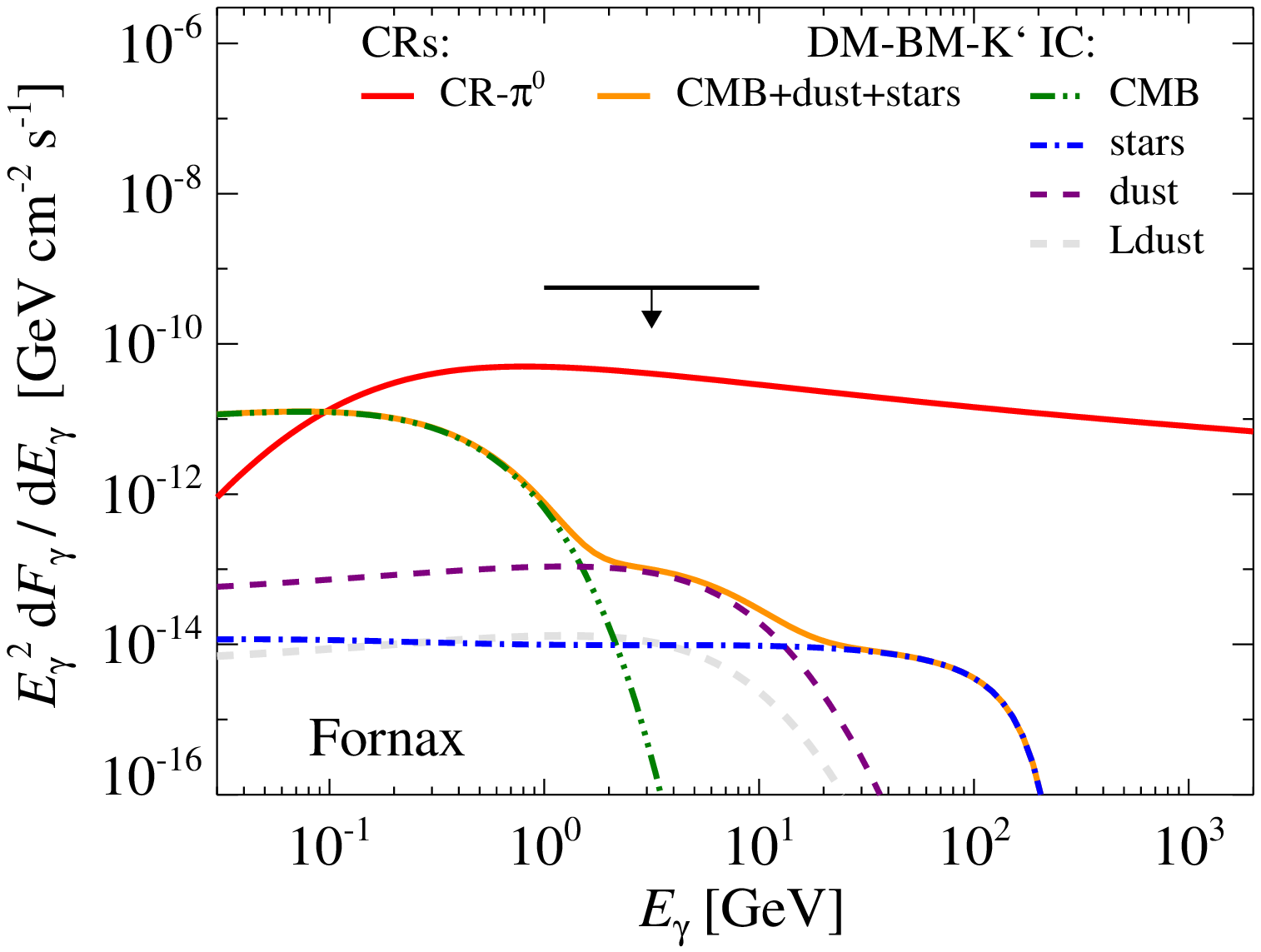}
\caption{\colo Comparing the flux from different inverse
  Compton upscattered radiation fields. We show the differential
  inverse Compton emission induced by leptophilic DM in the left panel
  and by the $\Kp$ benchmark model in the right panel. The
  contribution from each individual radiation field from top line to
  bottom line; CMB (green dashed-triple-dotted), dust (purple dashed),
  low dust model (grey dashed), and stars (blue dashed-dotted). The
  sum of the three components are shown with the orange solid
  line. The red solid lines show the CR induced gamma-ray flux. The
  black arrow shows the spatially extended differential upper limit
  from \Fermi \protect \cite{2010ApJ...717L..71A} indicating that the
  LP model assumptions such as Sommerfeld and/or substructure boost
  are in conflict with the upper limit. All fluxes are calculated for
  the Fornax cluster within $\rvir$ using a point spread function of
  $0.1\degs$. For this cluster, the enhancement due to substructures
  from IC upscattered CMB and SD photons is 890. The saturated
  Sommerfeld boost is 530.}
 \label{fig:IR_comp}
\end{minipage}
\end{figure*}

In Fig.~\ref{fig:IR_comp} we show the total IC emission as well as its
individual contributions from different IC upscattered radiation
fields. The left panel shows the gamma-ray emission from the LP model
and the right panel from the $\Kp$ BM model, and for comparison we
overplot the emission expected from the CRs. Due to the flat electron
and positron spectrum resulting from the LP DM model and the smaller
mean energy of the CMB compared to the SD, we find that the
upscattered CMB photons dominate the total DM IC emission in the
energy regime below 100~GeV, while the SD dominate above this
energy. For the BM models, this transition energy is shifted towards
smaller energies since the electron and positron spectrum has a
steeper spectral distribution (see Fig.~\ref{fig:q_DM}). At the
highest gamma-ray energies of about 100~GeV and above, the IC from
starlight steepens because it probes the high energy tail of the
electrons and more importantly, it suffers from the Klein-Nishina
suppression. When substructures are present, most of their flux
resides in the outer parts of clusters. However, if we remove the
boost from substructures, the density profile of electrons and
positrons is more centrally peaked, and the relative importance of the
IC upscattered SD increases by a factor $\sim 30$ (see
Fig.~\ref{fig:IR_comp})\footnote{Note that for the leptophilic DM
  model this increase is much smaller since Sommerfeld enhancement is
  no longer dominated by the low velocity DM particles living in the
  subhalos (see Sec.~\ref{sect:LP} for more details).}. A larger
fraction of electrons (in the core) will now also cool by
Compton-upscattering SD photons which suppresses the IC-upscattered
CMB light. For comparison, we include the contribution of IC
upscattered SD photons and to bracket the uncertainty in the SD model,
with a smaller amount of dust. In this low-dust model, we have reduced
the energy in dust by a factor 10. However, the resulting flux from
the IC upscattered dust in this model only decreases with a factor
that is slightly smaller than 10 since the IC cooling of the electrons
and positrons also decreases.

Because of the large total boost factor for the LP model ($\sim5\times
10^5$), we overproduce the upper gamma-ray limit in the $1-10$~GeV
energy interval set by \Fermi-LAT by about a factor 100. This strongly
constrains the boost from both substructures and the
SFE. Additionally, these constraints might improve with future more
sensitive Cherenkov telescopes such as CTA. However, considering that
an indirect detection of DM in clusters relies on the boost from
substructures whose main contribution comes from the periphery, we
conclude that the wide angular extent of clusters on the sky in
gamma-rays suggest that these sources are not ideal for Cherenkov
telescopes since their sensitivity drops linearly with source
extension.

\begin{figure}
 \includegraphics[width=0.99\columnwidth]{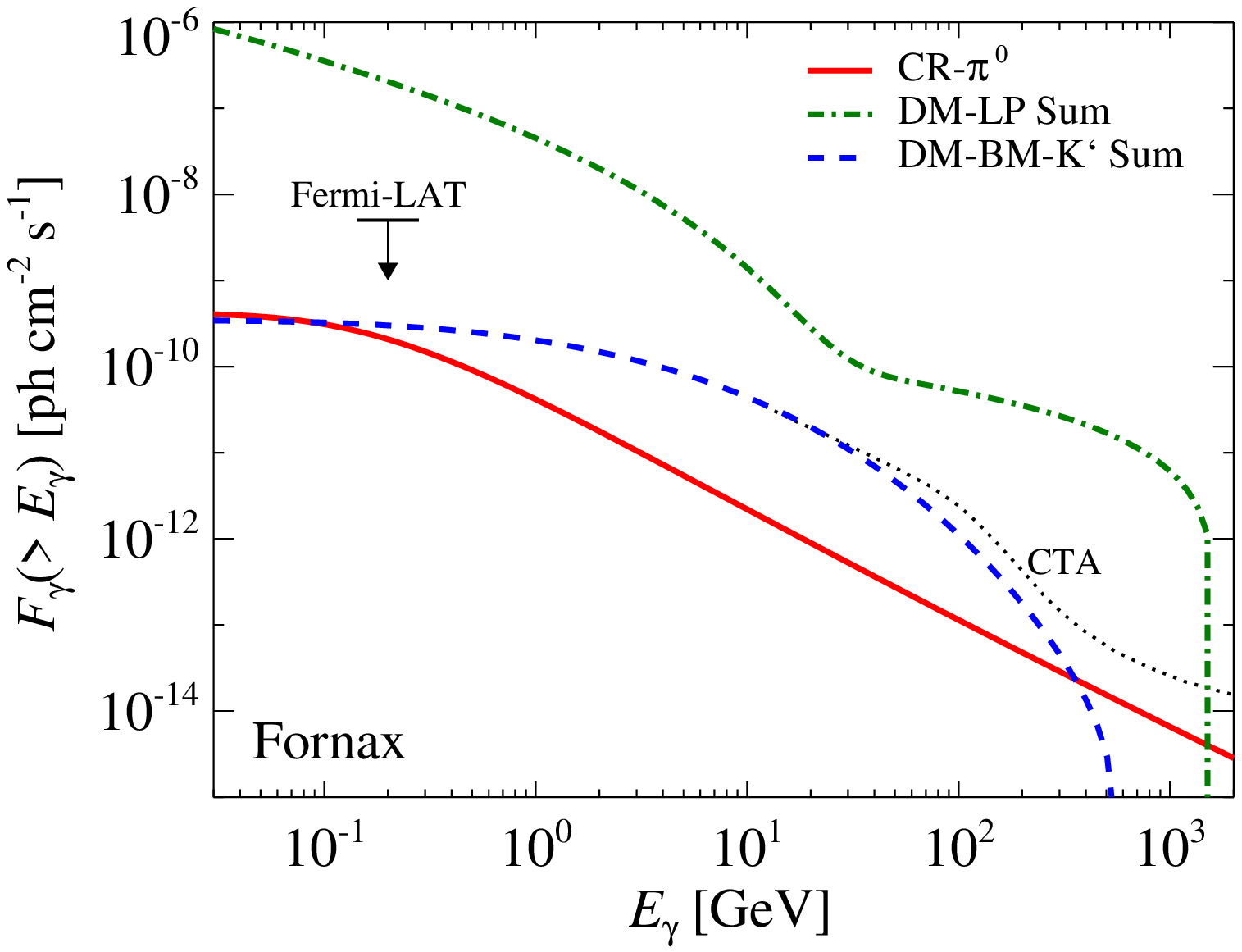}
 \caption{\colo Comparing the energy integrated flux from
   different models. We show the emission from CR induced emission
   (red solid), a leptophilic (LP) model that includes both final
   state radiation and IC upscattered CMB, dust and starlight (green
   dash-dotted), and the benchmark $\Kp$ model (BM) that includes
   continuum emission, and IC upscattered CMB, dust and starlight
   (blue dashed). The black arrow shows the spatially extended
   integrated flux upper limit set by \Fermi-LAT again indicating
   challenges for the assumptions underlying the LP model. We also
   show projected CTA point source sensitivities ($5\sigma$, 50h). The
   emission is calculated for the Fornax cluster using a point spread
   function of $0.1\degs$. The boost from Sommerfeld and substructures
   is about 530 and 890, respectively.}
 \label{fig:flux_int}
\end{figure}

It is also interesting to compare the total contribution from the LP
model, the brightest BM model ($\Kp$), and the CR induced emission. In
Fig.~\ref{fig:flux_int} we show the integrated flux from Fornax for
our different gamma-ray models and compare it to the unresolved
integrated flux upper limit on Fornax set by \Fermi-LAT where they
averaged the flux over the energy range $0.2-100\,\gev$ assuming a
spectral index of 2. Again, the annihilation flux in the LP model is
in conflict with the upper limits by \Fermi-LAT, although only by a
factor of 30 which is a factor few less constraining than the
differential flux in the energy range $1-10$~GeV. As shown in this
figure, the LP model is dominating the entire gamma-ray energy range
up to the DM rest mass energy in this model of about 1~TeV. Finally,
we note that the flux the DM BM $\Kp$ model is larger than the
predicted emission from the CRs in the $0.1-100$~GeV energy
regime. Hence, for an experiment with a high sensitivity even for
extended sources, the prospects for detecting the $\Kp$ BM model over
the expected gamma-ray background induced by CRs looks promising in
clusters. Present day Cherenkov telescopes, however, have a trigger
region that is smaller than the size of clusters, hence it has to be
increased to several degrees to overcome problems with background
estimation. In addition, even though the projected CTA point source
sensitivity ($5\sigma$, 50h) shows the potential of this experiment in
constraining leptophilic models as well as BM models with a very large
neutralino mass $m_\chi c^2 \gtrsim 1$~TeV, we find that analysis
techniques have to be developed that enable the detection of extended
sources without too much degradation of sensitivity.

\begin{figure*}
\begin{minipage}{2.0\columnwidth}
 \includegraphics[width=0.49\columnwidth]{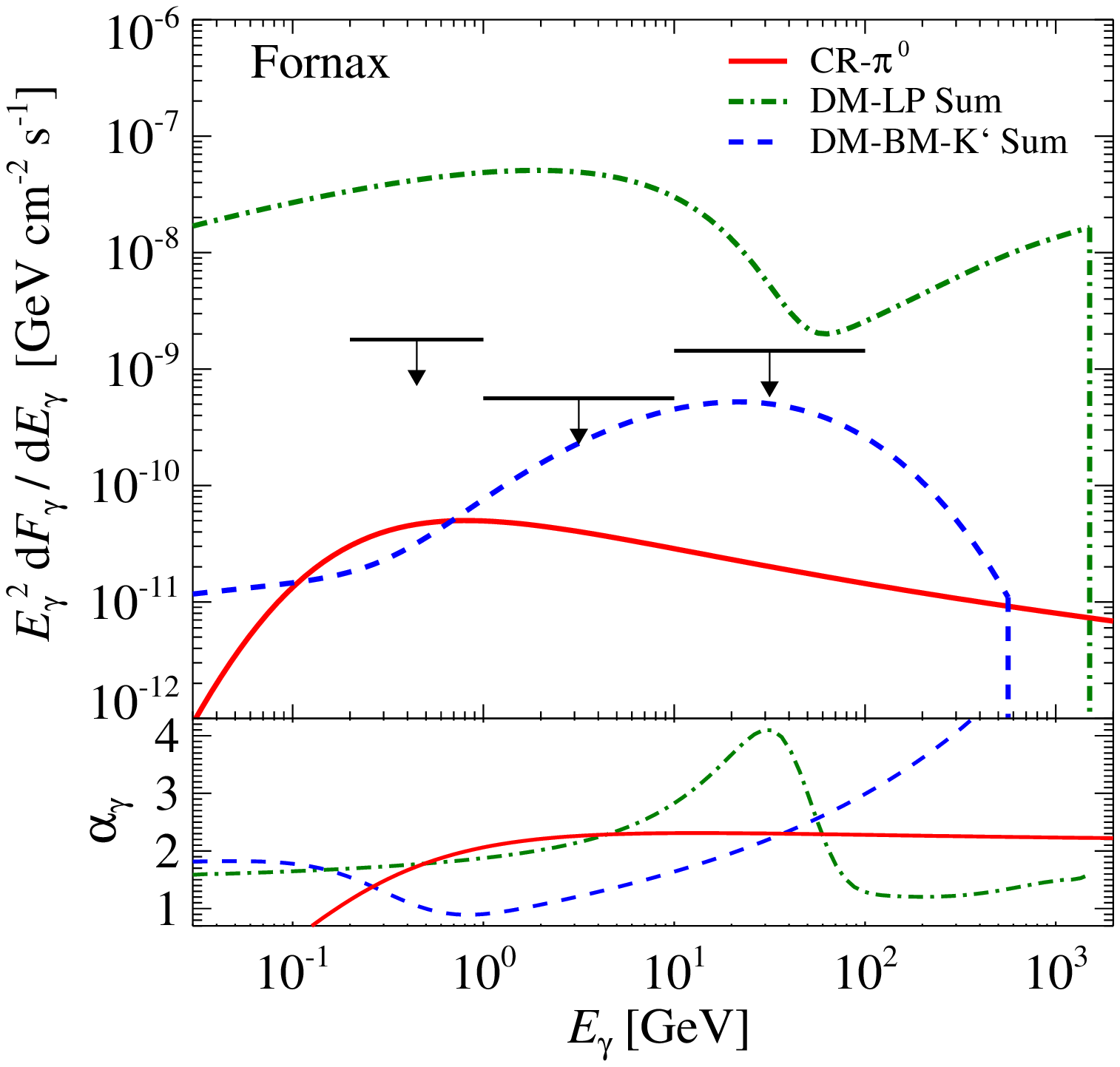}
\includegraphics[width=0.49\columnwidth]{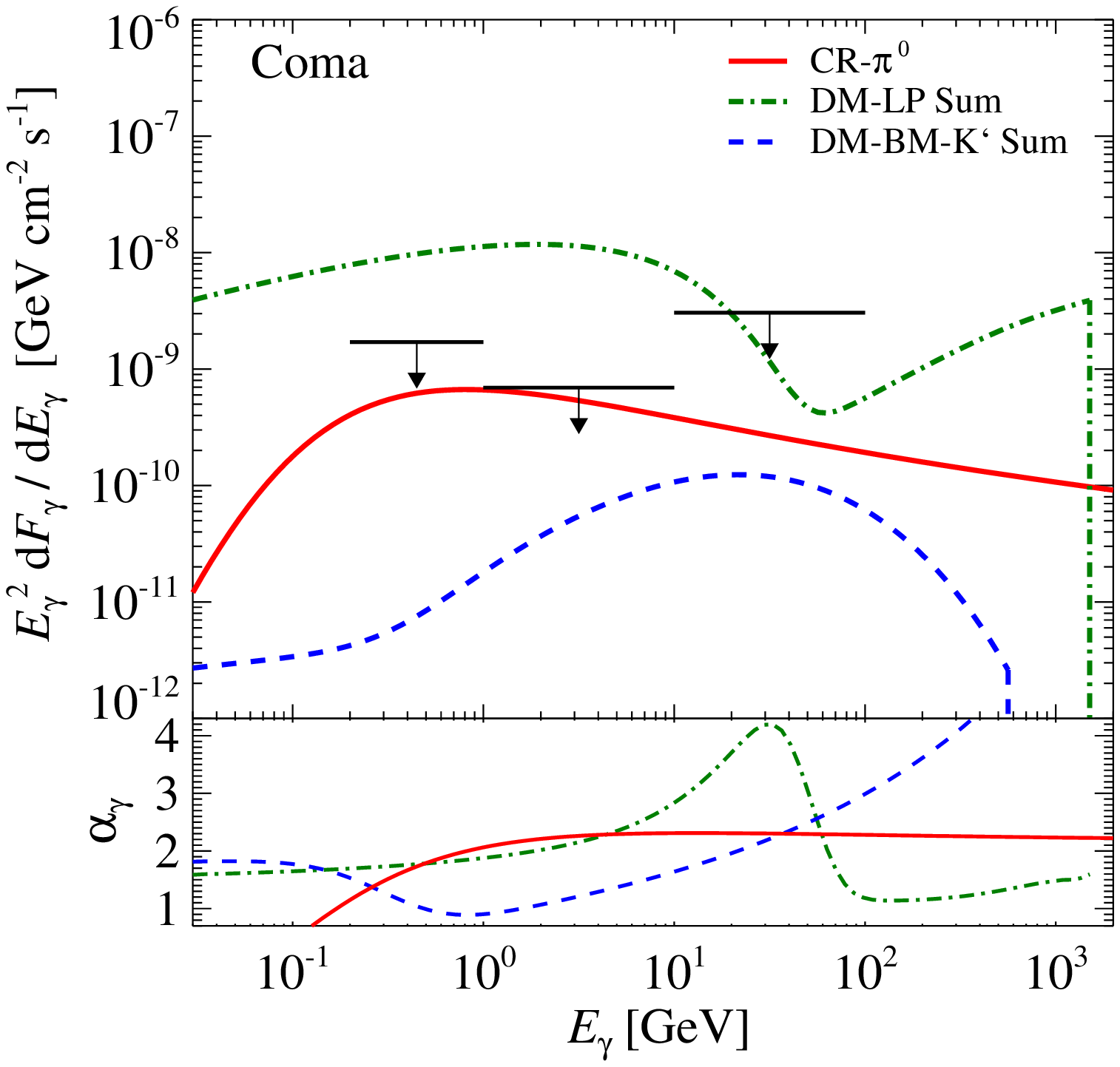}
\includegraphics[width=0.49\columnwidth]{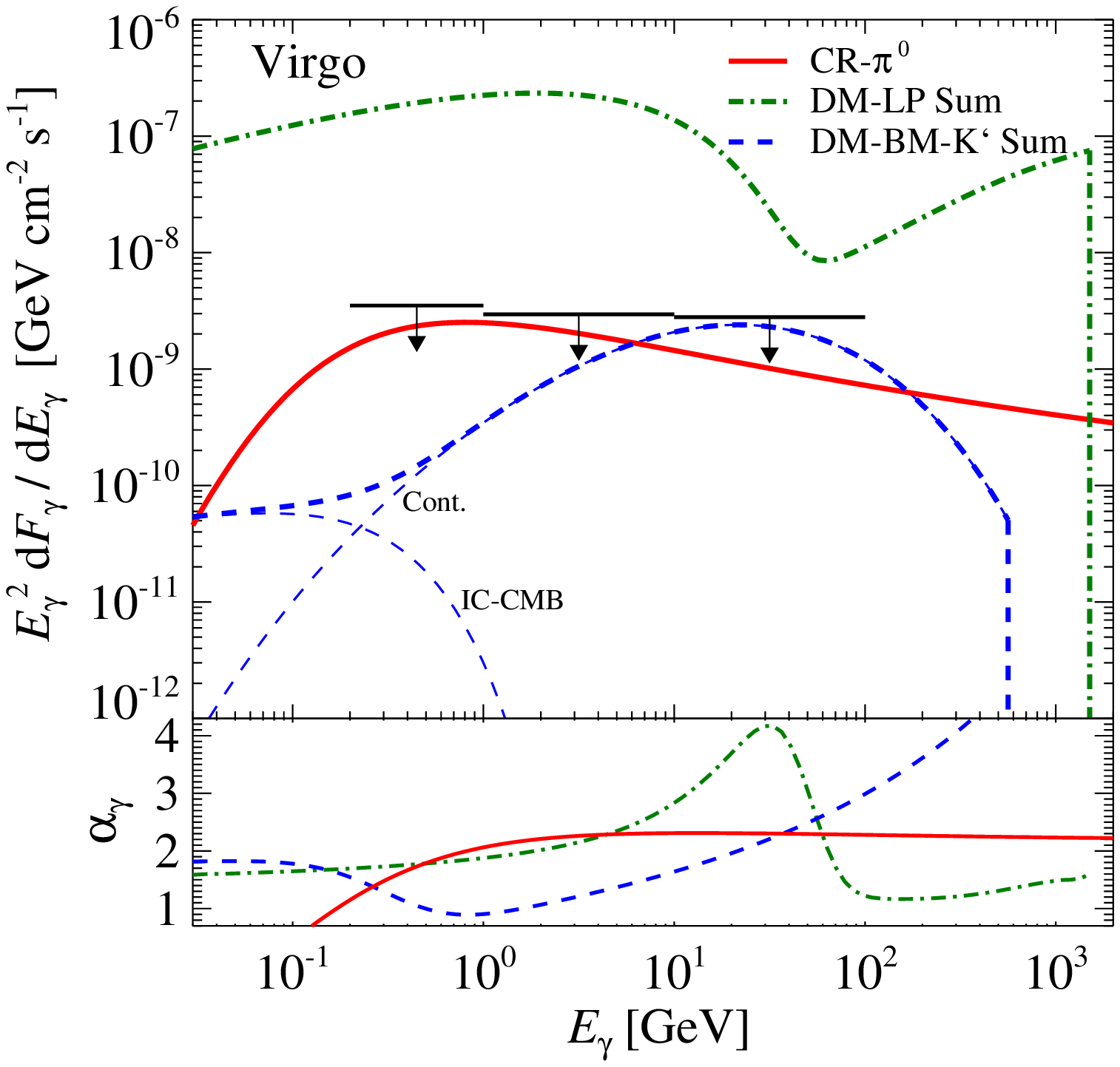}
\includegraphics[width=0.49\columnwidth]{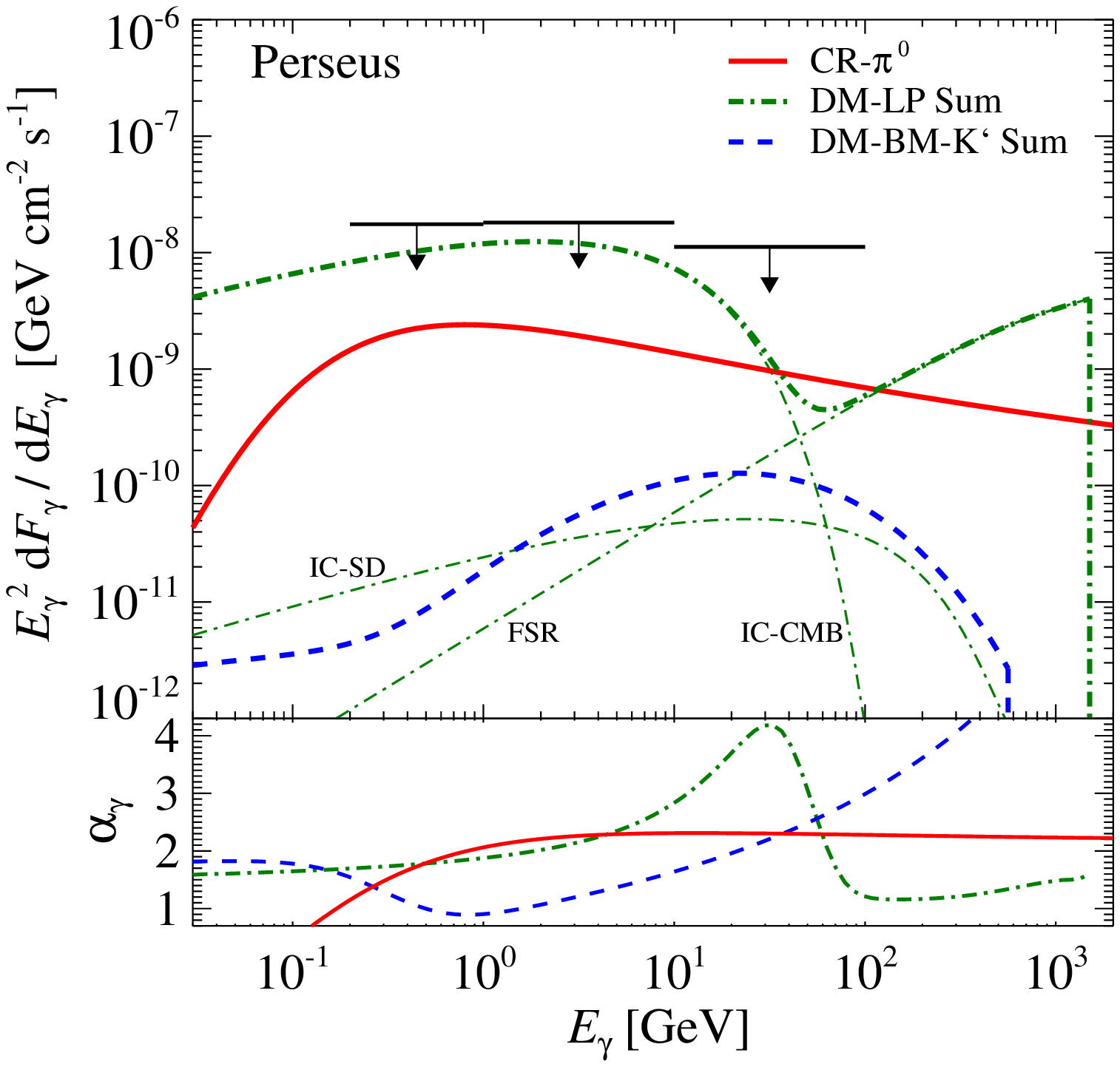}
\caption{\colo Comparing the gamma-ray emission from
  different clusters. We show the differential flux of clusters in the
  upper panel of each figure: Fornax (upper left), Coma (upper right),
  Virgo (lower left), and Perseus (lower right). We show CR induced
  emission (red solid), a leptophilic (LP) model that includes both
  final state radiation and IC upscattered CMB, dust and starlight
  (green dash-dotted), and the benchmark $\Kp$ model (BM) that
  includes continuum emission, and IC upscattered CMB, dust and
  starlight (blue dashed). The arrows show the spatially extended
  differential upper limits set by \Fermi-LAT in the energy ranges
  $0.2-1$~GeV, $1-10$~GeV, and $10-100$~GeV from left to right (using
  Gauss profiles for the source extension with $\sigma=(1.2,1.0,0.8)$
  for (Virgo, Coma, Fornax) and an X-ray-inferred King profile for
  Perseus) \cite{2010ApJ...717L..71A}. We show the individual
  components for the BM model and LP model for Virgo and Perseus,
  respectively. In the lower panel of each figure we show differential
  spectral indices, $\alpha_\gamma$, where $\dd F_\gamma/\dd E_\gamma
  \sim E_\gamma^{-\alpha_\gamma}$. The flux from the clusters is
  integrated out to $\rvir$ using a point spread function of $0.1\degs$
  and includes the boost from substructures. We find that the lower
  GeV-energy regime is most constraining due to the peak sensitivity
  of \Fermi at these energies and use this regime to get upper limits
  on boost factors. Note that the CR-induced emission in Coma is close
  to the upper limits set by \Fermi and will be tested in the upcoming
  years.}
 \label{fig:clu_comp}
\end{minipage}
\end{figure*}

We continue by comparing the estimated differential flux from Fornax
to three other clusters in Fig.~\ref{fig:clu_comp}; the close-by and
well studied Virgo cluster, the X-ray bright Perseus clusters, and the
massive merging Coma cluster. We summarize in Table~\ref{tab:flux_tab}
the expected gamma-ray flux in our DM and CR models and the
corresponding boost factors.  We find high DM induced gamma-ray fluxes
from the Fornax and Virgo clusters, which confirms them as promising
targets for indirect DM searches. Especially in the $1-10$~GeV energy
range it is quite striking how constraining the upper limits are for
some of the clusters. At these energies, \Fermi-LAT has its peak
sensitivity due to a combination of increasing effective area and
decreasing source spectra as a function of energy. The upper limits
for Virgo and Perseus are unfortunately background contaminated by AGN
activity from M87 and NGC 1275, respectively, and do not gain much
from the increased sensitivity. In addition, the angular radius of
Virgo is about $6^\circ$, while the extended upper limit from
\Fermi-LAT is calculated assuming a $\sigma=1.2^\circ$ radius. In
fact, the ratio of the virial radius to the assumed extension for the
upper limits $\rvir/\sigma$ for $($Fornax, Coma, Virgo$)$ are
$(3.7,\,1.6,\,5.3)$, respectively. Similarly, it is interesting to
compare the radius that contains 68\% of the flux to the assumed
extension, $r_{68}/\sigma$, which is given by $(2.3,\,1.0,\,3.3)$ for
$($Fornax, Coma, Virgo$)$, respectively. This further motivates a
re-calculation of the \Fermi upper limits to account for the full
extension of the galaxy clusters. We also find that the theoretical
expectation of a high CR-induced flux in Coma and Perseus may
significantly complicate indirect DM searches in those clusters if the
boost factors are much lower than assumed in this work. The Fornax
cluster, however, is a great target for indirect DM studies because
the relative low gamma-ray flux from CRs, absents of an active AGN,
and high DM gamma-ray flux.

What is the figure of merit for selecting the most promising cluster
targets for indirect DM searches? To this end, we employ the
luminosity-to-mass scaling relations. The gamma-ray luminosity from
the smooth density profile is given by \cite{2009PhRvL.103r1302P}
\begin{equation}
L_{\gamma,\sm} \propto \int \dd V \rho(r)^2 \propto \frac{M_{200}\,c^3}
{\left[\log\left(1+c\right)-c/(1+c)\right]^2} \propto \mvir^{0.83}\,,
\end{equation}
and the total DM luminosity that includes boost factors for the LP and
the BM models is given by
\begin{eqnarray}
\label{eq:DM_scaling}
L_{\gamma} &=& L_{\gamma,\rmn{sm}} \B_\rmn{sub} \propto \frac{\mvir^{1.06}}{D_\rmn{lum}^2},\\
\rmn{where} & &\B_\sub \propto \mvir^{0.23}\,.
\end{eqnarray}
We note that the IC from upscattered SD photons scales slightly softer
with mass, hence gives rise to a larger fraction of gamma-rays in low
mass clusters compared to the total gamma-ray flux. Also note that we
have not included the SFE for the LP model in
Eq.~(\ref{eq:DM_scaling}) since it saturates to a constant value in
the substructures. However, if only a small fraction of the DM resides
in the subhalos, i.e. the scaling of the LP DM becomes shallower due
to the mass dependence inferred from the velocity dispersion:
\begin{eqnarray}
L_{\gamma,\rmn{LP~nosub}} &=& L_{\gamma,\rmn{sm}} \B_\rmn{sfe}
\propto \frac{\mvir^{0.5}}{D_\rmn{lum}^2},\\
\rmn{where} &\quad&\B_\rmn{sfe} \propto \mvir^{-1/3}.
\end{eqnarray}

\begin{table*}\hst\hst\hst\hst\hst\hst\hst
\begin{minipage}{2.0\columnwidth}
  \caption{Gamma-ray flux from various clusters within $\rvir$.}
\begin{tabular}{l c c c c c c c c c c c}
\hline
\hline
 Cluster &
\multicolumn{3}{c}{$F_{\gamma}(>100\,\rmn{MeV})$ $[\rmn{ph}\,\rmn{cm}^{-2}\,\rmn{s}^{-1}]$:} & &
\multicolumn{3}{c}{$F_{\gamma}(>100\,\rmn{GeV})$ $[\rmn{ph}\,\rmn{cm}^{-2}\,\rmn{s}^{-1}]$:} &
$\mvir$$^{(1)}$ & $D_\rmn{lum}$$^{(1)}$ & $\B_\rmn{sfe}$$^{(2)}$ & $\B_\rmn{sub}$$^{(3)}$ \\
         & DM-LP$^{(4)}$ & DM-BM-$\Kp$$^{(5)}$ & CR-$\pi^0$$^{(6)}$
         & & DM-LP$^{(4)}$ & DM-BM-$\Kp$$^{(5)}$ & CR-$\pi^0$$^{(6)}$ 
         & $[10^{14}\,\msun]$ & [Mpc] && \\
 \hline
 Coma & $8.2\mytimes10^{-8}$ & $7.6\mytimes10^{-11}$ & $4.1\mytimes10^{-9}$
 & \,\,\,\,\, & $1.2\mytimes10^{-11}$ & $2.8\mytimes10^{-13}$ & $1.5\mytimes10^{-12}$
 & $12.9$ & $101$ & $530/65$ & $1290$ \\
 Perseus \,\,\,\,\,\, & $8.6\mytimes10^{-8}$ & $7.8\mytimes10^{-11}$ & $1.5\mytimes10^{-8}$
 & \,\,\,\,\, & $1.2\mytimes10^{-11}$ & $2.9\mytimes10^{-13}$ & $5.5\mytimes10^{-12}$
 & $8.6$ & $79.5$ & $530/75$ & $1190$ \\
 Virgo & $1.6\mytimes10^{-6}$ & $1.5\mytimes10^{-9}$ & $1.5\mytimes10^{-8}$
 & \,\,\,\,\, & $2.3\mytimes10^{-10}$ & $5.4\mytimes10^{-12}$ & $5.7\mytimes10^{-12}$
 & $6.9$ & $17.2$ & $530/80$ & $1120$ \\
 Fornax & $3.5\mytimes10^{-7}$ & $3.2\mytimes10^{-10}$ & $3.1\mytimes10^{-10}$
 & \,\,\,\,\, & $5.1\mytimes10^{-11}$ & $1.2\mytimes10^{-12}$ & $1.1\mytimes10^{-13}$
 & $2.4$ & $19.8$ & $530/110$ & $890$ \\
\hline
\hline
\end{tabular}
\begin{quote}
  Notes: \\
  (1) The mass of Fornax, Coma, and Perseus are derived from \cite{2007A&A...466..805C},
  while the mass of Virgo is derived from \cite{1984ApJ...281...31T}. The luminosity 
  distance to Fornax, Coma, and Perseus are derived from \cite{2007A&A...466..805C},
  while the distance to Virgo is derived from \cite{2007ApJ...655..144M}. All distances 
  and masses assume $H_0= 70\,[\rmn{km}/\rmn{s}/\rmn{Mpc}]$.\\
  (2) The boost due to Sommerfeld enhancement. The first value shows the saturated
  Sommerfeld boost realized when substructures are present, the latter is the Sommerfeld
  boost without substructures (see Sec. II A 1).\\
  (3) The boost due to substructures relative a the smooth DM
  distribution. We integrate the emission in a cylinder with a radial
  extent of $2.5 R_{200}$ along the line-of-sight and an angular size
  corresponding to $\rvir$ (both measured from the cluster center).\\
  (4) The total gamma-ray flux from leptophilic (LP) DM where both the
  boost from substructures and Sommerfeld enhancement are
  included. Note that all these values are in conflict with upper limits from
  \Fermi (see the following sections for detail).\\
  (5) The total gamma-ray flux from the supersymmetric $\Kp$ benchmark (BM)
  model where the boost from substructures is included. See Secs. II A 2 and II B 3.\\
  (6) Gamma-ray flux induced by CR protons. See Sec. V C.
 \label{tab:flux_tab}
  \end{quote}
\end{minipage}
\end{table*}

In the lower panels of Fig.~\ref{fig:clu_comp}, we also provide the
differential spectral index, $\alpha_\gamma=-\dd \log(\dd F/\dd E)/\dd
\log(E)$, for the different emission models. We find very similar
spectral indices for different clusters assuming a substantial boost
from substructures, while for models without substructures the
relative contribution from the upscattered dust and starlight breaks
the spectral universality for the DM models. In the $1~\gev-1~\tev$
energy regime the CR spectral index is approximately constant $\sim
(2.1-2.3)$, while the spectral index for the LP model varies
substantially between $1.2-4.0$, hence implying that gamma-ray upper
limits are more sensitive to the specific energy regime for these
models. Similarly, the index for the BM $\Kp$ model is about 1.0 at
1~GeV and increases monotonically toward higher energies. To maximize
photon count statistics, experiments calculate band-integrated fluxes
which implies a fixed spectral index over that energy range. Hence, we
show $\alpha_\gamma$ in Table~\ref{tab:spectral_index} for our
emission models in four different energy bands. We find smaller
variations for the banded $\alpha_\gamma$-values, hence reducing the
importance of the specific energy regimes.

\begin{table}
\begin{tabular}{ccccc}
\hline\hline
      Model & $\alpha_{100\,\rmn{MeV}}^{1\,\rmn{GeV}}$ &
              $\alpha_{1\,\rmn{GeV}}^{10\,\rmn{GeV}}$ &
              $\alpha_{10\,\rmn{GeV}}^{100\,\rmn{GeV}}$ &
              $\alpha_{100\,\rmn{GeV}}^{1\,\rmn{TeV}}$ \\
\hline
CR-$\pi^0$ & 1.44 & 2.24 & 2.30 & 2.26 \\
DM-LP Sum & 1.74 & 2.20 & 3.06 & 1.30 \\
DM-BM-$\Kp$ Sum & 1.28 & 1.22 & 2.23 & - \\
\hline\hline
\end{tabular}
\caption{Banded gamma-ray spectral index $\alpha_{E_1}^{E_2}$ between
  energies $E_1$ and $E_2$. We show the spectral index for three
  different emission models where the boost from substructures is
  included; the CR induced emission, the leptophilic (LP) model that
  includes both final state radiation and IC upscattered CMB, dust and
  starlight, and the benchmark (BM) $\Kp$ model that includes
  continuum emission, and IC upscattered CMB, dust and
  starlight. $\alpha_\gamma$-values are derived for the Fornax
  cluster, although the variance between clusters is very small when
  we account for the enhancement from substructures. This suppresses
  the relative contribution from the upscattered dust and starlight
  component that depends on cluster mass. \label{tab:spectral_index}}
\end{table}

\begin{figure}
 \includegraphics[width=0.99\columnwidth]{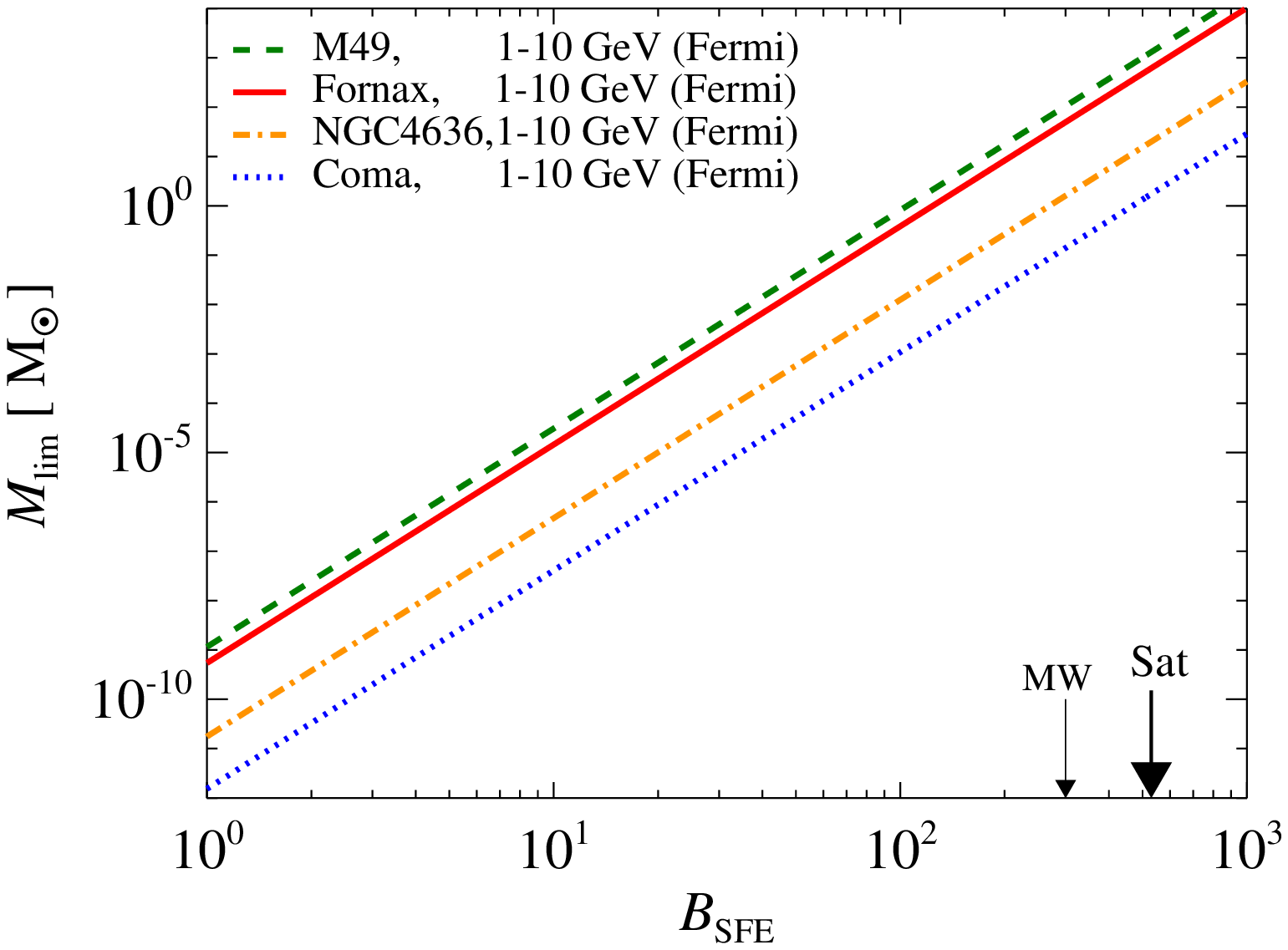}
 \caption{\colo Constraining boost factors using flux
   upper limits. The leptophilic (LP) gamma-ray emission is derived
   within $\rvir$ using a point spread function of $0.1\degs$. In order
   not to overproduce the \Fermi-LAT differential flux upper limits in
   the energy interval $1-10$~GeV the boost from substructures and
   Sommerfeld enhancement (SFE) is constrained for four clusters; M49
   (green dashed), Fornax (red solid), NGC4636 (orange dash-dotted),
   and Coma (blue dotted). We indicate both the saturated SFE of 530
   (right black arrow) as well as the local boost in the Milky Way of
   300 (c.f. Eq.~\ref{eq:B_sfe}) that is required to explain the
   electron and positron excess observed at Earth with LP DM (left
   black arrow). If the DM interpretation is correct, we can constrain
   the smallest size of halos to be larger than
   $10^4\,\msun$. Contrarily, if the smallest size of DM halos is
   $10^{-6}\,\msun$, we can constrain the SFE to $\lesssim 5$ in M49
   and Fornax. This corresponds to a maximum SFE in the MW of
   $\lesssim 3$, which would be too small to support the DM
   annihilation hypothesis for the PAMELA/\Fermi excess.}
 \label{fig:boost_const}
\end{figure}

In this section, we have found that the LP DM models overproduces both
the spatially extended differential and integrated gamma-ray flux
upper limits set by the \Fermi-LAT 18 month data for several
clusters. We can use these upper limits to constrain the boosts due to
SFE and substructures. Furthermore, as more sensitive experiments and
better upper limits emerge, we can start ruling out models that give
rise to boost factors that are manifested in the gamma-ray flux. It is
especially interesting to constrain the SFE, where a boost of the
particle physics cross section of $\sim$300 is required to explain the
excess of electrons and positrons observed in the vicinity of Earth by
PAMELA/\Fermi/H.E.S.S. In Fig.~\ref{fig:boost_const} we use the
estimated flux within $\rvir$ of four bright and maximal constraining
clusters (Fornax, M49, NGC4636, and Coma, where we have excluded Virgo
due to its large angular extent) to limit the SFE as well as the
minimum mass of substructures where
$M_\rmn{lim}\propto\mvir^{0.226}$. We find that M49 and the Fornax
cluster are the most constraining clusters, although the current upper
limits are not good enough to rule out the boost from substructures or
SFE. Assuming that measured $e^+/e^-$ excess is due to SFE DM, then in
order not to overproduce the differential gamma-ray upper limits set
by \Fermi-LAT in the energy range\footnote{We average the LP DM flux
  in the energy range $1-10$~GeV before we compare it to the upper
  limits.} $1-10$~GeV, we can constrain the smallest mass of
substructures to $M_\rmn{lim}\gtrsim 10^4\,\msun$. Instead, assuming
the standard mass of the smallest substructures of $M_\rmn{lim}\approx
10^{-6}\,\msun$, we can constrain the saturated boost from SFE to
$\lesssim 5$ in M49 and Fornax, which corresponds to a maximum SFE in
the MW of $\lesssim3$. Hence, we conclude that without substantial
improvement in the modelling of these extended sources, \Fermi-LAT
will not be able to rule out LP models in their current form using
clusters but can put impressive constraints on them.

\section{Surface brightness profiles}
\label{sect:spatial}

The large angular extent of clusters on the sky in combination with
the small PSF of most gamma-ray probes ($\sim 0.1\degs$) suggest that
we would be able to probe the different spatial regimes of a
cluster. Especially important is the spatial distribution of the
gamma-ray emission as it biases upper limits derived for spatially
extended clusters where a simple profile is often assumed. In fact, we
will show that if DM substructures are present with the discussed
abundances, the gamma-ray brightness profile is almost flat and very
different from the gamma-ray distribution following a smooth density
profile with a steep outer surface brightness slope of $-5$.

\begin{figure}
 \includegraphics[width=0.99\columnwidth]{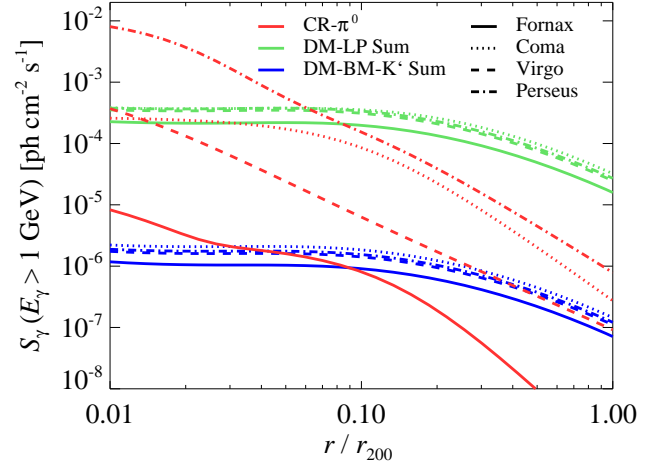}
 \caption{\colo Comparing the intrinsic surface
   brightness from different clusters without taking PSF effects into
   account. We show the CR induced emission (red), leptophilic
   emission (light green), and emission from the $\Kp$ benchmark model
   (dark blue). The different line styles each represent a cluster;
   Fornax (solid), Coma (dotted), Virgo (dashed), and Perseus
   (dash-dotted). We include the boost from both substructures and
   Sommerfeld effect. The shape and normalization of the different DM
   models are very similar. In contrast, the CR-induced emission
   profiles have a much larger scatter due to the large variation of
   the gas fraction and expected CR fraction in these clusters.}
 \label{fig:SB_clu}
\end{figure}

In this section we investigate the gamma-ray surface brightness
profiles induced by both CRs and different models of DM in more
detail. We start by comparing the intrinsic surface brightness that
includes substructures from different clusters in
Fig.~\ref{fig:SB_clu}. As expected, we find that the DM flux in all
clusters follows the same universal shape imposed by the
substructures. Already at $0.01\rvir$, we are dominated by the
substructures due to the projection of their peaked distribution the
outer cluster regions (see Sec.~\ref{sect:subst} for a longer
discussion). This implies that the spatial distribution of
annihilating DM is independent of the type of DM model, and can not be
used to separate different models. The surface brightness, $S_\gamma$,
of the DM scales as $S_\gamma\propto
L_{\gamma,\rmn{sm}}\bsub\mvir^{-2/3} \propto \mvir^{0.40}$, which
explains the factor two difference in normalization between the most
massive cluster Coma and the least massive cluster Fornax with a mass
ratio of about 10. The surface brightness induced by CRs is
proportional to the projected squared gas profile with an additional
enhancement in the center of cool core clusters due to the adiabatic
compression of CRs during the formation of the cool core. Hence, we
see a much larger variation in these profiles driven by the large
variation of the gas mass fraction. It is noticeable that Fornax shows
a small surface brightness. This is explained by the low gas density
outside the BCG and the expected small abundance of CRs in a small
cluster in comparison to a massive cluster
\cite{2010MNRAS.409..449P}. This is the reason why Fornax is a good
target for indirect DM searches, and superior to DM searches in dwarf
galaxies.

\begin{figure}
 \includegraphics[width=0.99\columnwidth]{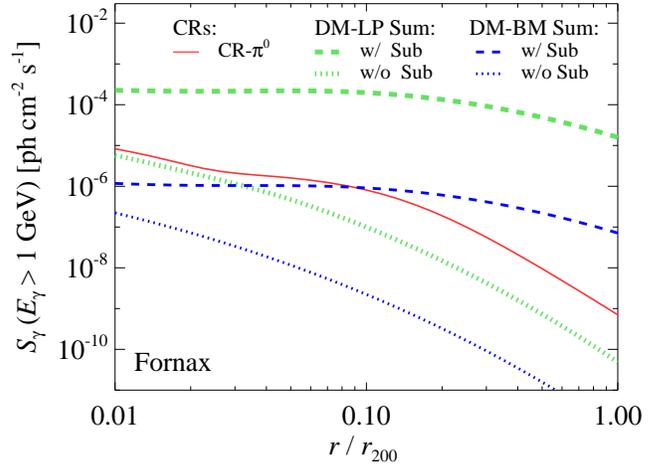}
 \caption{\colo The impact of substructures on the
   intrinsic surface brightness profile o the Fornax cluster at
   1~GeV. The emission induced by CRs is denoted by the thin red solid
   line, the leptophilic model by thick light green lines, and the DM
   $\Kp$ benchmark model by dark blue lines. The dashed lines show the
   brightness profiles where the boost from substructures is included
   while the substructures are excluded for the dotted lines. The
   boost from Sommerfeld enhancement with and without substructures is
   110 and 530, respectively. The boost due to substructures is about
   890. Notice the nearly flat profiles when substructures are
   included, and the relative large boost at $0.01\rvir$ that is due
   to projection of peripheral substructures onto the cluster center.}
 \label{fig:SB_sub}
\end{figure}

To quantify the impact of substructures on the spatial profiles in
more detail we again turn to the Fornax cluster and show in
Fig.~\ref{fig:SB_sub} a comparison of surface brightness profiles with
and without substructures. It is quite remarkable how flat the
profiles become when substructures are present compared to the case
without. This implies that the relative flux within the PSF in the
center is comparable to the outer parts, which should increase the
signal-to-noise significantly. If we assume that -- for
some reason -- substructures are suppressed in the Fornax cluster, then the expected
signal from DM is swamped by astrophysical backgrounds over the entire
extent of the cluster, even though Fornax has one of the highest
ratios of DM to CR induced gamma-ray brightness.

\begin{figure*}
\begin{minipage}{2.0\columnwidth}
  \includegraphics[width=0.49\columnwidth]{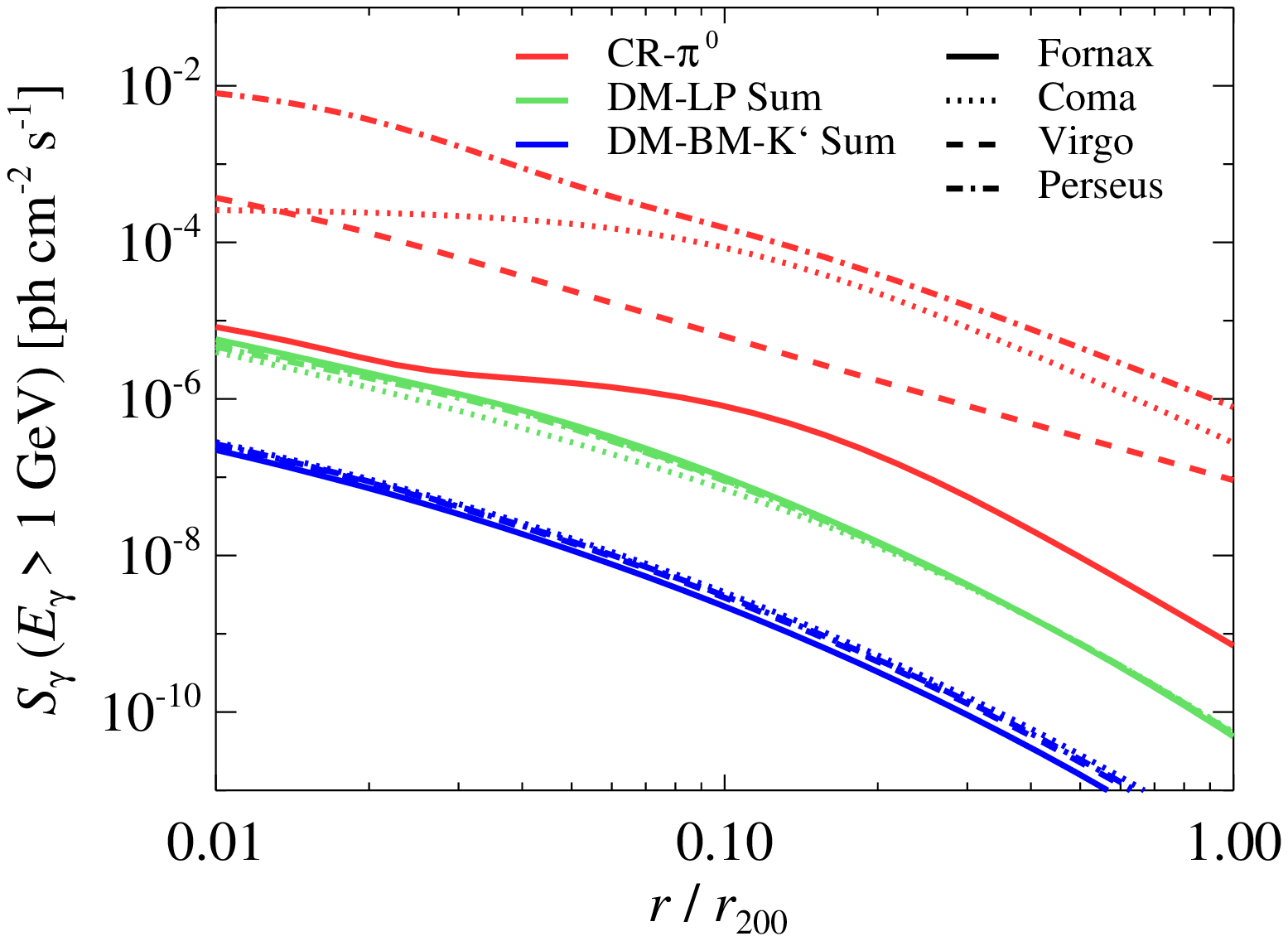}
  \includegraphics[width=0.49\columnwidth]{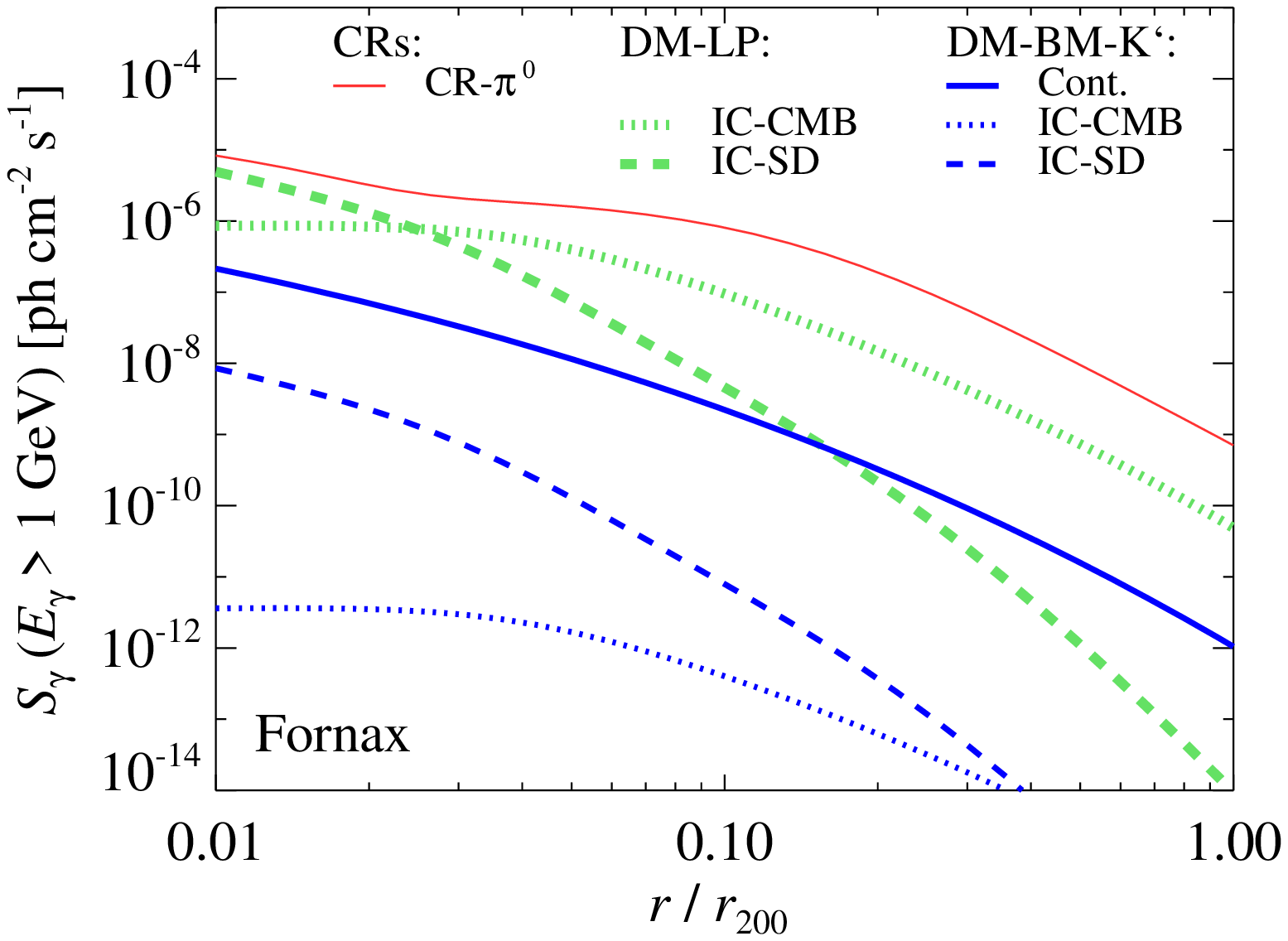}
  \caption{\colo The intrinsic surface brightness profiles without
    substructures at energies $E>1\ GeV$. We show the CR induced
    emission (red), leptophilic emission (light green), and emission
    from the $\Kp$ benchmark model (dark blue). Left panel: comparison
    of the surface brightness of different clusters; Fornax (solid),
    Coma (dotted), Virgo (dashed), and Perseus (dash-dotted). Right
    panel: comparison of different emission components in Fornax;
    dotted lines show the inverse Compton (IC) upscattered CMB
    photons, dashed lines show the IC upscattered photons from stars
    and dust, and blue solid lines show the continuum emission from
    the $\Kp$ benchmark model.}
 \label{fig:SB_clu_nosub}
\end{minipage}
\end{figure*}

In order to learn about the spatial distribution of the DM flux in
various clusters for different DM models and the associated gamma-ray
components, we show in the left panel of Fig.~\ref{fig:SB_clu_nosub}
the brightness profiles for the smooth DM distribution for the same
four representative clusters that were shown previously in
Fig.~\ref{fig:SB_clu}. We find that the surface brightness, $S_\gamma$
of the DM emission above 1~GeV in the outer parts of all clusters have
the same shape. In addition, we show the spatial dependence of the
individual gamma-ray components for the Fornax cluster in the right
panel. We see that in the BM model, the emission is dominated by the
continuum emission. In the LP model, the emission at intermediate and
large radii is dominated by IC-upscattered CMB photons while at small
radii, IC-upscattered SD photons dominate the emission. The fact that
the sum of both IC components resembles the profile of the BM model is
not surprising: in this energy range, all the energy of the
annihilation is imparted on leptons which radiate all their energy
away through IC emission and the sum of the IC components represents a
calorimeter for these radiating leptons. We note that the surface
brightness due to SD upscattered photons dominates over larger central
regions for energies $E_\gamma>1$~GeV (cf. Fig.~\ref{fig:IR_comp}).
Furthermore, since there is no enhancement from substructures, the
overall mass normalization of the LP DM model has a marginally
negative trend $\sim\mvir^{-0.17}$. Hence, we only see a very small
difference in outer parts of the different clusters from these type of
models. The DM BM models scale as $S_\gamma\sim\mvir^{0.16}$, hence
more massive clusters show a slightly higher surface
brightness. Finally, we note that the flux from DM without
substructures is dominated by the CR induced emission for all
clusters.

\begin{figure*}
\begin{minipage}{2.0\columnwidth}
  \includegraphics[width=0.49\columnwidth]{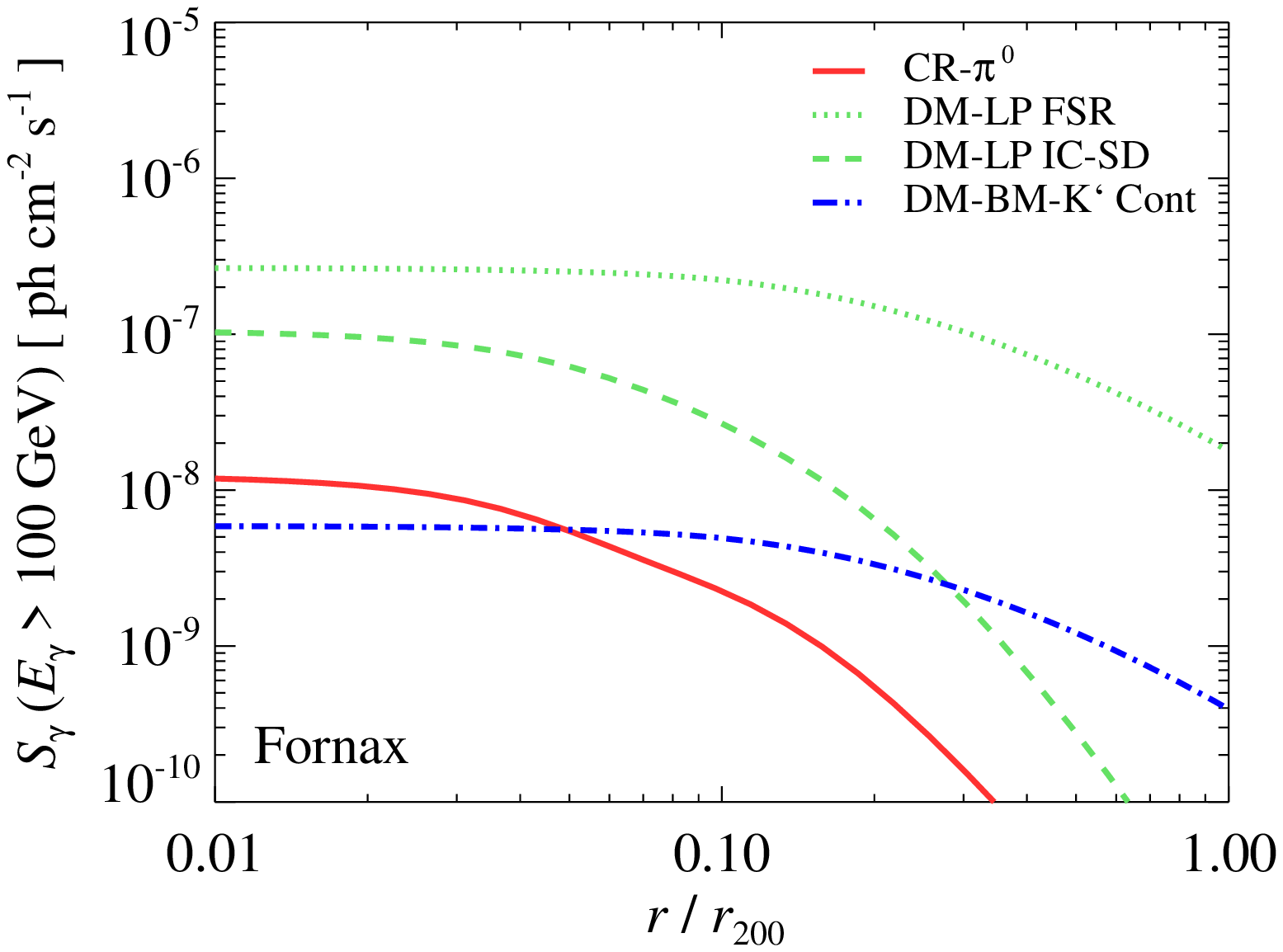}
  \includegraphics[width=0.49\columnwidth]{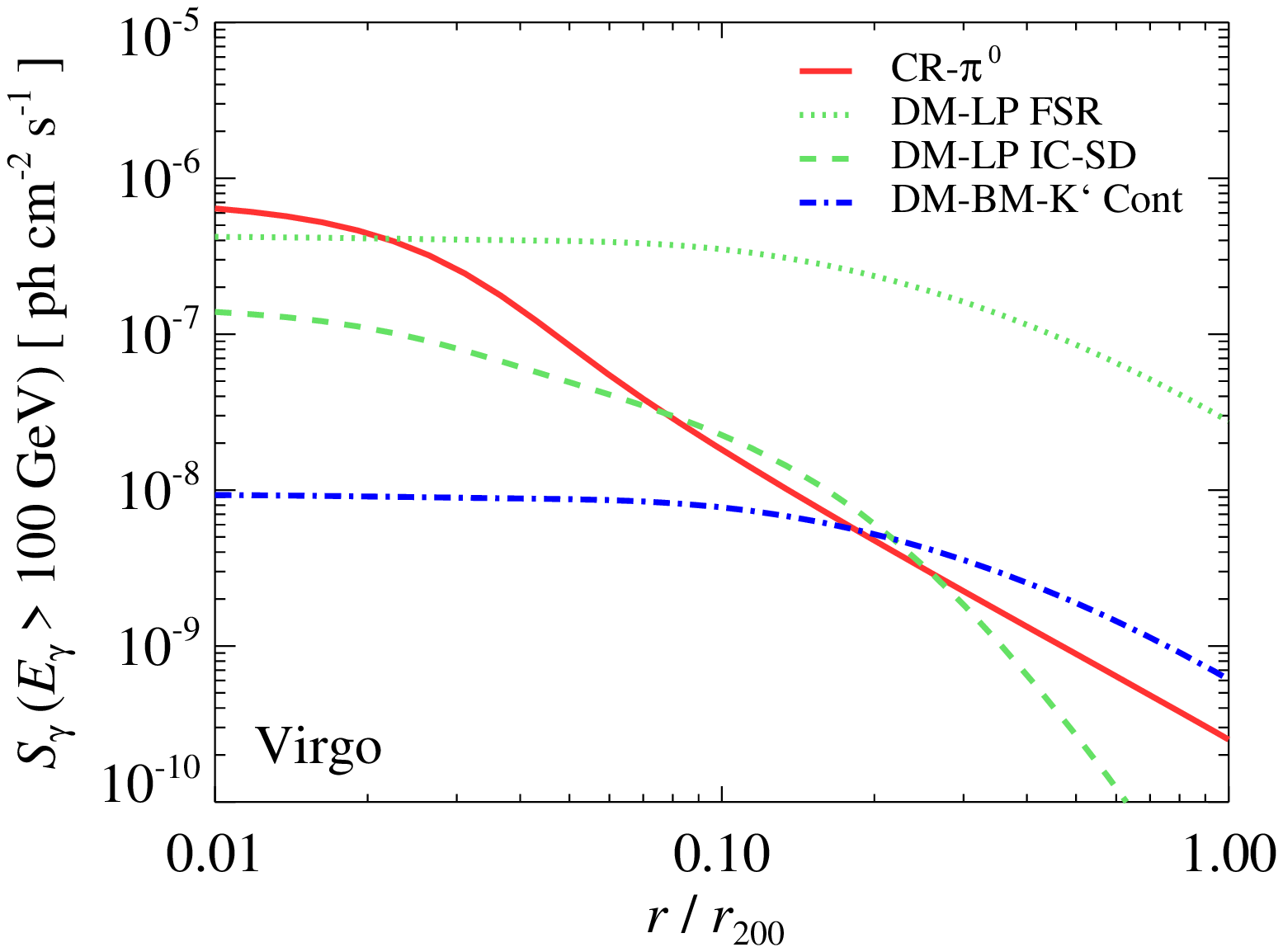}
\caption{\colo Surface brightness predicted for Cherenkov
  telescopes at high energies. We show the emission above $100$~GeV
  and include the boost from substructures. We use a point spread
  function of $\psf=0.1\degs$ that is typical for Cherenkov telescopes
  as well as the \Fermi-LAT at this energy. Left panel shows the
  Fornax cluster and right panel the Virgo cluster. The gamma-ray
  emission is derived for the following components; CRs (red solid),
  continuum emission from the DM $\Kp$ benchmark model (dark blue
  dash-dotted), as well as final state radiation (light green dashed)
  and inverse Compton upscattered dust and starlight (light green
  dotted) from leptophilic DM.}
 \label{fig:SB_IACTs}
\end{minipage}
\end{figure*}

In Sec.~\ref{sect:spectral} we saw that the spectral distribution of
gamma-ray flux from the LP model was dominated at high energies of
$E_\gamma\gtrsim 100\,$~GeV by final state radiation and IC upscattered
starlight and dust, while the BM models were mainly dominated by the
continuum emission. These high energies are observationally important
for both \Fermi-LAT and Cherenkov telescopes. In
Fig.~\ref{fig:SB_IACTs}, we thus show the surface brightness above
$100\,$~GeV for a realistic PSF of $0.1\degs$ where we include the
boost from substructures. We investigate two clusters with an expected
high DM signal; the Fornax and Virgo cluster. Both clusters show a
very flat brightness profile $r<0.1\rvir$, due to both the effect from
substructures and the convolution with the PSF. The CR induced emission
also show a smoothing of the central parts due to the PSF. In the
outer part of the Fornax cluster the CR flux is suppressed compared to
the DM flux boosted by substructures, where the LP model gives rise to
a gamma-ray flux that is more than three orders of magnitude
larger. However, it should be noted that there are large uncertainties
in the gas density profile which we use to estimate the CR induced
gamma-ray surface brightness (see Fig.~\ref{fig:dens_fornax} for more
details).

\section{Population studies: flux predictions and observational limits}

In this section, we compute the expected gamma-ray fluxes from DM
annihilation and CR interactions of the brightest clusters of the
X-ray flux-complete sample in the local universe. All fluxes in this
section are derived within an angle corresponding to $\rvir$; in
addition we neglect the convolution with the PSF if nothing else is
stated.  We confront these predictions to upper limits obtained by
\Fermi 1.5-year data and conclude on the viability of the underlying
models and perspectives for the next years of \Fermi observations.

\subsection{Scaling relations}
\begin{figure*}
  \includegraphics[width=0.99\columnwidth]{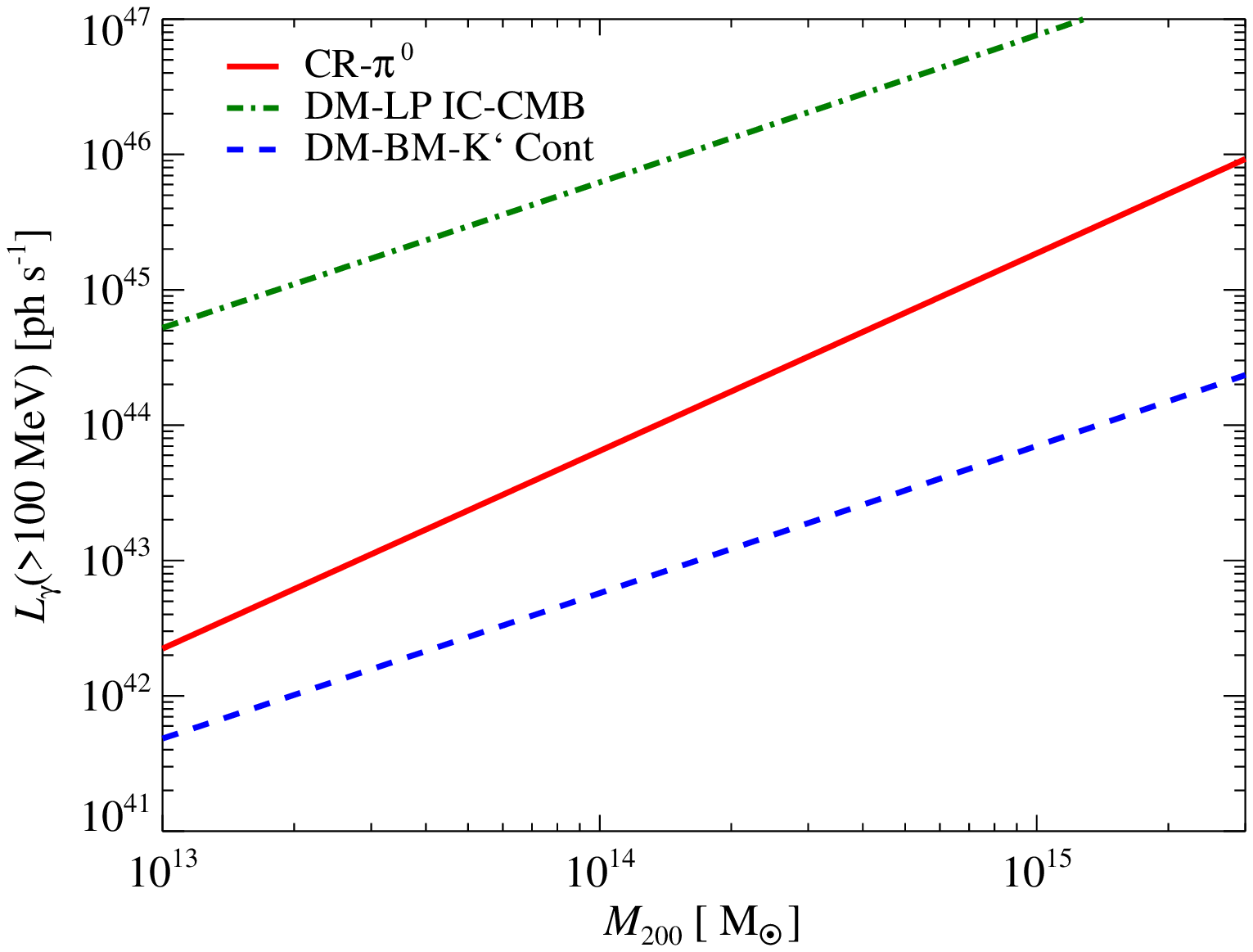}
  \includegraphics[width=0.99\columnwidth]{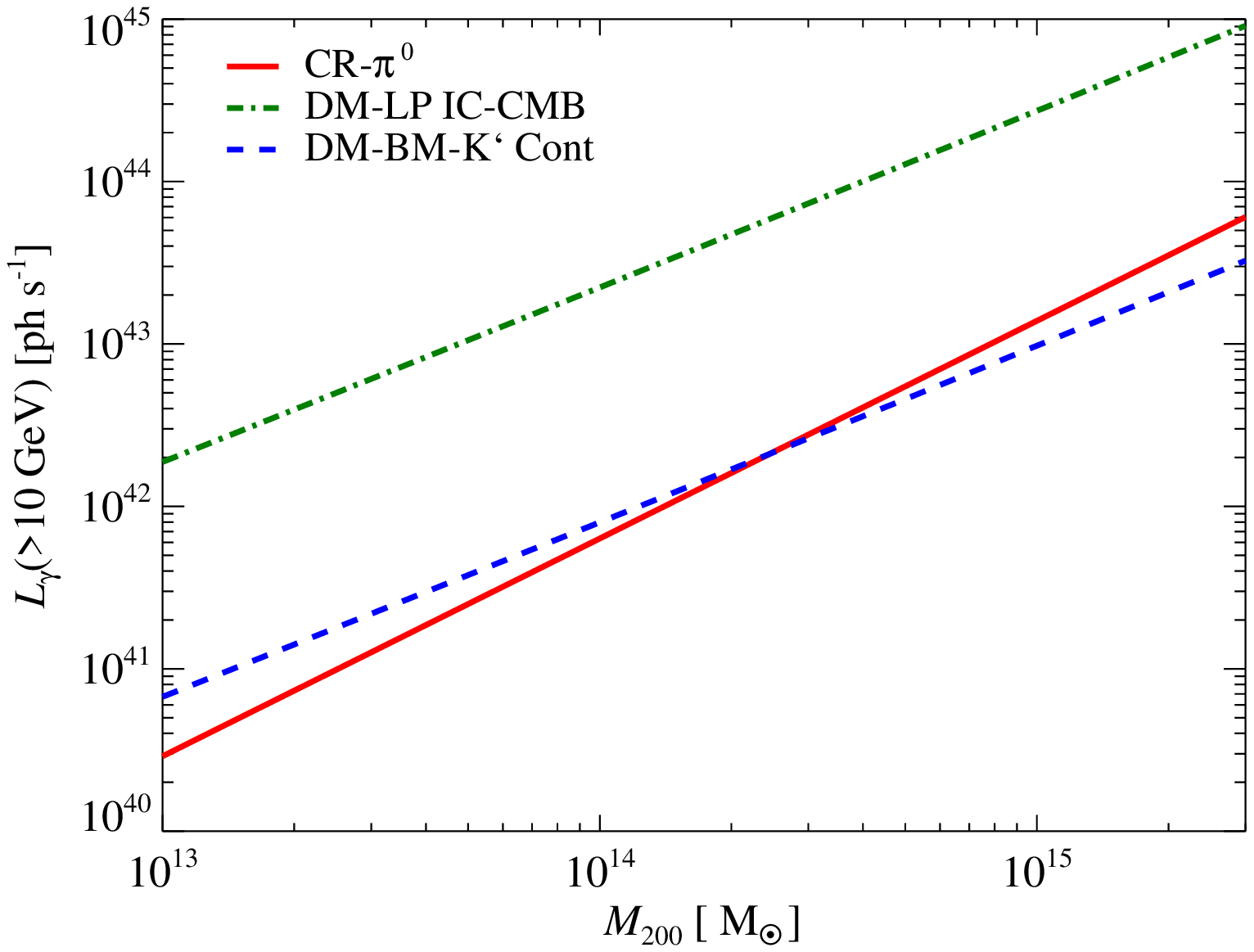}
  \caption{\colo Scaling relations of the cluster's
    virial mass $\mvir$ and the gamma-ray luminosity for energies
    $E_\gamma>100$~MeV (left) and $E_\gamma>10$~GeV (right). Shown are
    the relations for the CR-induced emission that is dominated by
    pion decay resulting from hadronic CR interactions (solid), the
    leptophilic model of DM annihilations (dash-dotted) as well as the
    benchmark model $K'$ of DM annihilation. Both DM models include
    the scaling of the substructure boost with cluster mass of
    Eq.~(\ref{eq:DM_scaling}) (dashed). Note that the scaling of the
    DM models are indicative for all DM annihilation models (as long
    as there is no additional mass scaling from, e.g., non-saturated
    SFE models) while the normalization depends on the particular
    cross section and neutralino mass.}
\label{fig:lum_mass_scaling}
\end{figure*}

First we focus on gamma-ray flux-cluster mass scaling relations for DM
annihilation and CR induced emission in
Fig.~\ref{fig:lum_mass_scaling}.
The mass scaling of the substructure boost steepens the intrinsically
shallower DM annihilation relation in Eq.~(\ref{eq:DM_scaling}) to
$L_\gamma\propto \mvir^{1.06}$. However, the CR scaling relation is
still considerably steeper as shown by reference
  \cite{2010MNRAS.409..449P}:
\begin{eqnarray}
L_{\gamma}(>100\,\mev) = 1.8\times10^{45}
\left(\frac{\mvir}{10^{15}\,\msun}\right)^{1.46}\mbox{ph~s}^{-1}\nonumber\\
L_{\gamma}(>10\,\gev) = 1.4\times10^{43}
\left(\frac{\mvir}{10^{15}\,\msun}\right)^{1.34}\mbox{ph~s}^{-1}\,.\nonumber\\
\end{eqnarray}
The CR luminosity scaling relations\footnote{The scaling relations
  show the flux inside $\rvir$ and do not include the contribution
  from galaxies. Also note that in this paper the scaling relations
  are normalized at $10^{15}\,h_{70}^{-1}\,\msun$. In contrast, the
  corresponding scaling relations in Table 5 in
  \cite{2010MNRAS.409..449P} are normalized at
  $10^{15}\,h^{-1}\,\msun$ instead of the mentioned
  $10^{15}\,h_{70}^{-1}\,\msun$ in the caption. However, all
  figures/tables/equations including the scaling relation figures in
  \cite{2010MNRAS.409..449P} are derived for masses in units of
  $10^{15}\,h_{70}^{-1}\,\msun$.} include the IC gamma-ray
contribution from shock-accelerated primary electrons as well as
secondary electrons created in CR-proton interactions. However, the
gamma-ray flux is dominated by the decaying neutral pions. The
difference in the mass scaling arises from the larger contribution of
primary IC emission for higher-mass clusters at 100~MeV that have a
larger fraction of radio relics due to their greater mass accretion
rates in comparison to galaxy groups---a direct consequence of
hierarchical growth of structure \protect
\cite{2009ApJ...707..354Z,2011arXiv1106.5505P}. The steeper mass
scaling of CR emission compared to DM annihilation already implies a
general strategy to minimize the CR-induced foreground for DM
annihilation and argues for very nearby groups. We note that these
simulations do not include AGN feedback that is thought to furthermore
reduce the baryon fraction in groups relative to that in clusters
\cite{2008ApJ...687L..53P}. The resulting smaller target density for
hadronic CR interactions steepens the $L_\gamma-M$ relation of CR
induced emission, making the case for groups even stronger.

\subsection{DM annihilation}

\begin{figure*}
\begin{minipage}{2.0\columnwidth}
  \includegraphics[width=0.99\columnwidth]{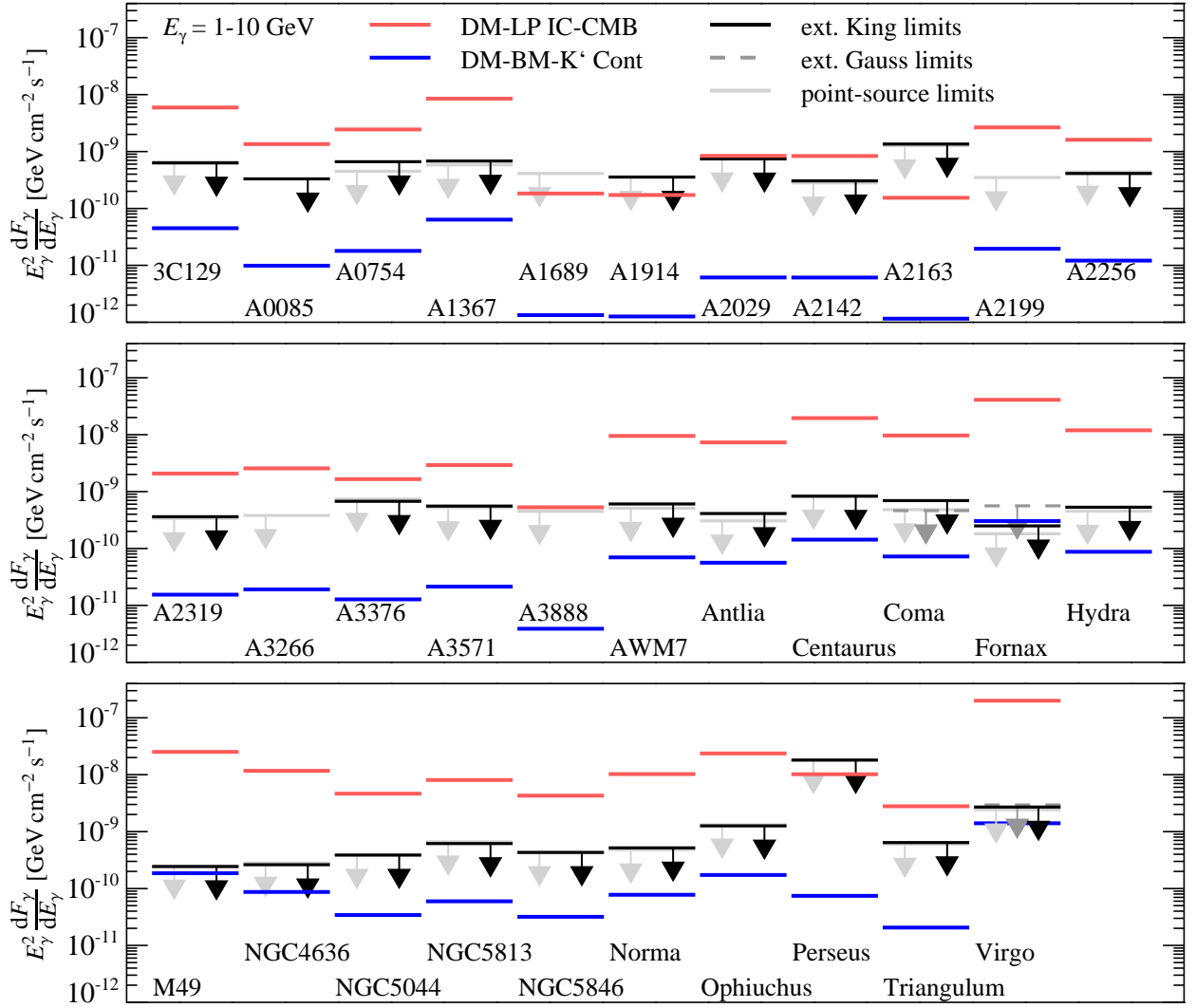}
  \caption{\colo \Fermi gamma-ray flux upper limits are contrasted to
    predicted DM gamma-ray fluxes. We show the mean differential flux
    in the energy range $E_\gamma=1-10$~GeV for 32 clusters. The
    extended \Fermi-LAT upper limits are shown with black arrows (King
    profiles) and grey arrows (Gaussian profiles -- Virgo, Coma,
    Fornax), while the point source limits are shown in light
    grey. The predicted fluxes are derived from the dominant inverse
    Compton-scattering of CMB photons in a leptophilic DM model (light
    red), and the continuum emission from the DM $\Kp$ benchmark model
    (blue). Assuming a boost factor due to substructures that has a
    mass spectrum extending down to an Earth mass, the expected
    leptophilic fluxes are ruled out by upper limits in 28 of the
    clusters, with the strongest constraints set by M49 and Fornax. At
    the present time, we cannot constrain the benchmark models with
    \Fermi-LAT data, although improved modelling of extended sources
    as well as stacking of clusters appears promising in testing
    benchmark models in the future.}
 \label{fig14}
\end{minipage}
\end{figure*}

\begin{figure*}
\begin{minipage}{2.0\columnwidth}
 \includegraphics[width=0.99\columnwidth]{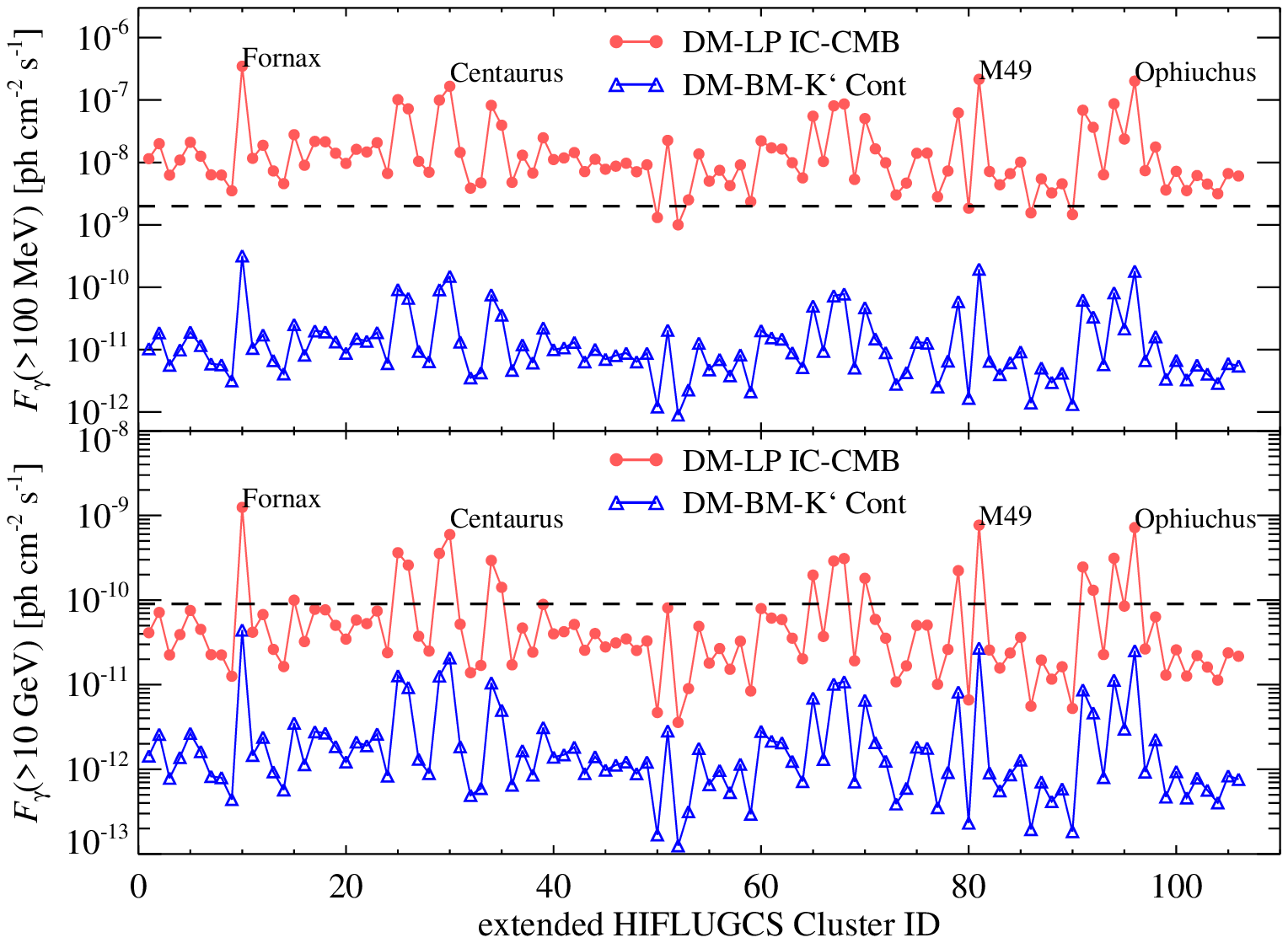}
 \caption{\colo Comparing the DM annihilation flux from clusters in
   the extended HIFLUGCS catalogue. We show the energy integrated
   gamma-ray fluxes derived from both leptophilic DM that result in
   inverse Compton upscattered CMB photons (light red), and the
   continuum emission from the DM $\Kp$ benchmark model (blue). The
   fluxes are calculated within $\rvir$ for each of the 106 clusters
   included in the extended HIFLUGCS catalogue. The upper panel shows
   the energy integrated flux above 100~MeV and the lower panel that
   above 10~GeV, both as a function of HIFLUGCS cluster ID. For
   comparison we show the estimated point source sensitivity of the
   \Fermi-LAT after two years of data taking
   $F_\gamma(>100\,\rmn{MeV})\sim 2\times10^{-9}\,\rmn{ph}\,
   \rmn{cm}^{-2}\,\rmn{s}^{-1}$ and $F_\gamma(>10\,\rmn{GeV})\sim
   9\times10^{-11}\,\rmn{ph}\, \rmn{cm}^{-2}\,\rmn{s}^{-1}$ (dashed
   lines). The four brightest clusters are labelled yielding M49 and
   Fornax as the brightest targets.}
 \label{fig21}
\end{minipage}
\end{figure*}

Figure~\ref{fig14} compares \Fermi upper limits on the gamma-ray flux
with predictions of DM annihilation fluxes in the LP and BM models. We
assume a boost factor due to substructures that has a constant
contribution per decade in substructure mass and has a mass spectrum
extending down to Earth masses for our LP and BM DM annihilation
models. The gamma-ray fluxes are calculated for those clusters where
18 months \Fermi-LAT upper limits are derived (see
\cite{2010ApJ...717L..71A}). These limits rule out the LP models in
their present form with the mentioned assumptions in 28 clusters, and
limit the boost from SFE to less than 5 in M49 and Fornax. Assuming
universality, this limits the Sommerfeld boost in the MW to less than
3 (see Fig.~\ref{fig:boost_const}). The flux level of the \Fermi
limits are an indirect measure of the ambient background flux in the
gamma-ray sky and/or the presence of strong point sources such as AGN
in the immediate vicinity of the cluster position. Hence, the ratio of
the \Fermi upper limits to the predicted annihilation fluxes,
$F_{\mathrm{UL}}/F_{\mathrm{DM}}$, is a good indication of the best
cluster candidates for indirect DM experiments. We identify Fornax,
M49, NGC4636, and Virgo to be prime candidates. Note however, that
Virgo extends $12^\circ$ over the sky which implies a lower surface
brightness and lower signal-to-noise. \Fermi limits on individual
clusters are expected to improve as $\sqrt{T/1.5 \mathrm{yr}}$, where
$T$ is the total elapsed time of the \Fermi mission. We emphasize that
the very inhomogeneous distribution of $F_{\mathrm{UL}} /
F_{\mathrm{DM}}$ makes it unlikely to dramatically improve the limits
through a likelihood stacking analysis as there are only a few
clusters with comparably good flux ratios.

In Figure~\ref{fig21}, we show the DM annihilation flux predictions of
all the clusters in the extended HIFLUGCS catalogue
\cite{2007A&A...466..805C}. We find M49 and Fornax to be that
brightest clusters in the entire sample. While the LP model predicts
that almost all clusters in the sample should be observed by \Fermi
after two years of data taking, the fluxes from the BM models are too
low to be detected in the near future without improved
modelling. Assuming the projected point source sensitivity ($5\sigma$,
50h) of the future Cherenkov telescope array (CTA) of
$F_\gamma(>E_\gamma) = (4\times10^{-11}, 2\times10^{-12},
2\times10^{-14})\,\rmn{ph}\,\rmn{cm}^{-2}\,\rmn{s}^{-1}$ at energies
of $E_\gamma=(10~\rmn{GeV}, 100~\rmn{GeV}, 1~\rmn{TeV})$
\cite{Doro:2009qs}, we note that it will be very challenging to detect
the DM annihilation signal from clusters without a Sommerfeld boost by
Cherenkov telescopes. This is because the boost from substructures is
extended while the sensitivity of Cherenkov telescopes scales
approximately linearly with source extension. Hence the fluxes quoted
in Fig.~\ref{fig21} will have to be compared to a sensitivity that is
scaled down by the ratio of cluster radius to angular resolution of
$0.1^\circ$ assuming current background subtraction techniques of
Cherenkov telescopes. This important finding should encourage the
development of new methods to overcome the degradation of sensitivity
for diffuse and very extended sources. Such a break-through would be
needed to probe and potentially detect the presented class of BM
models with a large investment of observational time.

\subsection{CR-induced emission}
\label{sec:CRemission}
\begin{figure*}
\begin{minipage}{2.0\columnwidth}
  \includegraphics[width=0.99\columnwidth]{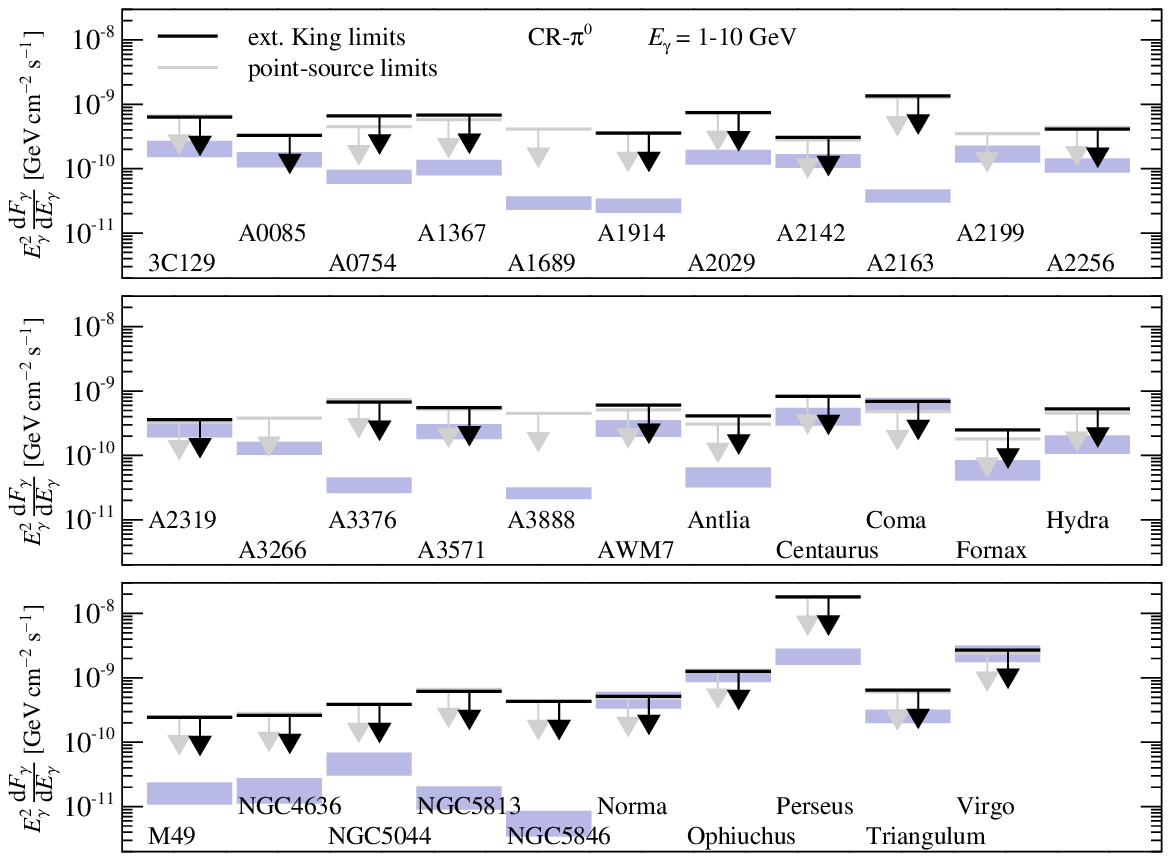}
  \caption{\colo We contrast predictions of the CR
    induced gamma-ray emission using an analytical CR model \protect
    \cite{2010MNRAS.409..449P} to \Fermi-LAT flux upper limits. We
    show the mean differential flux in the energy range
    $E_\gamma=1-10$~GeV for 32 clusters. The extended \Fermi upper
    limits assume King profiles and are shown with black arrows, while
    the point source limits are show with grey arrows. The blue boxes
    show the gamma-ray emission from CR-induced $\pi^0$-decay, where
    the upper (lower) bounds show the estimates for an optimistic
    (conservative) model (see Sec.~\ref{sec:CRemission} and
    \cite{2010MNRAS.409..449P} for details). Note that our models are
    in perfect agreement with the derived upper limits from
    \Fermi. Interestingly, the limits of Virgo, Norma, and Coma are
    closing in on our conservative predictions and will enforce
    constraints on the parameters of hadronic models such as shock
    acceleration efficiencies or CR transport properties in the coming
    years.}
 \label{fig15}
\end{minipage}
\end{figure*}

\begin{figure*}
\begin{minipage}{2.0\columnwidth}
  \includegraphics[width=0.99\columnwidth]{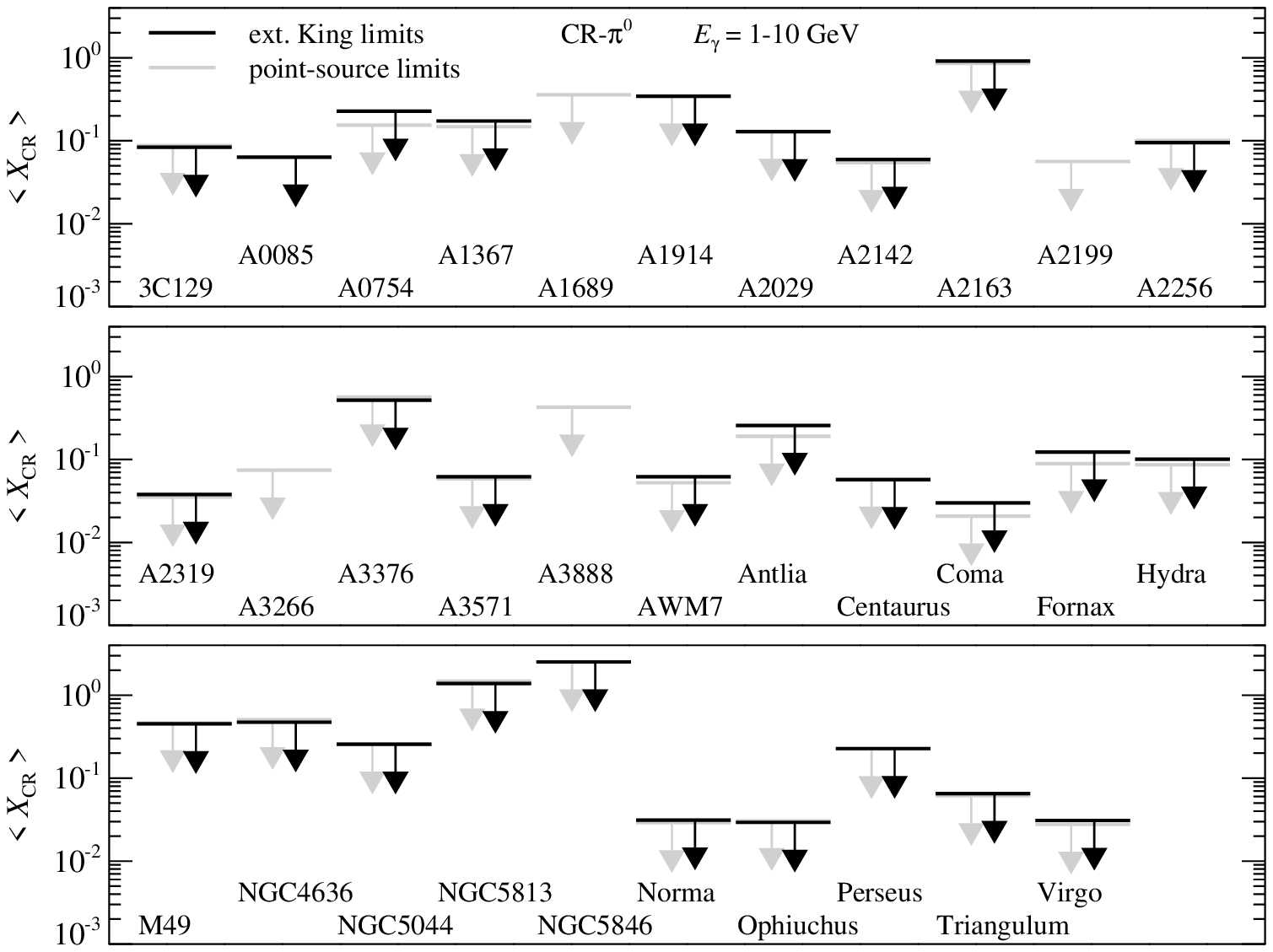}
  \caption{Limits on the relative CR pressure, $\bra
    X_\rmn{CR}\ket = \bra P_\rmn{CR} \ket / \bra P_\rmn{th}\ket$
    averaged across the cluster. Our constraints are obtained with the
    extended \Fermi-LAT limits in the energy range $E_\gamma=1-10$~GeV
    that assume a King profile which matches the extension of the
    thermal X-ray emission (black) and for comparison with \Fermi-LAT
    point-source limits (grey) \cite{2010ApJ...717L..71A}. In
    computing those limits, we adopted our (conservative) analytic
    model \cite{2010MNRAS.409..449P} and an averaged relative CR flux
    as obtained from our simulations.}
 \label{fig:XCR}
\end{minipage}
\end{figure*}

\begin{figure*}
\begin{minipage}{2.0\columnwidth}
 \includegraphics[width=0.99\columnwidth]{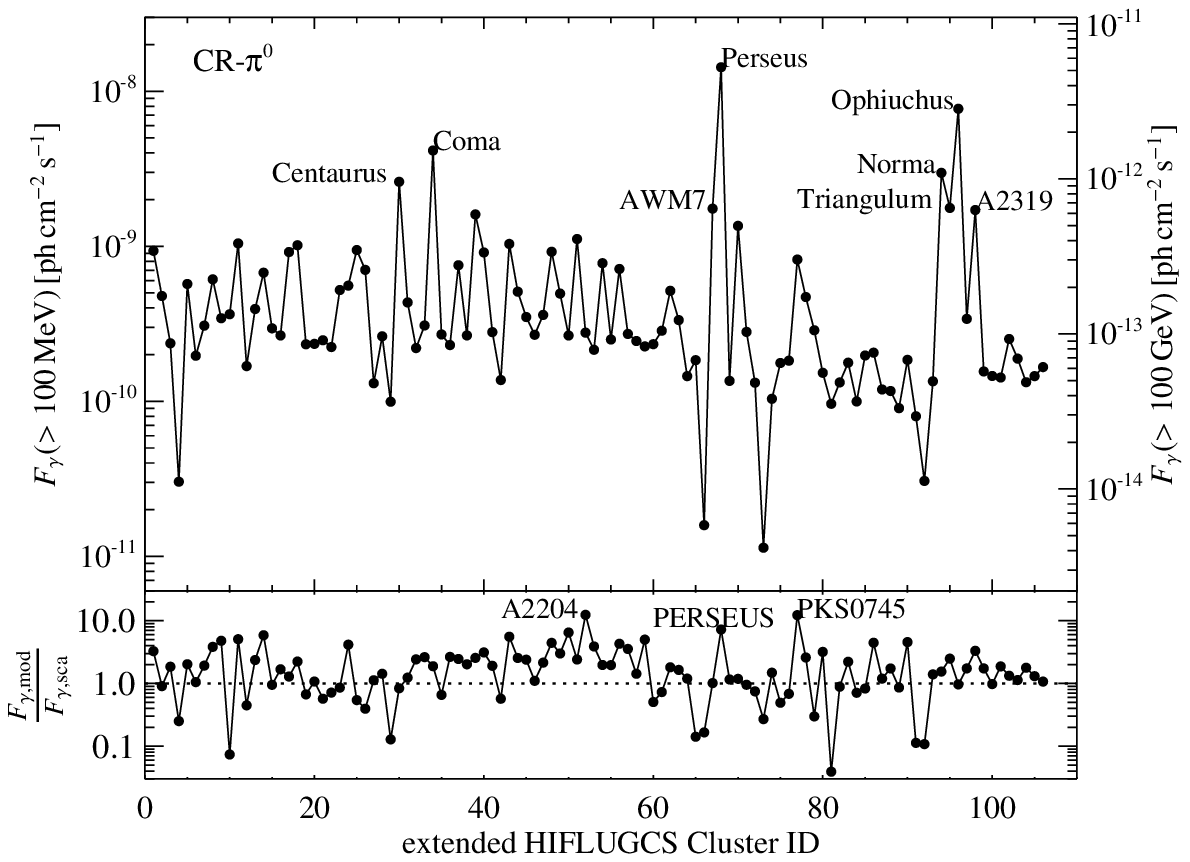}
 \caption{Comparison of the gamma-ray flux induced by CRs for the
   106 clusters included in the extended HIFLUGCS catalogue. The CR
   distribution follows the analytical model derived through
   hydrodynamical cosmological cluster simulations, where the spectral
   shape is independent of cluster mass \protect
   \cite{2010MNRAS.409..449P} (see text for details). Gamma-ray fluxes
   are calculated within $\rvir$ and are derived using a single or
   double beta profile for each cluster's gas density profile as
   obtained by ROSAT X-ray observations
   \cite{2007A&A...466..805C}. The upper panel shows the energy
   integrated flux above 100~MeV (left side) and above 100~GeV (right
   side), both as a function of HIFLUGCS cluster ID. The eight
   brightest clusters are labelled. The lower panel shows the relative
   difference between the flux above 100~MeV computed using the
   analytical model and the gamma-ray flux predicted by the
   mass-luminosity scaling relation \cite{2010MNRAS.409..449P}. The
   clusters with the largest positive offset are labelled. As
   expected, they all are cool core clusters.}
 \label{fig19}
\end{minipage}
\end{figure*}

Figure~\ref{fig15} shows flux predictions by an analytical CR model
\cite{2010MNRAS.409..449P} for CR induced gamma-rays and compare those
to the extended \Fermi-LAT flux upper limits that assume King profiles
which trace the X-ray emitting gas. For consistency, we use single-
and double-beta model fits to ROSAT data of X-ray bright clusters for
our gamma-ray flux estimates as an input
\cite{2007A&A...466..805C}. The deprojection is performed following
the formalism developed in e.g. \cite{2004A&A...413...17P}. These
gamma-ray flux predictions are superior to those that use
luminosity-mass scaling relation (a method used by reference
\cite{2010ApJ...717L..71A} to compare to \Fermi limits) which do not
account for the substantial scatter in the scaling relation. The
analytic CR model that we employ here follows CR-acceleration at
structure formation shocks and their advective transport over cosmic
time. This leads to a CR distribution with a universal spectral form
and an almost universal spatial distribution. Depending on whether or
not we account for the bias of ``artificial galaxies'' in cosmological
simulations, we derive an optimistic or conservative limit of the
expected gamma-ray emission from decaying pions (bracketing our
ignorance of the contribution of cluster galaxies and ram-pressure
stripped gas parcels from infalling galaxies to the overall gamma-ray
flux from a cluster). As discussed in Sec.~\ref{sect:CRs}, the model
predictions are subject to two main uncertainties, the acceleration
efficiency at formation shocks and CR transport parameters and hence
may be scaled down depending on specific values realized in clusters.

The tightest \Fermi limits are obtained in the 1-10 GeV regime due to
the highest sensitivity of \Fermi-LAT there. While the effective area
of LAT increases up to these energies, the typical photon spectra
decrease as a function of energy: combining those effects selects this
energy range to be most sensitive. In this energy band, the \Fermi
limits close in on the conservative flux predictions of Virgo, Coma
and Norma. Using integrated flux limits from \Fermi, we arrive at very
similar results. The insensitive limit on Perseus is due to the
bright central source NGC1275 with an AGN in the center
\cite{2010ATel.2916....1M} making it hard to determine a comparable
limit as obtained for the other clusters. Additionally, recent
gamma-ray observations of the head-tail radio galaxy IC310 complicate
matters furthermore \cite{2010ApJ...723L.207A,2010A&A...519L...6N}.

A quantity that is of great scientific interest is the CR pressure
($P_\CR$) relative to the thermal pressure ($P_\rmn{th}$) as it may
bias galaxy cluster observables, e.g., the Sunyaev-Zel’dovich signal
or hydrostatic mass estimates, and the resulting cosmological
constraints if the analysis does not marginalize over a potential
pressure contribution from CRs. Before we turn to the newly derived
constraints on the relative CR pressure, $X_\CR = P_\CR/P_\rmn{th}$,
we study the Bayesian prior that different models (a simulation based
and a simple analytic model) impose on the scaling of $X_\CR$ with
cluster mass.

In the simulation based analytic CR model \cite{2010MNRAS.409..449P},
the CR pressure derives from the CR distribution function, $f(p)\dd p =
C\,p^{-\alpha}\dd p$, where $C$ is the normalization, $p=P/m_\rmn{p}
c$ is the normalized CR proton momentum, and $\alpha$ is the CR
spectral index. The pressure is given by $P_\CR \propto C\propto
\tilde{C} \,\rho_\rmn{gas}$, where the volume weighted, dimensionless
normalization scales as $\bra\tilde{C}\ket\propto \mvir^{0.44}$ for clusters
with a mass $\mvir\gtrsim 10^{14}\,\msun$
\cite{2010MNRAS.409..449P}. The thermal pressure $P_\rmn{th} =
n_\rmn{gas} k_\B T$ and is derived from the temperature in virial
equilibrium that scales as $k_\B T \propto G \mvir/R_{200} \propto
\mvir^{2/3}$. Hence, the volume weighted relative CR pressure scales
as
\begin{equation}
    \bra X_{\CR,\,\rmn{sim}}\ket = \frac{\bra P_\CR\ket}{\bra
      P_\rmn{th}\ket} \propto \frac{\bra\tilde{C}\ket}{\bra k_\B
      T\ket} \propto M^{-0.22}\,,
\end{equation}
which depends only weakly on cluster virial mass. This suggests that
CRs, which end up in a cluster, are accelerated at the strongest
formation shocks, i.e. during proto-cluster formation at high$-z$ or
in between voids and pre-collapsed IGM. These CRs experience a
similar transport history in the form of adiabatic compression or
encounter shocks of similarly (weak) strength so that their
distribution becomes only weakly dependent upon cluster mass.

In contrast, the analytic (isobaric) CR model
\cite{2004A&A...413...17P}, that was used by the \Fermi collaboration
to derive CR pressure constraints \cite{2010ApJ...717L..71A}, assumes
$P_\CR = X_\CR P_\rmn{th}$ and constrains $X_\CR$ by assuming no
intrinsic cluster mass dependence of the CR pressure. While
apparently very natural, this implies a strong cluster mass dependence
of $X_\CR$ as the following estimate shows:
\begin{equation}
    \bra X_{\CR,\,\rmn{iso}}\ket = \frac{\bra P_\CR\ket}{\bra P_\rmn{th}\ket}
    \propto \frac{1}{\bra n_\rmn{gas} k_\B T\ket} \propto M^{-0.8}\,.
\end{equation}
Here, we determine the mass scaling of the average gas density through
its relation to the enclosed gas fraction, $f_\rmn{gas}$, and find
that $n_\rmn{gas}\propto f_\rmn{gas} \propto M^{0.135}$
\cite{2009ApJ...693.1142S}.

Figure~\ref{fig:XCR} shows the volume weighted relative CR pressure
for 32 clusters, for which extended \Fermi-LAT upper limits are
derived (that assume King profiles). We adopt a constant $\bra
X_\CR\ket\approx 0.02$ derived from the average of the large clusters
in our simulation sample \cite{2008MNRAS.385.1211P,
  2010MNRAS.409..449P}. Using the extended gamma-ray upper limits on
most of the clusters in the \Fermi sample, we can constrain $\bra
X_\CR \ket< 0.1$. The most constraining clusters are Norma, Ophiuchus,
and Coma with $\bra X_\CR\ket < 0.03$. Most of the 32 clusters have
limits on $\bra X_\CR\ket$ within a factor few from the best
constraining clusters suggesting that stacking of clusters will be
able to improve these limits (provided these analysis adopt the
simulation-based analytic model \cite{2010MNRAS.409..449P} as a
prior).

In a recent work by the \Fermi collaboration
\cite{2010ApJ...717L..71A}, where the 1.5 year \Fermi data was used to
derive limits on $X_\CR$, they found that the most massive clusters
are on average more constraining yielding the best constraints of the
order of 5\% (by adopting, however, point-like upper limits). In this
work we find that medium-sized clusters provide similarly strong
constraints. This difference is explained by the different priors on
$P_\CR$. We adopt a prior inferred from cosmological simulations that
dynamically trace the CRs during cluster assembly suggesting that
adiabatic transport is dominant in shaping the CR population in
clusters and implying a weak mass dependence of $X_\CR$. Their
analysis assumes a CR pressure that is independent of cluster mass,
i.e., constraints on $X_\CR$ inherit the inverse mass dependence of
the thermal pressure, $X_\CR\propto 1/P_\rmn{th}\propto
M^{-0.8}$. While it is not clear which mass scaling of $X_\CR$ is
realized in Nature, these considerations suggest that it will be
critical for future work to account for the Bayesian prior in deriving
limits on their pressure.

To complete this section, we present our analytical model predictions
for the CR-induced gamma-ray flux for all clusters in the extended
HIFLUGCS sample in Fig.~\ref{fig19}. In the lower panel of this
figure, we show the relative difference between the gamma-ray flux
computed using the analytical CR model to the flux predicted by the
mass-luminosity scaling relation. Statistically, one can view the
analytical modeling as explicitly accounting for the scatter of the expected
gamma-ray flux at a given mass. Cool core clusters have a denser core
which increases the target density of gas protons for the hadronic
reaction which in turn leads to a systematic increase of the expected
gamma-ray fluxes. We confirm this effect since all clusters with
increased flux ratio of our analytical CR model relative to the scaling relation
expectation are in fact cool core clusters. This suggests that the X-ray
emission should be a good proxy for the expected gamma-ray
emission. Note however that there could be the counter-acting effect
of CR streaming in relaxed clusters which would tend to decrease the
gamma-ray luminosity \cite{2011A&A...527A..99E}. Future careful
modelling is needed to quantify this effect.

\section{Conclusions}

In this paper, we study the possibility for detecting gamma-ray
emission in galaxy clusters. We consider benchmark (BM) and
leptophilic (LP) models of supersymmetric dark matter (DM) as well as
cosmic ray (CR) induced pion decay which is thought to dominate the
astrophysical signal from clusters.

{\em Dark matter annihilation.} Once supersymmetric cold DM decouples
from the expanding universe it streams and erases any potentially
present smaller scales. When gravitational instability causes halos to
collapse, this free-streaming scale imprints as a characteristic
minimum halo mass $M_\mathrm{min}$ into the hierarchy of structure
formation and should still survive until the present time as a mass
cutoff in the substructure mass function within larger halos. We show
that the boost due to substructures dominates over the smooth
component for halo masses $\mvir>10^3 \msun$ and contributes a
constant luminosity for each decade in subhalo mass. The ratio of
halo-to-minimum subhalo mass $\mvir/M_\mathrm{min}$ is maximized for
the largest virialized systems in the Universe---galaxy clusters. The
corresponding substructure boost can reach values exceeding $10^3$
which should make nearby clusters the brightest DM annihilation
sources after the MW center and hypothetical very close-by DM
substructures. We stress that this conclusion relies on two hypotheses
that are still open problems in particle physics and numerical
cosmology. We assume (i) the existence of cold DM, i.e., the density
fluctuation power spectrum extends down to the DM free streaming scale
of about an Earth mass, and (ii) assume that structure formation at
scales smaller than currently resolved ones ($\sim 10^5\,\msun$)
proceeds scale-invariant, i.e., there is no non-linear mode coupling
between different scales. This could potentially erase structures
smaller than a characteristic one as the linear power spectrum
approaches $\Delta^2(k) \propto k^3 P(k)=\rmn{const.}$, implying that
structure on different scales may not form any more hierarchically,
but may collapse at the same time.

Since the mass density of substructures peaks at radii close to
$\rvir$, the contribution to the annihilation emission from
substructures is also dominated from these outer cluster regions.
This implies an almost flat surface brightness profile of the
annihilation emission which makes it necessary to have the entire
cluster in the field-of-view to take advantage of the total
substructure boost. This property makes it difficult to detect DM
annihilation emission without an enhancement of the particle physics
cross section over its standard value of $\sigma v\sim 3\times
10^{-26} ~\mathrm{cm}^3~\mathrm{s}^{-1}$ with imaging air Cherenkov
telescopes due to the spatially extended boost from substructures over
an angular extent of $\sim 0.5-1\degs$ (as independently found by
reference \cite{2011arXiv1104.3530S}). This large angular extent is
not well matched with the sensitivity of Cherenkov telescopes which
drops approximately linearly with radius outside the point spread
function (with a characteristic scale of $\sim 0.1\degs$), unless new
background subtraction schemes can be found, e.g., for the survey mode
of CTA.

In this work, we thoroughly study the spectral emission
characteristics of the DM emission. In general there are different
radiation mechanism that emit in the gamma-ray regime, namely (i)
continuum emission following the hadronization of the annihilating
neutralinos in BM models which leads to the production of charged and
neutral mesons which decay finally into electrons, neutrinos, and
gammas, (ii) inverse Compton (IC) emission by the leptons in the final
state which up-scatter radiation fields from the cosmic microwave
background (CMB), dust, and starlight, and (iii) final state radiation
or internal bremsstrahlung which dominates the highest energy emission
close to the rest mass of the self-annihilating neutralinos (if
present in the model under consideration).\footnote{While this final
  state radiation is sometimes not distinguished from the continuum
  emission contribution, we keep it separate as it dominates the very
  high-energy tail of the gamma-ray spectrum in LP models while there
  the continuum emission is negligibly low.}

For the first time, we compute the IC emission component not only from
the upscattered CMB photons but also from a realistically modeled
photon field due to dust and starlight extending from infra-red to
optical wavelengths. To this end, we construct a rather simple
analytic model for the spatial and spectral distribution of dust and
starlight in clusters (which can also be used for the annihilation
emission in galaxies after adopting a different spatial profile). We
find that in BM models, the continuum emission dominates over the IC
emission of any available radiation field. In the LP models, where the
continuum emission is low, the IC up-scattering of the CMB dominates
below 60 GeV. For larger energies, the final state radiation dominates
the emission up to the cutoff at $m_\chi c^2 \sim 1$~TeV. The IC
up-scattering of the dust and starlight is always subdominant compared
to the upscattered CMB (below 60~GeV) and the final state radiation
(above 60~GeV). We stress, that the inclusion of the dust and
starlight radiation fields are negligible in the cooling function for
the leptons. The main reason why the IC emission from starlight and
dust remains subdominant is the small overlap of this component with
the peripherally peaked substructure mass density profiles. If for
some reason, DM has less substructure, the relative importance of this
IC component increases considerably.

We identify Virgo, Fornax and M49 to be the best cluster/group
candidates for indirect DM studies as they are expected to emit the
brightest annihilation flux.\footnote{We excluded Virgo from our
  further analysis due to its large angular extent that needs a
  separate treatment.} More importantly, Fornax has a comparably low
CR-induced gamma-ray flux which may enable an indirect DM detection or
a particularly tight limit on DM properties. Other nearby bright
sources of the extended HIFLUGCS sample are Ophiuchus, and Centaurus.

The non-detection of gamma-ray emission by \Fermi in a total of 28
clusters/groups considerably constrains LP models for DM annihilation
which were introduced to explain the increasing positron-to-lepton
ratio beyond 10 GeV. Assuming that the minimum DM substructure mass
extends down to an Earth mass and that the DM annihilation flux is
dominated by small scale substructure, we limit the saturated
Sommerfeld enhancement factor in M49 and Fornax to $\lesssim 5$. This
corresponds to a value in the Milky Way (MW) of $\lesssim3$, due to
the larger Sommerfeld boost factor in smaller mass halos which have a
lower velocity dispersion that enhances the particle cross section
(while assuming universality of the DM model). This would rule out LP
models in their current form based on the non-observation of
gamma-rays in any of the fore-mentioned clusters/groups and hence
strongly challenge the DM interpretation of the increasing positron
fraction with energy as seen by PAMELA. Alternatively, assuming the LP
models to be correct, this would limit the minimum substructure mass
to $>10^4~\msun$ in M49 and Fornax which presents a problem for
structure formation in most particle physics models
\cite{2009NJPh...11j5027B}.

{\em Cosmic ray-induced emission.} We substantially improve the
modelling of the expected gamma-ray signal from pion-decay resulting
from hadronic CR interactions with gas protons over previous work
\cite{2010ApJ...717L..71A}. We employ the analytic model of the CR
spectrum and spatial distribution that is based on cosmological
hydrodynamical simulations of galaxy clusters that self-consistently
follow the evolution of CRs in cosmic structure formation
\cite{2010MNRAS.409..449P}. Adopting the model for all clusters in the
HIFLUGCS sample of flux-limited X-ray clusters, we compare the
expected gamma-ray emission to that inferred from using an
$L_\gamma-\mvir$ scaling relation. We note that the predictions of the
``optimistic'' model in \cite{2010MNRAS.409..449P} are in tension with
\Fermi upper limits for Norma, Coma, and Ophiuchus. In the next 2-3
years \Fermi will be able to probe the predictions of their
``conservative'' CR model. This will enable us to put realistic limits
on a combination of the acceleration efficiency of CRs in structure
formation shocks and CR transport coefficients in clusters.

As expected, the analytical CR model accounts for the ``scatter'' in
the scaling relation and biases the gamma-ray flux high for prominent
cool core clusters by up to a factor of 10 relative to the expectation
from the scaling relation. This is due to the high target gas
densities and associated CR densities in those cool cores that shape
the spatial emission characteristic to be very similar to that
observed in thermal X-rays. We caution that these predictions only
apply for the case of negligible active CR transport such as CR
streaming and diffusion relative to the gas. Gamma-ray fluxes might be
considerably smaller in those cluster that do not show an extended
diffuse radio(-mini) halo that might point to a centrally concentrated
CR distribution \cite{2011A&A...527A..99E}. The CR spectrum $E^2
dN/dE$ shows the characteristic pion bump at energies around 1~GeV. It
is expected to dominate the DM annihilation signal of BM models for
clusters/groups, although we find that the gamma-ray flux induced by
the BM is higher for about 1/5 of the clusters in the HIFLUGCS
catalogue, where Fornax and M49 have the highest signal-to-noise
ratio. (Note that we only include the cluster-intrinsic CR foreground
and do not consider the dominant galactic and instrumental noise
sources here.)

Combining extended \Fermi upper limits and our model flux predictions,
we limit the relative CR pressure, $X_\CR = P_\CR/P_\rmn{th}$, in 32
nearby bright galaxy clusters of the \Fermi sample. The best limits
are found in Norma and Coma of the order of 3\%, with typical limits
around 10\%. This is comparable to those inferred by the \Fermi
collaboration \cite{2010ApJ...717L..71A} and mainly due to the
differently assumed Bayesian prior on the CR pressure which implies a
different cluster mass dependence of the resulting limits on $X_\CR$.


\smallskip We would like to thank Volker Springel, Megan Donahue,
Stefano Profumo, Tesla Jeltema and Fabio Zandanel for helpful
discussions. We express our thanks to Neal Weiner and Tracy Slayter
for many comments and suggestions that improved an earlier version of
this manuscript. A.P. acknowledges NSF grant AST 0908480 for
support. We would furthermore like to thank KITP for their hospitality
during the galaxy cluster workshop. This research was supported in
part by the National Science Foundation under Grant No. NSF
PHY05-51164. C.P. gratefully acknowledges financial support of the
Klaus Tschira Foundation. The work of L.B. was supported by the
Swedish Research Council (VR), under contracts no. 621-2009-3915 and
349-2007-8709 (the Oskar Klein Centre).

\vspace{-0.7cm}
\bibliography{bibtex/paper}

\appendix

\section{Detailed modelling of starlight and dust in clusters}
\label{sect:SD}
In this section, we present the details of our model of stars and dust
(SD) in clusters of galaxies. In the first part, we use the measured
spectral distribution of light from SD in a typical galaxy to derive a
simple spectral model for SD in galaxy clusters. This is done by
renormalizing the galactic SD distribution to match the observed luminosity
from a cluster in the far infra-red (IR) and ultra-violet (UV) bands
separately. In the second part, we derive the spatial distribution of
the SD light from a stacked sample of clusters.

\subsection{Energy spectrum}
The cluster emission from far-IR to optical/UV is due to dust emission
and stellar light, respectively. We distinguish three components: the
brightest cluster galaxy (BCG), the intra-cluster light (ICL) and
individual galaxies. While the latter distribution is highly clumped,
the first two components are smoothly distributed. We use the far-IR
to UV spectrum derived in reference \cite{2006ApJ...648L..29P} for a
Milky Way (MW)-type galaxy to characterize the spectral distribution
of SD in a cluster. Note that we only keep the spectral shape, and
renormalize the amplitude of the spectral distribution using the
luminosity from SD in clusters.

In Fig.~\ref{fig:SD_spectra}, we show the cosmic microwave background
(CMB) black-body distribution together with the spectral distribution
of SD light of a $6.0\times10^{14}\,\msun$ galaxy cluster. We fit the
spectral shape of SD individually; the dust that peaks at about
$10^{-2}\,\ev$ is fitted using a double power-law, while the broader
spectral distribution of the stars that peaks at about $1\,\ev$ is
fitted using a triple power-law. The best fit to the galactic spectrum
is given by:
\begin{eqnarray}
  u_\stars^\gal(\eph) &=& \frac{23\,\rmn{eV}}{\rmn{cm}^3}
  \left(\frac{1.23\,\rmn{eV}}{\eph}\right)^{1.9} \nonumber \\
  &\times&\left[1+\left(\frac{2.04\,\rmn{eV}}{\eph}\right)^{20}\right]
  ^{-\frac{1.9}{20}}\nonumber \\
  &\times& \left[1+\left(\frac{0.78\,\rmn{eV}}{\eph}\right)^{20}\right]^{-\frac{2.6}{20}}\,, \\
  u_\dust^\gal(\eph) &=&
  \frac{40\,\rmn{eV}}{\rmn{cm}^3}
  \left(\frac{0.0144\,\rmn{eV}}{\eph}\right)^{4.9}\nonumber \\
  &\times& \left[1+\left(\frac{0.0144\,\rmn{eV}}{\eph}\right)^{1.9}\right]^{-4.9}\,.
\end{eqnarray}
The specific energy density of the light from SD in a cluster can now
be derived from
\begin{eqnarray}
u_\sd(\eph, r) &=& j(r)\,u_\sd(\eph)\,,
\label{eq:u_SD_er}
\end{eqnarray}
where the spectral distribution of SD is given by
\begin{eqnarray}
  u_\sd(\eph) &\equiv& \eph^2\frac{\dd^2 N_\ph}{\dd \eph \dd V}
  = \sum_i N_i(\mvir) u_i^\gal(\eph)\,,\nonumber \\
\rmn{for}\quad i&=&\{\rmn{stars,dust}\}\,.
\end{eqnarray}
Here $\eph$ denotes the energy of the SD photons, $j(r)$ describes the
unitless spatial distribution of SD (derived in
Sec.~\ref{sect:SD_radial}), and $N_i(\mvir)$ denotes the mass
dependent normalization for the SD. This function is determined using
the total energy of the light from SD within $\rvir$, which is
represented by $E_{i,\vir}$. We relate this energy to the luminosity
$L_i$ in each respective wavelength band (far-IR light for the
dust, and optical/UV light for the stars) using
\begin{eqnarray}
  E_{i,\vir} &=& L_i \frac{\rvir}{c} \nonumber \\
  &=&N_i(\mvir)\int_{\rvir} \int_i \,j(r)
  \frac{u_i^\gal(\eph)}{\eph}\,\dd V\dd \eph\,,\nonumber \\
 \rmn{for}\quad i&=&\{\rmn{stars,dust}\}\,.
\label{eq:E_SD}
\end{eqnarray}
Here we approximate the total energy of the photons within $\rvir$
with the SD luminosity multiplied with the typical timescale that it
takes for a photon to propagate through the cluster (we assume that
the cluster is optically thin).

The {\em luminosity from starlight} is given by
\begin{eqnarray}
L_\stars(M_{200\rmn{av}})&=&10^{\frac{m_\odot-m_\rmn{cl}-2.5\log\left(1+z\right)}{2.5}}
\left(\frac{D_\clu}{D_\odot}\right)^2 L_\odot\nonumber\\
&\approx& 5.3\times10^{44}\,\rmn{erg}\,\rmn{s}^{-1}\,,
\label{eq:L_stars}
\end{eqnarray}
where the average apparent magnitude of a cluster at $z\approx 0.25$
in the $(r+i)$-band is given by $m_\rmn{cl}\approx 15.5$
\cite{2005MNRAS.358..949Z}. This magnitude is derived from stacked
clusters observed by the SDSS with the average mass
$M_{200\rmn{av}}=4.0\times10^{13}\msun$. We also use an apparent
magnitude of the sun $m_\odot\approx -27.7$, distance to the sun
$D_\odot=4.85\times10^{-6}\,\rmn{pc}$, luminosity of the sun
$L_\odot=1.17\times10^{33}\,\rmn{erg\,s}^{-1}$, and distance to the
cluster average of $D_\clu=1.26\times10^9\,\rmn{pc}$. Furthermore, to account for
different cluster masses, we employ a simple mass scaling for the
luminosity from stars in Eq.~(\ref{eq:L_stars}). Here we assume
that the total starlight has the same halo mass scaling as the
brightest cluster galaxy (BCG) \cite{2010ApJ...713.1037H}, which is a
reasonable assumption since a large contribution of the smoothly distributed
starlight in a cluster comes from the BCG. The starlight luminosity as
a function of mass is given by
\begin{equation}
L_\stars(\mvir)=L_\stars(M_{200\rmn{av}})\,
\left(\frac{\mvir}{M_{200\rmn{av}}}\right)^{0.18}\,.
\label{eq:L_stars_m}
\end{equation}
We can now fix the unitless normalization constant for the stars in
Eq.~(\ref{eq:E_SD}) by integrating over the cluster volume and
spectral distribution of the stars:
\begin{equation}
 N_\stars(\mvir) =
\left(\frac{\mvir}{M_{200\rmn{av}}}\right)^{0.18}\,
\frac{\rvir\, 6.0\times10^{-9}\,\kpc^2}{\int_{\rvir} j(r) \,\dd V}\,.
\label{eq:N_stars}
\end{equation}

The {\em luminosity from dust} is derived from the
luminosity-richness scaling relation found in
\cite{2008A&A...490..547G}
$L_\dust=10^{44.8}\left(\frac{N_{200}}{10}\right)^{0.8}$. We pick a
high richness cluster with $N_{200}=82$ to normalize the luminosity
since these clusters are less likely biased by chance
coincidences. This richness corresponds to a virial mass of approximately
$M_\rmn{200dust}=6.0\times10^{14}\,\msun$
\cite{2010ApJ...713.1037H}. We assume that the dust scales with halo
mass in the same way as the stars, and determine the normalization
constant for the dust in Eq.~(\ref{eq:E_SD}) to
\begin{equation}
 N_\dust(\mvir) =
\left(\frac{\mvir}{M_\rmn{200dust}}\right)^{0.18}\,
\frac{\rvir\,4.0\times10^{-7}\,\kpc^2}{\int_{\rvir} j(r) \,\dd V}\,.
\label{eq:N_dust}
\end{equation}

\subsection{Radial profiles}
\label{sect:SD_radial}
\begin{figure}
 \includegraphics[width=0.99\columnwidth]{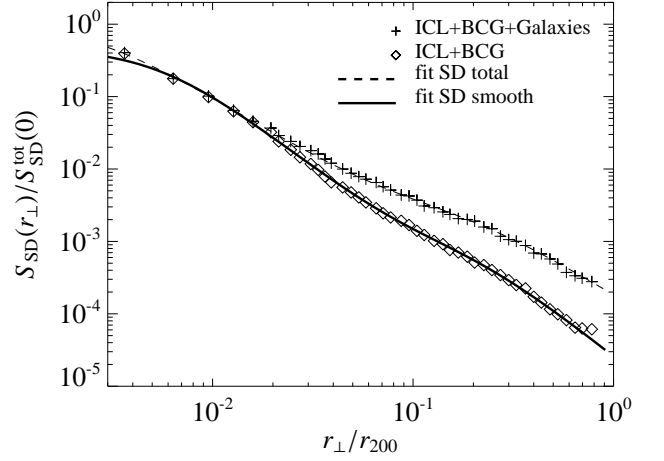}
 \caption{Surface brightness profiles of the $r$-band $(h\nu\sim
   1\,\ev)$ obtained from stacked clusters in the Sloan Digital Sky
   Survey (SDSS) at the redshift $z \sim 0.25$ \protect
   \cite{2005MNRAS.358..949Z}. The crosses show the total observed
   starlight including the diffuse intracluster light (ICL), galaxies,
   and the brightest cluster galaxy (BCG) in the center of the
   cluster. The diamonds denote the smooth part of the observed
   starlight of the ICL and the BCG. The solid line shows the fit to
   the data of the total light, while the smooth component is
   represented by the dashed line. Note that we use total SD
   brightness in the cluster center, $S_\sd^\rmn{tot}(0)$, to
   normalize the SD surface brightness profiles.}
 \label{fig:SD_spatial}
\end{figure}

Our goal in this section is to derive a simple spatial model for the
distribution of the light from SD in galaxy clusters. For this reason
we use stacked cluster observations from the Sloan Digital Sky Survey
(SDSS) in the $r$ and $i$-band that trace the starlight. For
simplicity we assume that the dust traces the stars in the
clusters. This is justified for sufficiently young stellar populations
(blue BCG and spiral galaxies) or, in the case of the ICL, if the dust
got stripped alongside the stars without having been destroyed by
spallation processes thereafter. The stacked surface brightness
profiles in \cite{2005MNRAS.358..949Z} are measured at redshift $z
\approx 0.25$ with an average mass of the clusters of
$4.0\times10^{13}\,\msun$. As already mentioned, three components
contribute to the SD: diffuse ICL, galaxies, and the BCG in the center
of clusters. However, the overlapping volume of the galactic light
with the DM distribution of a cluster is very small compared to the
overlapping volume of ICL- and BCG-light with the DM distribution;
thus, the relative contribution of the galaxies to the IC emission is
suppressed (c.f. Eq.~\ref{eq:SD_overlap}). In our benchmark model for
SD, we remove the contribution from galaxies which corresponds to
roughly $70$ percent of the total SD light. In
Fig.~\ref{fig:SD_spatial} we show the SDSS stacked brightness profiles
as well as the fitted profiles\footnote{The measured brightness is
  converted into units of $\rmn{erg}\,\rmn{s}^{-1}$ using
  \cite{2010...book} $S(\rmn{mag}\,''^{-2}) =
  \mathcal{M}_\odot+21.6-2.5\log_{10}[S(L_\odot\,\rmn{pc}^{-2})]$,
  where the sun's absolute magnitude in the $r$-band is given by
  $\mathcal{M}_\odot=27.1$ \cite{1998gaas.book.....B} and the
  luminosity of the sun by $L_\odot=3.85\times10^{33}\,
  \rmn{erg}\,\rmn{s}^{-1}$.}, where both the data and fits are
normalized with the central surface brightness of the total SD
component, $S_\sd^\rmn{tot}(0)$. Our spatial benchmark model is shown
with the solid black line.

Instead of modelling the surface brightness with a de Vaucouleur
profile and a power-law, we use a double beta profile model to
simplify deprojection. It is given by
\begin{equation}
\tilde{S}_\sd (r_\bot) = \frac{S_\sd (r_\bot)}{S_\rmn{SD}^\rmn{tot}(0)} =
\sum_{i=1}^2 \,\tilde{S}_i\,
\left[1 + \left( \frac{r_\bot}{r_{\mathrm{c}_i}}\right)^2\right]
^{-3\beta_i + 1/2}.
\label{double_beta}
\end{equation}
Our fit parameters for the normalized central brightness,
$\tilde{S}_i$, the core radius, $r_{\mathrm{c}_i}$, and slope
$\beta_i$ are given for the total model by
\begin{align}
&\tilde{S}_1^\rmn{tot} = 1.0\,,
&&r_{\mathrm{c}_1}^\rmn{tot} = 1.8\,\rmn{kpc}\,,
&&&\beta_1^\rmn{tot} = 0.45\, \nonumber\\
&\tilde{S}_2^\rmn{tot} = 2.3\times10^{-3}\,,
&&r_{\mathrm{c}_2}^\rmn{tot} = 0.19\,\rvir\,,
&&&\beta_2^\rmn{tot} = 0.44\,.\nonumber\\
& && &&&
\label{fit_spatial_IR}
\end{align}
For the smooth model (our benchmark SD model), we find:
\begin{align}
&\tilde{S}_1^\rmn{sm}=0.47,\,
&&r_{\mathrm{c}_1}^\rmn{sm}=3.9\,\rmn{kpc}\,,
&&&\beta_1^\rmn{sm}=0.53 \,\nonumber\\
&\tilde{S}_2^\rmn{sm}=8.3\times10^{-4}\,,
&&r_{\mathrm{c}_2}^\rmn{sm}=0.19\,\rvir\,,
&&&\beta_2^\rmn{sm}=0.54\,.\nonumber\\
& && &&&
\label{fit_spatial_IR_sm}
\end{align}
The three dimensional spatial profile is derived by deprojecting the
surface brightness in Eq.~(\ref{double_beta}) (see
e.g. \cite{2004A&A...413...17P} for details about the deprojection):
\begin{eqnarray}
  j(r) = \sum_{i=1}^2 \frac{\tilde{S}_i}{2\pi\,r_{\mathrm{c}_i}}\,
  \frac{6 \beta_i - 1}{\left(1 + r^2/r^2_{\mathrm{c}_i}\right)^{3\beta_i}}\,
  \mathcal{B}\left(\frac{1}{2},3\beta_i\right)\,,\nonumber\\
\end{eqnarray}
where $\mathcal{B}(a,b)$ denotes the beta-function
\cite{1965hmfw.book.....A}.

The radial distribution of the energy density of SD governs the seed
photon distribution for IC emission (together with the CMB). The
energy density from starlight and dust in a galaxy cluster is given by
\begin{eqnarray}
\label{eq:u_SD_r}
u_\sd(r) &=& \int \dd \eph \frac{\dd^2 N_\ph}{\dd \eph \dd V}\,\eph
=\int \dd \eph \frac{u_\sd(\eph, r)}{\eph}
\nonumber \\
&=& j(r) \int \dd \eph \sum_i
N_i(\mvir) \frac{u_i^\gal(\eph)}{\eph}\,, \nonumber \\
\end{eqnarray}
where we have used the specific energy density of the light from SD
given by Eq.~(\ref{eq:u_SD_er}). We compare the energy
density $u_\sd(r)$ to other radiation background fields in
Fig.~\ref{fig:SD_Edens} for a galaxy cluster with the mass
$\mvir=6.0\times10^{14}\msun$. We find that inside a few percent of of
$\rvir$, the total energy density is dominated by the light from SD
while the CMB is dominating elsewhere. This implies that if the boost
from dark matter (DM) substructures is significant, then the overlap of
light from SD with the electron and positron distribution that trace
the substructures is small. Hence the resulting flux from inverse
Compton (IC) upscattered SD photons is suppressed compared to the IC
upscattered CMB photons.

\section{Source functions of different DM models}
We use {\sc DarkSUSY} to derive the spectral distribution of the
continuum emission from our four DM benchmark (BM) models. These
spectra are shown in the left panel of Fig.~\ref{fig:q_DM}. We choose
to fit the source function in the energy regime $30\,\rmn{MeV}<\eg<
m_\chi c^2$ using the following functional form
\begin{eqnarray}
q_\rmn{BM} (\eg,r)&=&\left[\frac{\rho(r)}{10^{-29}\rmn{g}\,\cm^{-3}}\right]^2
\times\frac{a_1}{\egt}\times\nonumber\\
&\times&\left[\frac{\exp\left(-\egt^{-a_2}\right)}{1+\exp\left(\egt^{a_3}\right)}
+\left(\frac{\egt}{a_4}\right)^{4.5}\right]\,, \nonumber\\
\rmn{where}&\quad& \egt = \frac{\eg}{\gev}\,,
 \label{eq:bm_cont}
\end{eqnarray}
where the model specific parameters are given in
Table~\ref{tab:bm_cont}.

\begin{table}[h]
\begin{tabular}{c c c c c }
\hline
\hline
 BM model & $a_1$ & $a_2$ & $a_3$ & $a_4$ \\
          & $[\gev^{-1}\,\cm^{-2}\,\rmn{s}^{-1}]$ & & & \\
 \hline
$\Ip$ & $2.0\times10^{-4}$ & $0.36$ & $0.51$ & $701$ \\
$\Jp$ & $4.1\times10^{-6}$ & $0.40$ & $0.43$ & $1178$ \\
$\Kp$ & $1.7\times10^{-4}$ & $0.42$ & $0.37$ & $-$ \\
$\Js$ & $2.1\times10^{-6}$ & $0.34$ & $0.51$ & $490$ \\
\hline
\hline
\end{tabular}
\caption{Fit parameters to the continuum emission from our
  supersymmetric benchmark models.
 \label{tab:bm_cont}}
\end{table}

We also use {\sc DarkSUSY} to derive the number of electrons and
positrons resulting from each annihilation. These spectra are shown in
the right panel of Fig.~\ref{fig:q_DM}. The electron and positron
yield resulting from our supersymmetric BM model is fitted in the
regime $30\,\rmn{MeV}<\ee< m_\chi c^2$ using
\begin{eqnarray}
N_{\e,\rmn{BM}}(>\ee) &=& \frac{b_1}{\eet^{b_2}}\times
\frac{\left[1+\left(\frac{\eet}{b_6}\right)^{b_7}\right]^{b_8}}
{\left[1+\left(\frac{\eet}{b_3}\right)^{b_4}\right]^{b_5}}\,,\nonumber\\
\rmn{where}&\quad&\eet=\frac{\ee}{\gev}\,.
\label{eq:bm_elec}
\end{eqnarray}
Fit parameters are given in Table~\ref{tab:bm_elec}.
\begin{table}[h]
\begin{tabular}{c c c c c c c c c}
\hline
\hline
 BM model & $b_1$ & $b_2$ & $b_3$ & $b_4$ & $b_5$ & $b_6$ & $b_7$ & $b_8$ \\
 \hline
$\Ip$ & $28$ & $0.06$ & $1.8$ & $0.9$ & $2.0$ & $60$ & $5$ & $-1$ \\
$\Jp$ & $34$ & $0.08$ & $5.5$ & $0.8$ & $2.6$ & $250$ & $5$ & $-1$ \\
$\Kp$ & $44$ & $0.05$ & $1.7$ & $0.85$ & $2.1$ & $10$ & $2$ & $0.62$ \\
$\Js$ & $30$ & $0.14$ & $5.0$ & $1.0$ & $1.5$ & $300$ & $5$ & $-1$ \\
\hline
\hline
\end{tabular}
\caption{Fit parameters to the electron and positron yield
  above the electron energy $\ee$ from the supersymmetric benchmark
  models.
 \label{tab:bm_elec}}
\end{table}

Finally, we use the data from
\cite{2011JCAP...03..019C,2011JCAP...03..051C} to derive the electron
and positron yield resulting from leptophilic (LP) DM annihilating
indirectly into decaying muons or electrons and positrons. In the
energy regime $30\,\rmn{MeV}<\ee< m_\chi c^2$ we find the following
best fit for the annihilation into muons:
\begin{eqnarray}
N_{\e,\rmn{LP}\mu}(>\ee) &=& \exp\left[c_1+c_2\,x+c_3\,(2-x)^{c_4}\right]
\,,\nonumber\\
\rmn{where}\quad x&=&\frac{\ee}{m_\chi\,c^2}\,\quad\rmn{and}\nonumber\\
c_1=1.49,\,c_2&=&-5.00,\,c_3=-2.46,\,c_4=-4.40\,.\nonumber\\
\label{eq:lp_mu}
\end{eqnarray}
Similarly, for the annihilation into electrons and positrons we find the following best fit:
\begin{eqnarray}
N_{\e,\rmn{LP}\e}(>\ee) &=& \exp\left(c_1+c_2\,y+c_3\,y^2\right)
\,,\nonumber\\
\rmn{where}\quad y&=&\,\log\left(1-x\right)\,,\quad
x=\frac{\ee}{m_\chi\,c^2}\,\quad\rmn{and}\nonumber\\
c_1&=&1.40,\,c_2=1.11,\,c_3=0.058\,.
\label{eq:lp_ep}
\end{eqnarray}

The LP model, that we are using in this work has a branching factor
branching ratio of ($1/4:1/4:1/2$) into
($\mu^+\mu^-:\e^+\e^-:\pi^+\pi^-$). The contribution from the
electrons from the decaying pions is relative small, hence the
resulting electron and positron yield for the LP model is
approximately given by
\begin{equation}
N_{\e,\rmn{LP}}(>\ee) =
N_{\e,\rmn{LP}\e}(>\ee)/4+N_{\e,\rmn{LP}\mu}(>\ee)/4\,.
\end{equation}

\section{Gas density profiles}
\begin{table*}[t]
\begin{minipage}{1.99\columnwidth}
\begin{tabular}{ccccccccccc}
\hline
\hline
Cluster & \multicolumn{9}{c}{$\rho_\rmn{gas} = \left\{\sum_i n_i^2(0)\left[1+\left(\frac{r}{r_{c,i}}\right)^2\right]^{-3\beta_i}\right\}^{1/2}$} & Reference \\
& $n_\rmn{1}(0)\,[\el\,/\cm^3]$ & $r_\rmn{c,1}\,[\kpc]$ & $\beta_1$ &
$n_\rmn{2}(0)\,[\el\,/\cm^3]$ & $r_\rmn{c,2}\,[\kpc]$ & $\beta_2$ &
$n_\rmn{3}(0)\,[\el\,/\cm^3]$ & $r_\rmn{c,3}\,[\kpc]$ & $\beta_3$ & \\
 \hline
Fornax & 0.35 & 0.36 & 0.54 & $2.2\times 10^{-3}$ & 190 & 41 & $5.4\times 10^{-4}$ & 183 & 0.8 & \protect{\cite{2002ApJ...565..883P}} \\
Coma & $3.55\times 10^{-3}$ & 245 & 0.65 & & & & & & & \protect{\cite{2007A&A...466..805C}} \\
Virgo & 0.15 & 1.6 & 0.42 & $1.2\times 10^{-2}$ & 20 & 0.47 & & & & \protect{\cite{2002A&A...386...77M}} \\
\hline
Perseus & \multicolumn{9}{c}{$46\times10^{-3}\,[1+(r/57\,\rmn{kpc})^2]^{-1.8}+
            4.8\times10^{-3}\,[1+(r/200\,\rmn{kpc})^2]^{-0.87} \,[\el\,/\cm^3]$} & \protect{\cite{2003ApJ...590..225C}} \\
\hline
\hline
\end{tabular}
\caption{Electron number density profiles for selected
  clusters. \label{tab:dens_prof}}
\end{minipage}
\end{table*}
The production of gamma-rays and secondaries from hadronic CRs in
clusters depend both on the gas density and the CR number density,
which roughly trace the gas outside the core regime and is slightly
enhanced in the center. These gas density profile of galaxy clusters
are derived from X-ray observations, where the measurements are mainly
sensitive to the central parts of the clusters where there X-ray flux
is high, while the core and outer parts are more difficult to measure
due to the low signal to noise and often involves extrapolation from
the center. The uncertainties in measurements and modelling, in
combination with different specifications of active X-ray satellites,
give rise to important differences in the gas profiles. Since the
gamma-ray flux from a galaxy cluster is sensitive to the details of
the gas profile, we choose to model the density profile of four bright
clusters (Fornax, Virgo, Coma, Perseus) in more detail where we use
the most recent and detailed modelled gas profile available in the
literature. We show these profiles in table~\ref{tab:dens_prof}.

The X-ray emitting gas in Fornax does not follow a simple
$\beta$-profile. Instead, based on deep Rosat data and supported by
Chandra data, it is best modeled by a multicomponent bidimensional
model \cite{2002ApJ...565..883P}, consisting of: (1) a central
component ($r<5\,\kpc$); (2) a ``galactic'' component
($5\,\kpc<r<40\,\kpc$); and (3) an elliptical ICM component
($r>40\,\kpc$).

\begin{figure}
 \includegraphics[width=0.99\columnwidth]{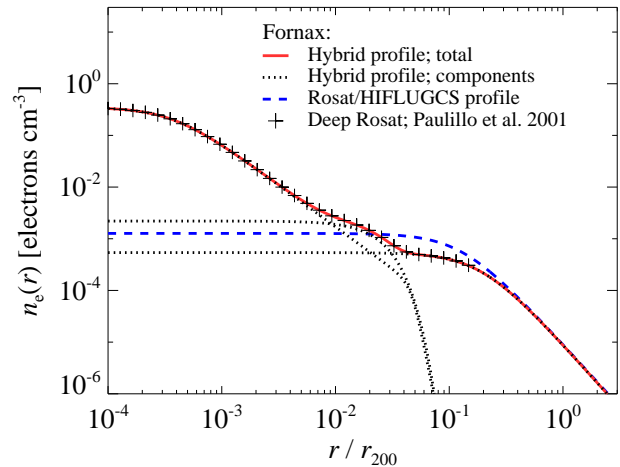}
 \caption{\colo Comparing different electron number
   density profiles for the Fornax cluster. Black crosses show the
   density profile inferred from deprojected deep Rosat X-ray surface
   brightness observations \protect \cite{2002ApJ...565..883P}. The
   total hybrid profile shown by the red solid line represent the best
   fit to the data, where the fitted individual density components are
   shown by the black dotted lines. The blue dashed line shows the
   single beta density profile inferred from the HIFLUGCS
   catalogue. Due to insufficient sensitivity of the data to the outer
   cluster part, we use the outer slope from the HIFLUGCS catalogue
   \cite{2007A&A...466..805C}.}
 \label{fig:dens_fornax}
\end{figure}

In Fig.~\ref{fig:dens_fornax} we show the data points for the electron
number density ($n_\e$) in Fornax, derived from the deep Rosat data
presented in reference \cite{2002ApJ...565..883P}, and the best fit
density profile together with the individual components. The total
$n_\e$ profile is derived from their fitted central and galactic
surface brightness components while we re-fit the ICM component. The
reason for the re-fitted outer component is the large uncertainty in
the data points outside $(0.2-0.3)\rvir$ which we exclude in our
fit. Instead we assume that the outer slop of $n_\e$ follows the outer
slope of Fornax in the HIFLUGCS catalogue
\cite{2007A&A...466..805C}. Deprojecting the fitted surface brightness
components yields the following electron number density profile:
\begin{equation}
n_\e(r) = \left\{\sum_i n_i^2\left[1+\left(\frac{r}{r_{c,i}}\right)^2\right]^{-3\beta_i}\right\}^{1/2}
\end{equation}
where
\begin{align}
&n_1 = 0.35\, \el\,\rmn{cm}^{-3}\,,
&&r_{c,1} = 0.36\,\kpc\,,\,\,\,
\beta_1 = 0.54\,, \nonumber\\
&n_2 = 2.2\times10^{-3}\,\el\,\rmn{cm}^{-3}\,,
&&r_{c,2} = 190\,\kpc\,,\,\,\,\,
\beta_2 = 41\,, \nonumber\\
&n_3 = 5.4\times10^{-4}\,\el\,\rmn{cm}^{-3}\,,
&&r_{c,3} = 183\,\kpc\,,\,\,\,\,
\beta_3 = 0.8\,. \nonumber\\
& &&
\label{fit_fornax}
\end{align}
Note that we remove the contribution of the flat central component of
the fit ($i=1$) in the outer parts of the cluster, thus we neglect the
central contribution for $r>40\,\kpc$.

For 106 X-ray bright clusters extended HIFLUGCS catalogue we adopt
more general density profiles derived in the recent paper by Chen et
al. \cite{2007A&A...466..805C}. They provide single beta-model density
fits for all clusters, as well as the double beta-model fits to the
surface profiles. We follow the deprojection procedure in
\cite{2004A&A...413...17P}, and only choose the fitted double beta
profiles with a $\chi^2$ that is smaller compared to the single beta
profile. \footnote{We adopt the single beta density profile for A2255
  since it harbors a giant radio halo and most likely has a disturbed
  morphology that might have given the fit the impression of a cool
  core. Furthermore, the difference in $\chi^2$ between the single-
  and double-beta profiles is marginal for this cluster.}

\section{Flux tables for the HIFLUGCS catalogue}

In this section we present the gamma-ray flux predictions using the
clusters in the extended HIFLUGCS catalogue. We show the brightest 50
clusters in descending order; CR proton induced $\pi^0$ that decays
into gamma-rays in Table~\ref{tab:flux_tab_CRs}, the supersymmetric DM
BM-$\Kp$ model where the Neutralino emit a continuum as well as final
state radiation in Table~\ref{tab:flux_tab_BM}, and LP DM that emit
final state radiation and annihilate either indirectly to
$\rmn{e}^+/\rmn{e}^-$ or through $\mu^+/\mu^-$ to
$\rmn{e}^+/\rmn{e}^-$ that IC upscatter CMB photons in
Table~\ref{tab:flux_tab_LP}. The flux from DM has been boosted by the
substructures, where we in addition include the Sommerfeld enhancement
for the LP model. For completeness, we show in
Table~\ref{tab:flux_tab_CLp} the parameters that we use to derive the
fluxes for the clusters in the HIFLUGCS sample.

\begin{table*}
\begin{minipage}{2.0\columnwidth}
  \caption{Gamma-ray CR-$\pi^0$ flux within $\rvir$ from brightest 50 clusters in HIFLUGCS catalogue.}
\begin{tabular}{l c c c c c c c}
\hline
\hline
 Cluster & ID$^{(1)}$ & $F_{\gamma}$$^{(2)}$ & $F_{\gamma}$$^{(3)}$&
 $F_{\gamma}$$^{(4)}$ & $\eg^2\,\dd F_{\gamma}/\dd \eg$$^{(5)}$ &
 $\eg^2\,\dd F_{\gamma,0.1}/\dd \eg$$^{(5,6)}$ &
 $\eg^2\,\dd F_{\gamma,1.0}/\dd \eg$$^{(5,7)}$\\
  & & $(>100\,\rmn{MeV})$ & $(>1\,\rmn{GeV})$ & $(>100\,\rmn{GeV})$ &
 $(5\,\rmn{GeV})$ & $(5\,\rmn{GeV})$ & $(5\,\rmn{GeV})$\\
 \hline
PERSEUS  &  68 &  14.28 &  19.15 &  52.49 &  16.26 &  16.25 &  10.06 \vstt \\
OPHIUCHU &  96 &   7.71 &  10.34 &  28.33 &   8.78 &   8.78 &   6.01 \vstt \\
COMA     &  34 &   4.15 &   5.56 &  15.24 &   4.72 &   4.72 &   2.55 \vstt \\
NORMA    &  94 &   2.98 &   3.99 &  10.94 &   3.39 &   3.38 &   2.19 \vstt \\
CENTAURU &  30 &   2.61 &   3.50 &   9.58 &   2.97 &   2.97 &   2.62 \vstt \\
TRIANGUL &  95 &   1.77 &   2.37 &   6.50 &   2.01 &   2.01 &   0.37 \vstt \\
AWM7     &  67 &   1.75 &   2.35 &   6.43 &   1.99 &   1.99 &   1.24 \vstt \\
A2319    &  98 &   1.71 &   2.30 &   6.30 &   1.95 &   1.94 &   0.28 \vstt \\
A3571    &  39 &   1.61 &   2.15 &   5.90 &   1.83 &   1.82 &   0.41 \vstt \\
3C129    &  70 &   1.35 &   1.82 &   4.98 &   1.54 &   1.54 &   0.71 \vstt \\
A2199    &  51 &   1.11 &   1.49 &   4.10 &   1.27 &   1.27 &   0.33 \vstt \\
2A0335   &  11 &   1.04 &   1.40 &   3.83 &   1.19 &   1.19 &   0.19 \vstt \\
A2029    &  43 &   1.04 &   1.39 &   3.81 &   1.18 &   1.17 &   0.08 \vstt \\
A0496    &  18 &   1.02 &   1.36 &   3.73 &   1.16 &   1.15 &   0.27 \vstt \\
HYDRA    &  25 &   0.95 &   1.27 &   3.48 &   1.08 &   1.08 &   0.84 \vstt \\
A0085    &   1 &   0.93 &   1.25 &   3.44 &   1.06 &   1.06 &   0.12 \vstt \\
A2142    &  48 &   0.92 &   1.24 &   3.39 &   1.05 &   1.04 &   0.06 \vstt \\
A3266    &  17 &   0.92 &   1.23 &   3.37 &   1.05 &   1.04 &   0.16 \vstt \\
A1795    &  40 &   0.91 &   1.22 &   3.35 &   1.04 &   1.04 &   0.11 \vstt \\
PKS0745  &  77 &   0.82 &   1.10 &   3.02 &   0.94 &   0.88 &   0.03 \vstt \\
A2256    &  54 &   0.78 &   1.04 &   2.86 &   0.89 &   0.88 &   0.11 \vstt \\
A3558    &  37 &   0.75 &   1.01 &   2.77 &   0.86 &   0.86 &   0.12 \vstt \\
A3667    &  56 &   0.71 &   0.96 &   2.62 &   0.81 &   0.80 &   0.07 \vstt \\
A1367    &  26 &   0.70 &   0.94 &   2.59 &   0.80 &   0.80 &   0.43 \vstt \\
A0478    &  14 &   0.68 &   0.91 &   2.48 &   0.77 &   0.75 &   0.03 \vstt \\
A0401    &   8 &   0.61 &   0.82 &   2.25 &   0.70 &   0.68 &   0.05 \vstt \\
A0262    &   5 &   0.57 &   0.77 &   2.10 &   0.65 &   0.65 &   0.26 \vstt \\
HYDRA-A  &  24 &   0.56 &   0.75 &   2.05 &   0.63 &   0.63 &   0.05 \vstt \\
A0754    &  23 &   0.52 &   0.70 &   1.92 &   0.59 &   0.59 &   0.10 \vstt \\
A4038    &  62 &   0.52 &   0.69 &   1.90 &   0.59 &   0.59 &   0.13 \vstt \\
A2052    &  44 &   0.51 &   0.68 &   1.87 &   0.58 &   0.58 &   0.09 \vstt \\
A2147    &  49 &   0.50 &   0.66 &   1.82 &   0.56 &   0.55 &   0.11 \vstt \\
A0119    &   2 &   0.48 &   0.64 &   1.76 &   0.54 &   0.54 &   0.10 \vstt \\
A0644    &  78 &   0.47 &   0.63 &   1.73 &   0.54 &   0.53 &   0.04 \vstt \\
A1644    &  31 &   0.43 &   0.58 &   1.60 &   0.49 &   0.49 &   0.07 \vstt \\
A3158    &  13 &   0.39 &   0.53 &   1.45 &   0.45 &   0.44 &   0.04 \vstt \\
FORNAX   &  10 &   0.36 &   0.49 &   1.34 &   0.42 &   0.42 &   0.41 \vstt \\
A2063    &  47 &   0.36 &   0.48 &   1.33 &   0.41 &   0.41 &   0.06 \vstt \\
MKW3S    &  45 &   0.35 &   0.47 &   1.29 &   0.40 &   0.40 &   0.04 \vstt \\
A3112    &   9 &   0.34 &   0.46 &   1.26 &   0.39 &   0.38 &   0.02 \vstt \\
ZwCl1742 &  97 &   0.34 &   0.46 &   1.25 &   0.39 &   0.38 &   0.03 \vstt \\
A4059    &  63 &   0.33 &   0.45 &   1.23 &   0.38 &   0.38 &   0.05 \vstt \\
A1651    &  33 &   0.31 &   0.41 &   1.13 &   0.35 &   0.34 &   0.02 \vstt \\
A0399    &   7 &   0.31 &   0.41 &   1.13 &   0.35 &   0.34 &   0.02 \vstt \\
NGC1550  &  15 &   0.30 &   0.40 &   1.09 &   0.34 &   0.34 &   0.16 \vstt \\
ANTLIA   &  79 &   0.29 &   0.39 &   1.06 &   0.33 &   0.33 &   0.22 \vstt \\
A2657    &  61 &   0.29 &   0.38 &   1.05 &   0.32 &   0.32 &   0.06 \vstt \\
A0539    &  71 &   0.28 &   0.38 &   1.03 &   0.32 &   0.32 &   0.07 \vstt \\
A3581    &  41 &   0.28 &   0.37 &   1.03 &   0.32 &   0.32 &   0.07 \vstt \\
A2204    &  52 &   0.28 &   0.37 &   1.02 &   0.32 &   0.21 &   0.01 \vstt \\
\hline
\hline
\end{tabular}
\begin{quote}
  Notes:
   (1) The HIFLUGCS cluster ID.
   (2) In units of $10^{-9} \rmn{ph}\,\rmn{cm}^{-2}\,\rmn{s}^{-1}$.
   (3) In units of $10^{-10} \rmn{ph}\,\rmn{cm}^{-2}\,\rmn{s}^{-1}$.
   (4) In units of $10^{-13} \rmn{ph}\,\rmn{cm}^{-2}\,\rmn{s}^{-1}$.
   (5) In units of $10^{-10} \rmn{erg}\,\rmn{cm}^{-2}\,\rmn{s}^{-1}$.
   (6) The flux within $0.1^\circ$, smoothed with a point spread function of $0.1^\circ$.
   (7) The flux within $1.0^\circ$, smoothed with a point spread function of $0.1^\circ$.
 \label{tab:flux_tab_CRs}
  \end{quote}
\end{minipage}
\end{table*}

\begin{table*}
\begin{minipage}{2.0\columnwidth}
  \caption{Gamma-ray BM-$\Kp$ Continuum flux within $\rvir$ from brightest 50 clusters in HIFLUGCS catalogue.}
\begin{tabular}{l c c c c c c c c}
\hline
\hline
 Cluster & ID$^{(1)}$ & $F_{\gamma}$$^{(2)}$ & $F_{\gamma}$$^{(2)}$&
 $F_{\gamma}$$^{(3)}$ & $\eg^2\,\dd F_{\gamma}/\dd \eg$$^{(4)}$ &
 $\eg^2\,\dd F_{\gamma,0.1}/\dd \eg$$^{(4,5)}$ &
 $\eg^2\,\dd F_{\gamma,1.0}/\dd \eg$$^{(4,6)}$ & S/N$^{(7)}$ \\
  & & $(>100\,\rmn{MeV})$ & $(>1\,\rmn{GeV})$ & $(>100\,\rmn{GeV})$ &
 $(5\,\rmn{GeV})$ & $(5\,\rmn{GeV})$ & $(5\,\rmn{GeV})$ & \\
 \hline
FORNAX   &  10 &  31.88 &  20.21 &  11.15 &  32.00 &  31.98 &  30.27 &   9.93 \vstt \\
M49      &  81 &  19.48 &  12.35 &   6.81 &  19.55 &  19.54 &  18.13 &   9.29 \vstt \\
OPHIUCHU &  96 &  18.13 &  11.49 &   6.34 &  18.19 &  18.14 &  12.31 &   1.66 \vstt \\
CENTAURU &  30 &  14.99 &   9.50 &   5.24 &  15.05 &  15.02 &  12.40 &   2.32 \vstt \\
HYDRA    &  25 &   9.19 &   5.82 &   3.21 &   9.22 &   9.20 &   6.89 &   2.32 \vstt \\
NGC4636  &  29 &   9.12 &   5.78 &   3.19 &   9.15 &   9.14 &   8.19 &   5.38 \vstt \\
NORMA    &  94 &   8.15 &   5.17 &   2.85 &   8.18 &   8.16 &   5.42 &   1.20 \vstt \\
PERSEUS  &  68 &   7.80 &   4.95 &   2.73 &   7.83 &   7.80 &   4.95 &   0.53 \vstt \\
COMA     &  34 &   7.58 &   4.81 &   2.65 &   7.61 &   7.58 &   4.41 &   0.95 \vstt \\
AWM7     &  67 &   7.31 &   4.64 &   2.56 &   7.34 &   7.31 &   4.66 &   1.39 \vstt \\
A1367    &  26 &   6.67 &   4.23 &   2.33 &   6.69 &   6.66 &   3.84 &   1.95 \vstt \\
NGC5813  &  91 &   6.25 &   3.96 &   2.19 &   6.28 &   6.27 &   4.97 &   4.26 \vstt \\
ANTLIA   &  79 &   5.90 &   3.74 &   2.06 &   5.92 &   5.90 &   4.02 &   2.58 \vstt \\
A2877    &  65 &   5.00 &   3.17 &   1.75 &   5.02 &   4.99 &   2.54 &   2.66 \vstt \\
3C129    &  70 &   4.71 &   2.99 &   1.65 &   4.73 &   4.70 &   2.42 &   1.02 \vstt \\
NGC5044  &  35 &   3.59 &   2.28 &   1.26 &   3.61 &   3.59 &   2.35 &   1.67 \vstt \\
NGC5846  &  92 &   3.34 &   2.12 &   1.17 &   3.36 &   3.35 &   2.40 &   3.40 \vstt \\
NGC1550  &  15 &   2.53 &   1.61 &   0.89 &   2.54 &   2.53 &   1.36 &   1.15 \vstt \\
A3571    &  39 &   2.24 &   1.42 &   0.78 &   2.25 &   2.22 &   0.67 &   0.45 \vstt \\
TRIANGUL &  95 &   2.16 &   1.37 &   0.76 &   2.17 &   2.13 &   0.56 &   0.42 \vstt \\
A2199    &  51 &   2.05 &   1.30 &   0.72 &   2.06 &   2.03 &   0.68 &   0.50 \vstt \\
A2634    &  60 &   2.02 &   1.28 &   0.71 &   2.03 &   2.01 &   0.66 &   1.03 \vstt \\
A3266    &  17 &   2.00 &   1.27 &   0.70 &   2.01 &   1.97 &   0.46 &   0.53 \vstt \\
A0496    &  18 &   1.93 &   1.22 &   0.67 &   1.94 &   1.91 &   0.60 &   0.49 \vstt \\
A0262    &   5 &   1.92 &   1.21 &   0.67 &   1.92 &   1.91 &   0.84 &   0.64 \vstt \\
A0754    &  23 &   1.88 &   1.19 &   0.66 &   1.89 &   1.85 &   0.45 &   0.66 \vstt \\
A0119    &   2 &   1.87 &   1.19 &   0.65 &   1.88 &   1.84 &   0.49 &   0.68 \vstt \\
IIIZw54  &  12 &   1.72 &   1.09 &   0.60 &   1.73 &   1.71 &   0.53 &   1.03 \vstt \\
A2319    &  98 &   1.62 &   1.03 &   0.57 &   1.62 &   1.59 &   0.35 &   0.32 \vstt \\
A2657    &  61 &   1.55 &   0.98 &   0.54 &   1.55 &   1.53 &   0.40 &   0.73 \vstt \\
A3395s   &  21 &   1.51 &   0.96 &   0.53 &   1.52 &   1.49 &   0.34 &   0.76 \vstt \\
A0539    &  71 &   1.50 &   0.95 &   0.52 &   1.50 &   1.48 &   0.45 &   0.71 \vstt \\
A4038    &  62 &   1.48 &   0.94 &   0.52 &   1.49 &   1.47 &   0.45 &   0.52 \vstt \\
A0576    &  22 &   1.37 &   0.87 &   0.48 &   1.37 &   1.35 &   0.34 &   0.72 \vstt \\
A3376    &  19 &   1.34 &   0.85 &   0.47 &   1.34 &   1.31 &   0.30 &   0.69 \vstt \\
A1644    &  31 &   1.33 &   0.84 &   0.47 &   1.34 &   1.31 &   0.29 &   0.51 \vstt \\
A3395n   &  75 &   1.32 &   0.84 &   0.46 &   1.32 &   1.29 &   0.28 &   0.78 \vstt \\
MKW8     &  42 &   1.32 &   0.83 &   0.46 &   1.32 &   1.30 &   0.39 &   0.87 \vstt \\
UGC03957 &  76 &   1.28 &   0.81 &   0.45 &   1.28 &   1.26 &   0.33 &   0.74 \vstt \\
A2256    &  54 &   1.27 &   0.81 &   0.44 &   1.28 &   1.24 &   0.24 &   0.37 \vstt \\
A3558    &  37 &   1.20 &   0.76 &   0.42 &   1.20 &   1.17 &   0.25 &   0.35 \vstt \\
A0400    &   6 &   1.17 &   0.74 &   0.41 &   1.17 &   1.16 &   0.35 &   0.66 \vstt \\
A3581    &  41 &   1.07 &   0.68 &   0.37 &   1.07 &   1.06 &   0.33 &   0.51 \vstt \\
2A0335   &  11 &   1.05 &   0.67 &   0.37 &   1.05 &   1.03 &   0.25 &   0.26 \vstt \\
A0085    &   1 &   1.04 &   0.66 &   0.36 &   1.04 &   1.01 &   0.18 &   0.27 \vstt \\
A2052    &  44 &   1.01 &   0.64 &   0.35 &   1.02 &   1.00 &   0.24 &   0.36 \vstt \\
A1795    &  40 &   1.00 &   0.64 &   0.35 &   1.01 &   0.97 &   0.16 &   0.27 \vstt \\
NGC507   &   4 &   0.99 &   0.63 &   0.35 &   1.00 &   0.99 &   0.34 &   1.28 \vstt \\
MKW4     &  27 &   0.95 &   0.60 &   0.33 &   0.95 &   0.94 &   0.29 &   0.65 \vstt \\
NGC499   &  66 &   0.95 &   0.60 &   0.33 &   0.95 &   0.94 &   0.33 &   1.53 \vstt \\
\hline
\hline
\end{tabular}
\begin{quote}
  Notes:
   (1) The HIFLUGCS cluster ID.
   (2) In units of $10^{-11} \rmn{ph}\,\rmn{cm}^{-2}\,\rmn{s}^{-1}$.
   (3) In units of $10^{-13} \rmn{ph}\,\rmn{cm}^{-2}\,\rmn{s}^{-1}$.
   (4) In units of $10^{-11} \rmn{erg}\,\rmn{cm}^{-2}\,\rmn{s}^{-1}$.
   (5) The flux within $0.1^\circ$, smoothed with a point spread function of $0.1^\circ$.
   (6) The flux within $1.0^\circ$, smoothed with a point spread function of $0.1^\circ$.
   (7) Signal to noise (S/N) within $\rvir$ above $1\,$GeV. We estimate the signal to
  noise through $\sqrt{t_\rmn{obs}\,A_\rmn{eff}}\,F_{\gamma}/\sqrt{F_{\gamma}+F_{\gamma,\CR}}$,
  where $t_\rmn{obs}=3\,$yrs is the observation time and $A_\rmn{eff} = 7000\,\cm^2$
    is the effective area of Fermi-Lat at $1\,$GeV.
  \label{tab:flux_tab_BM}
  \end{quote}
\end{minipage}
\end{table*}

\begin{table*}
\begin{minipage}{2.0\columnwidth}
  \caption{Gamma-ray LP-IC-CMB and LP-BS flux within $\rvir$ from brightest 50 clusters in HIFLUGCS catalogue.}
\begin{tabular}{l c c c c c c c c}
\hline
\hline
 Cluster & ID$^{(1)}$ & $F_{\gamma}$$^{(2)}$ & $F_{\gamma}$$^{(3)}$&
 $F_{\gamma}$$^{(4)}$ & $\eg^2\,\dd F_{\gamma}/\dd \eg$$^{(5)}$ &
 $\eg^2\,\dd F_{\gamma,0.1}/\dd \eg$$^{(5,6)}$ &
 $\eg^2\,\dd F_{\gamma,1.0}/\dd \eg$$^{(5,7)}$ & S/N$^{(8)}$ \\
  & & $(>100\,\rmn{MeV})$ & $(>1\,\rmn{GeV})$ & $(>100\,\rmn{GeV})$ &
 $(5\,\rmn{GeV})$ & $(5\,\rmn{GeV})$ & $(5\,\rmn{GeV})$ & \\
 \hline
FORNAX   &  10 &  34.87 &  44.21 &  50.48 &  43.11 &  43.09 &  40.74 & 480.43 \vstt \\
M49      &  81 &  21.48 &  27.24 &  30.84 &  26.57 &  26.55 &  24.62 & 377.24 \vstt \\
OPHIUCHU &  96 &  20.12 &  25.51 &  28.70 &  24.88 &  24.80 &  16.83 & 358.32 \vstt \\
CENTAURU &  30 &  16.64 &  21.10 &  23.74 &  20.58 &  20.54 &  16.94 & 329.49 \vstt \\
HYDRA    &  25 &  10.14 &  12.86 &  14.54 &  12.54 &  12.52 &   9.37 & 258.08 \vstt \\
NGC4636  &  29 &   9.96 &  12.63 &  14.44 &  12.32 &  12.31 &  11.00 & 256.79 \vstt \\
NORMA    &  94 &   8.72 &  11.05 &  12.91 &  10.78 &  10.75 &   7.15 & 236.35 \vstt \\
PERSEUS  &  68 &   8.66 &  10.98 &  12.35 &  10.71 &  10.67 &   6.77 & 221.99 \vstt \\
COMA     &  34 &   8.22 &  10.43 &  12.00 &  10.17 &  10.12 &   5.90 & 227.75 \vstt \\
AWM7     &  67 &   8.08 &  10.24 &  11.58 &   9.99 &   9.96 &   6.34 & 228.93 \vstt \\
A1367    &  26 &   7.24 &   9.18 &  10.56 &   8.95 &   8.91 &   5.14 & 217.96 \vstt \\
NGC5813  &  91 &   6.88 &   8.72 &   9.90 &   8.50 &   8.49 &   6.73 & 213.36 \vstt \\
ANTLIA   &  79 &   6.23 &   7.90 &   9.34 &   7.71 &   7.68 &   5.24 & 202.74 \vstt \\
A2877    &  65 &   5.52 &   7.00 &   7.92 &   6.82 &   6.78 &   3.45 & 190.91 \vstt \\
3C129    &  70 &   5.05 &   6.41 &   7.46 &   6.25 &   6.21 &   3.21 & 180.60 \vstt \\
NGC5044  &  35 &   3.95 &   5.01 &   5.69 &   4.89 &   4.87 &   3.19 & 161.33 \vstt \\
NGC5846  &  92 &   3.65 &   4.63 &   5.29 &   4.52 &   4.50 &   3.23 & 155.50 \vstt \\
NGC1550  &  15 &   2.79 &   3.53 &   4.01 &   3.45 &   3.43 &   1.84 & 135.18 \vstt \\
A3571    &  39 &   2.48 &   3.15 &   3.55 &   3.07 &   3.03 &   0.92 & 124.33 \vstt \\
TRIANGUL &  95 &   2.37 &   3.01 &   3.42 &   2.93 &   2.88 &   0.76 & 120.94 \vstt \\
A2199    &  51 &   2.26 &   2.87 &   3.24 &   2.80 &   2.76 &   0.93 & 119.51 \vstt \\
A2634    &  60 &   2.22 &   2.81 &   3.20 &   2.74 &   2.71 &   0.90 & 120.59 \vstt \\
A3266    &  17 &   2.18 &   2.76 &   3.17 &   2.69 &   2.63 &   0.62 & 117.65 \vstt \\
A0496    &  18 &   2.14 &   2.72 &   3.06 &   2.65 &   2.62 &   0.83 & 116.49 \vstt \\
A0262    &   5 &   2.11 &   2.67 &   3.03 &   2.60 &   2.58 &   1.14 & 116.57 \vstt \\
A0754    &  23 &   2.08 &   2.63 &   2.98 &   2.57 &   2.52 &   0.61 & 115.87 \vstt \\
A0119    &   2 &   2.00 &   2.53 &   2.96 &   2.47 &   2.43 &   0.66 & 113.70 \vstt \\
IIIZw54  &  12 &   1.89 &   2.39 &   2.73 &   2.33 &   2.30 &   0.72 & 111.27 \vstt \\
A2319    &  98 &   1.77 &   2.24 &   2.56 &   2.19 &   2.13 &   0.47 & 103.33 \vstt \\
A2657    &  61 &   1.71 &   2.17 &   2.45 &   2.12 &   2.08 &   0.54 & 105.65 \vstt \\
A0539    &  71 &   1.65 &   2.09 &   2.37 &   2.04 &   2.01 &   0.62 & 103.75 \vstt \\
A4038    &  62 &   1.64 &   2.08 &   2.35 &   2.03 &   2.01 &   0.62 & 102.79 \vstt \\
A3395s   &  21 &   1.62 &   2.06 &   2.40 &   2.01 &   1.96 &   0.46 & 102.94 \vstt \\
A0576    &  22 &   1.47 &   1.87 &   2.17 &   1.82 &   1.79 &   0.46 &  98.07 \vstt \\
A1644    &  31 &   1.45 &   1.84 &   2.11 &   1.79 &   1.75 &   0.39 &  96.55 \vstt \\
MKW8     &  42 &   1.44 &   1.83 &   2.08 &   1.78 &   1.76 &   0.53 &  97.25 \vstt \\
UGC03957 &  76 &   1.41 &   1.79 &   2.02 &   1.75 &   1.71 &   0.45 &  96.10 \vstt \\
A3376    &  19 &   1.41 &   1.78 &   2.12 &   1.74 &   1.70 &   0.40 &  95.76 \vstt \\
A3395n   &  75 &   1.40 &   1.78 &   2.09 &   1.73 &   1.69 &   0.37 &  95.79 \vstt \\
A2256    &  54 &   1.37 &   1.73 &   2.01 &   1.69 &   1.64 &   0.32 &  92.60 \vstt \\
A3558    &  37 &   1.31 &   1.66 &   1.89 &   1.62 &   1.58 &   0.34 &  90.53 \vstt \\
A0400    &   6 &   1.26 &   1.60 &   1.85 &   1.56 &   1.54 &   0.47 &  90.68 \vstt \\
A3581    &  41 &   1.18 &   1.50 &   1.70 &   1.46 &   1.44 &   0.45 &  87.46 \vstt \\
2A0335   &  11 &   1.17 &   1.48 &   1.66 &   1.44 &   1.41 &   0.34 &  84.20 \vstt \\
A0085    &   1 &   1.15 &   1.46 &   1.64 &   1.42 &   1.38 &   0.25 &  83.97 \vstt \\
A2052    &  44 &   1.13 &   1.43 &   1.61 &   1.39 &   1.36 &   0.32 &  84.47 \vstt \\
A1795    &  40 &   1.11 &   1.41 &   1.59 &   1.38 &   1.33 &   0.23 &  82.62 \vstt \\
NGC507   &   4 &   1.09 &   1.39 &   1.57 &   1.35 &   1.34 &   0.46 &  85.00 \vstt \\
MKW4     &  27 &   1.04 &   1.32 &   1.50 &   1.29 &   1.27 &   0.39 &  82.67 \vstt \\
NGC499   &  66 &   1.04 &   1.32 &   1.50 &   1.28 &   1.27 &   0.45 &  82.88 \vstt \\
\hline
\hline
\end{tabular}
\begin{quote}
  Notes:
   (1) The HIFLUGCS cluster ID.
   (2) In units of $10^{-8} \rmn{ph}\,\rmn{cm}^{-2}\,\rmn{s}^{-1}$.
   (3) In units of $10^{-9} \rmn{ph}\,\rmn{cm}^{-2}\,\rmn{s}^{-1}$.
   (4) In units of $10^{-12} \rmn{ph}\,\rmn{cm}^{-2}\,\rmn{s}^{-1}$.
   (5) In units of $10^{-9} \rmn{erg}\,\rmn{cm}^{-2}\,\rmn{s}^{-1}$.
   (6) The flux within $0.1^\circ$, smoothed with a point spread function of $0.1^\circ$.
   (7) The flux within $1.0^\circ$, smoothed with a point spread function of $0.1^\circ$.
   (8) Signal to noise (S/N) within $\rvir$ above $1\,$GeV. We estimate the signal to
  noise through $\sqrt{t_\rmn{obs}\,A_\rmn{eff}}\,F_{\gamma}/\sqrt{F_{\gamma}+F_{\gamma,\CR}}$,
  where $t_\rmn{obs}=3\,$yrs is the observation time and $A_\rmn{eff} = 7000\,\cm^2$
    is the effective area of Fermi-Lat at $1\,$GeV.
 \label{tab:flux_tab_LP}
  \end{quote}
\end{minipage}
\end{table*}

\begin{table*}
\begin{minipage}{2.0\columnwidth}
  \caption{Galaxy cluster parameters for to all clusters in the HIFLUGCS catalogue.}
\begin{tabular}{l c c c c c c c c c c c c c}
\hline
\hline
Cluster & ID$^{(1)}$ & $D_\clu$$^{(1)}$ & $\rvir$$^{(2)}$ & $\rvir$ &
$\mvir$$^{(2)}$ & $n_\rmn{e,1}(0)$$^{(3)}$ & $r_\rmn{core,1}$$^{(3)}$ &
$\beta_1$$^{(3)}$ & $n_\rmn{e,2}(0)$$^{(3)}$ & $r_\rmn{core,2}$$^{(3)}$ &
$\beta_2$$^{(3)}$ & $r_\rmn{hlr,CR}$$^{(4)}$ & $r_\rmn{hlr,DM}$ $^{(5)}$ \\
& & [Mpc] & [Mpc] & [deg] & [$10^{14}\,\msun$]
& [$10^{-2}\,\rmn{cm}^{-3}$] & [kpc] & &
  [$10^{-2}\,\rmn{cm}^{-3}$] & [kpc] & & [deg] & [deg] \\
 \hline
A0085 & 1 & 248.29 & 2.11 & 0.49 & 10.90 & 3.02 & 41 & 0.60 & 0.32 & 275 & 0.73 & 0.04 & 0.22 \vst \\
A0119 & 2 & 194.82 & 2.18 & 0.64 & 11.90 & 0.17 & 285 & 0.76 & 0.06 & 1079 & 1.46 & 0.12 & 0.29 \vst \\
A0133 & 3 & 254.34 & 1.78 & 0.40 & 6.50 & 3.30 & 30 & 0.65 & 0.24 & 229 & 0.78 & 0.02 & 0.18 \vst \\
NGC507 & 4 & 71.57 & 0.97 & 0.78 & 1.10 & 0.59 & 29 & 0.76 & 1.24 & 52 & 4.29 & 0.01 & 0.35 \vst \\
A0262 & 5 & 69.81 & 1.17 & 0.96 & 1.90 & 0.96 & 29 & 0.44 & 0.00 & 0 & 0.00 & 0.17 & 0.44 \vst \\
A0400 & 6 & 104.69 & 1.29 & 0.71 & 2.50 & 0.24 & 110 & 0.53 & 0.00 & 0 & 0.00 & 0.13 & 0.32 \vst \\
A0399 & 7 & 323.00 & 2.09 & 0.37 & 10.50 & 0.26 & 320 & 0.71 & 0.00 & 0 & 0.00 & 0.07 & 0.17 \vst \\
A0401 & 8 & 338.70 & 2.13 & 0.36 & 11.20 & 0.69 & 170 & 0.69 & 0.17 & 375 & 0.66 & 0.05 & 0.16 \vst \\
A3112 & 9 & 339.66 & 1.78 & 0.30 & 6.50 & 4.12 & 36 & 0.63 & 0.62 & 117 & 0.62 & 0.01 & 0.14 \vst \\
FORNAX & 10 & 19.77 & 1.28 & 3.70 & 2.40 & 0.11 & 123 & 0.80 & 0.00 & 0 & 0.00 & 0.31 & 1.69 \vst \\
2A0335 & 11 & 153.49 & 1.58 & 0.59 & 4.50 & 6.47 & 23 & 0.57 & 0.00 & 0 & 0.00 & 0.02 & 0.27 \vst \\
IIIZw54 & 12 & 136.39 & 1.71 & 0.72 & 5.80 & 0.24 & 206 & 0.89 & 0.00 & 0 & 0.00 & 0.07 & 0.33 \vst \\
A3158 & 13 & 264.13 & 1.92 & 0.42 & 8.20 & 0.46 & 191 & 0.66 & 0.00 & 0 & 0.00 & 0.05 & 0.19 \vst \\
A0478 & 14 & 411.96 & 2.17 & 0.30 & 11.70 & 4.14 & 51 & 0.68 & 0.73 & 180 & 0.71 & 0.02 & 0.14 \vst \\
NGC1550 & 15 & 53.18 & 1.08 & 1.16 & 1.40 & 0.89 & 32 & 0.55 & 0.00 & 0 & 0.00 & 0.07 & 0.53 \vst \\
EXO0422 & 16 & 172.04 & 1.57 & 0.52 & 4.40 & 0.78 & 101 & 0.72 & 0.00 & 0 & 0.00 & 0.04 & 0.24 \vst \\
A3266 & 17 & 266.00 & 2.69 & 0.58 & 22.40 & 0.30 & 321 & 1.20 & 0.15 & 830 & 1.27 & 0.08 & 0.26 \vst \\
A0496 & 18 & 144.03 & 1.83 & 0.73 & 7.10 & 4.80 & 21 & 0.59 & 0.34 & 183 & 0.69 & 0.04 & 0.33 \vst \\
A3376 & 19 & 201.69 & 2.01 & 0.57 & 9.40 & 0.12 & 538 & 1.05 & 0.00 & 0 & 0.00 & 0.10 & 0.26 \vst \\
A3391 & 20 & 236.69 & 1.95 & 0.47 & 8.50 & 0.46 & 50 & 0.50 & 0.21 & 239 & 0.66 & 0.07 & 0.22 \vst \\
A3395s & 21 & 221.45 & 2.21 & 0.57 & 12.40 & 0.15 & 431 & 0.96 & 0.00 & 0 & 0.00 & 0.08 & 0.26 \vst \\
A0576 & 22 & 167.96 & 1.81 & 0.62 & 6.80 & 0.19 & 281 & 0.82 & 0.00 & 0 & 0.00 & 0.09 & 0.28 \vst \\
A0754 & 23 & 235.31 & 2.45 & 0.60 & 17.00 & 0.52 & 170 & 0.70 & 0.00 & 0 & 0.00 & 0.05 & 0.27 \vst \\
HYDRA-A & 24 & 239.94 & 1.75 & 0.42 & 6.20 & 4.15 & 70 & 1.84 & 0.85 & 130 & 0.73 & 0.02 & 0.19 \vst \\
HYDRA & 25 & 49.25 & 1.53 & 1.78 & 4.10 & 0.56 & 67 & 0.61 & 0.00 & 0 & 0.00 & 0.13 & 0.81 \vst \\
A1367 & 26 & 94.05 & 2.06 & 1.25 & 10.10 & 0.16 & 207 & 0.96 & 0.08 & 697 & 1.51 & 0.17 & 0.57 \vst \\
MKW4 & 27 & 86.98 & 1.08 & 0.71 & 1.40 & 3.45 & 7 & 0.44 & 0.00 & 0 & 0.00 & 0.07 & 0.33 \vst \\
ZwCl1215 & 28 & 339.66 & 2.21 & 0.37 & 12.40 & 0.32 & 307 & 0.82 & 0.00 & 0 & 0.00 & 0.05 & 0.17 \vst \\
NGC4636 & 29 & 15.89 & 0.76 & 2.74 & 0.50 & 1.99 & 4 & 0.49 & 0.00 & 0 & 0.00 & 0.07 & 1.25 \vst \\
CENTAURU & 30 & 44.46 & 1.67 & 2.15 & 5.30 & 2.16 & 23 & 0.57 & 0.16 & 194 & 0.70 & 0.15 & 0.98 \vst \\
A1644 & 31 & 210.40 & 2.06 & 0.56 & 10.00 & 0.33 & 195 & 0.83 & 0.06 & 1549 & 2.38 & 0.10 & 0.26 \vst \\
A1650 & 32 & 385.28 & 1.99 & 0.30 & 9.10 & 0.51 & 200 & 0.70 & 0.00 & 0 & 0.00 & 0.04 & 0.14 \vst \\
A1651 & 33 & 392.54 & 2.13 & 0.31 & 11.10 & 1.24 & 85 & 0.75 & 0.39 & 254 & 0.76 & 0.03 & 0.14 \vst \\
COMA & 34 & 101.14 & 2.24 & 1.27 & 12.90 & 0.36 & 245 & 0.65 & 0.00 & 0 & 0.00 & 0.20 & 0.58 \vst \\
NGC5044 & 35 & 38.81 & 0.99 & 1.46 & 1.10 & 4.08 & 7 & 0.52 & 0.00 & 0 & 0.00 & 0.03 & 0.67 \vst \\
A1736 & 36 & 204.44 & 1.47 & 0.41 & 3.70 & 0.15 & 267 & 0.54 & 0.00 & 0 & 0.00 & 0.12 & 0.19 \vst \\
A3558 & 37 & 213.16 & 2.01 & 0.54 & 9.30 & 0.54 & 165 & 0.68 & 0.07 & 855 & 1.17 & 0.08 & 0.25 \vst \\
A3562 & 38 & 221.91 & 1.68 & 0.43 & 5.50 & 0.69 & 73 & 0.52 & 0.04 & 957 & 1.26 & 0.07 & 0.20 \vst \\
A3571 & 39 & 175.22 & 2.16 & 0.71 & 11.60 & 1.14 & 67 & 0.82 & 0.60 & 182 & 0.68 & 0.07 & 0.32 \vst \\
A1795 & 40 & 276.29 & 2.23 & 0.46 & 12.80 & 3.39 & 56 & 0.72 & 0.32 & 308 & 0.89 & 0.02 & 0.21 \vst \\
A3581 & 41 & 93.17 & 1.17 & 0.72 & 1.80 & 1.91 & 25 & 0.54 & 0.00 & 0 & 0.00 & 0.04 & 0.33 \vst \\
MKW8 & 42 & 118.04 & 1.44 & 0.70 & 3.50 & 0.31 & 75 & 0.51 & 0.00 & 0 & 0.00 & 0.11 & 0.32 \vst \\
A2029 & 43 & 347.78 & 2.24 & 0.37 & 12.90 & 4.54 & 44 & 0.63 & 0.80 & 152 & 0.65 & 0.02 & 0.17 \vst \\
A2052 & 44 & 153.04 & 1.56 & 0.58 & 4.40 & 3.30 & 60 & 2.10 & 0.71 & 100 & 0.66 & 0.03 & 0.27 \vst \\
MKW3S & 45 & 199.40 & 1.64 & 0.47 & 5.10 & 2.12 & 65 & 1.42 & 0.71 & 108 & 0.68 & 0.02 & 0.22 \vst \\
A2065 & 46 & 325.85 & 2.31 & 0.41 & 14.20 & 0.24 & 492 & 1.16 & 0.00 & 0 & 0.00 & 0.05 & 0.19 \vst \\
A2063 & 47 & 155.75 & 1.51 & 0.56 & 4.00 & 1.12 & 39 & 0.49 & 0.18 & 457 & 2.02 & 0.06 & 0.25 \vst \\
A2142 & 48 & 411.47 & 2.48 & 0.35 & 17.50 & 1.90 & 100 & 0.67 & 0.18 & 637 & 1.01 & 0.03 & 0.16 \vst \\
A2147 & 49 & 154.39 & 1.50 & 0.56 & 3.90 & 0.20 & 169 & 0.44 & 0.00 & 0 & 0.00 & 0.16 & 0.25 \vst \\
A2163 & 50 & 990.91 & 2.55 & 0.15 & 19.20 & 0.63 & 370 & 0.80 & 0.00 & 0 & 0.00 & 0.02 & 0.07 \vst \\
A2199 & 51 & 132.35 & 1.77 & 0.77 & 6.40 & 0.98 & 99 & 0.65 & 0.00 & 0 & 0.00 & 0.06 & 0.35 \vst \\
A2204 & 52 & 727.47 & 1.93 & 0.15 & 8.30 & 6.00 & 47 & 0.60 & 0.00 & 0 & 0.00 & 0.01 & 0.07 \vst \\
A2244 & 53 & 446.19 & 1.90 & 0.24 & 7.90 & 1.42 & 89 & 0.61 & 0.00 & 0 & 0.00 & 0.02 & 0.11 \vst \\
A2256 & 54 & 269.27 & 2.36 & 0.50 & 15.20 & 0.31 & 419 & 0.91 & 0.00 & 0 & 0.00 & 0.07 & 0.23 \vst \\
A2255 & 55 & 363.60 & 2.10 & 0.33 & 10.60 & 0.21 & 423 & 0.80 & 0.00 & 0 & 0.00 & 0.06 & 0.15 \vst \\
\end{tabular}
 \label{tab:flux_tab_CLp}
\end{minipage}
\end{table*}
\newpage
\begin{table}
\begin{minipage}{2.0\columnwidth}
\begin{tabular}{l c c c c c c c c c c c c c}
\hline
\hline
Cluster & ID$^{(1)}$ & $D_\clu$$^{(1)}$ & $\rvir$$^{(2)}$ & $\rvir$ &
$\mvir$$^{(2)}$ & $n_\rmn{e,1}(0)$$^{(3)}$ & $r_\rmn{core,1}$$^{(3)}$ &
$\beta_1$$^{(3)}$ & $n_\rmn{e,2}(0)$$^{(3)}$ & $r_\rmn{core,2}$$^{(3)}$ &
$\beta_2$$^{(3)}$ & $r_\rmn{hlr,CR}$$^{(4)}$ & $r_\rmn{hlr,DM}$ $^{(5)}$ \\
& & [Mpc] & [Mpc] & [deg] & [$10^{14}\,\msun$]
& [$10^{-2}\,\rmn{cm}^{-3}$] & [kpc] & &
  [$10^{-2}\,\rmn{cm}^{-3}$] & [kpc] & & [deg] & [deg] \\
 \hline
A3667 & 56 & 250.15 & 1.88 & 0.43 & 7.60 & 0.39 & 287 & 0.89 & 0.06 & 1696 & 1.70 & 0.09 & 0.20 \vst \\
S1101 & 57 & 259.46 & 1.60 & 0.35 & 4.70 & 3.43 & 47 & 0.79 & 0.20 & 272 & 0.96 & 0.01 & 0.16 \vst \\
A2589 & 58 & 183.87 & 1.64 & 0.51 & 5.10 & 0.85 & 67 & 0.66 & 0.19 & 222 & 0.74 & 0.04 & 0.23 \vst \\
A2597 & 59 & 388.66 & 1.71 & 0.25 & 5.70 & 4.29 & 40 & 0.63 & 0.00 & 0 & 0.00 & 0.01 & 0.12 \vst \\
A2634 & 60 & 136.84 & 1.80 & 0.75 & 6.70 & 0.28 & 57 & 0.47 & 0.07 & 849 & 1.89 & 0.14 & 0.34 \vst \\
A2657 & 61 & 178.41 & 1.95 & 0.63 & 8.60 & 0.63 & 105 & 0.89 & 0.10 & 568 & 1.27 & 0.05 & 0.29 \vst \\
A4038 & 62 & 123.85 & 1.54 & 0.71 & 4.20 & 1.75 & 37 & 0.58 & 0.19 & 172 & 0.70 & 0.05 & 0.33 \vst \\
A4059 & 63 & 203.98 & 1.79 & 0.50 & 6.60 & 1.39 & 58 & 0.64 & 0.15 & 312 & 0.90 & 0.03 & 0.23 \vst \\
A2734 & 64 & 278.16 & 1.83 & 0.38 & 7.10 & 0.38 & 150 & 0.62 & 0.00 & 0 & 0.00 & 0.05 & 0.17 \vst \\
A2877 & 65 & 105.13 & 2.02 & 1.10 & 9.50 & 0.21 & 230 & 3.58 & 0.08 & 432 & 1.23 & 0.11 & 0.50 \vst \\
NGC499 & 66 & 63.67 & 0.89 & 0.80 & 0.80 & 1.11 & 16 & 0.72 & 0.00 & 0 & 0.00 & 0.01 & 0.37 \vst \\
AWM7 & 67 & 74.64 & 1.84 & 1.41 & 7.20 & 0.69 & 89 & 0.78 & 0.19 & 290 & 0.88 & 0.12 & 0.65 \vst \\
PERSEUS & 68 & 79.48 & 1.95 & 1.41 & 8.60 & 3.84 & 45 & 0.54 & 0.00 & 0 & 0.00 & 0.10 & 0.64 \vst \\
S405 & 69 & 274.88 & 1.81 & 0.38 & 6.80 & 0.14 & 327 & 0.66 & 0.00 & 0 & 0.00 & 0.09 & 0.17 \vst \\
3C129 & 70 & 97.15 & 1.89 & 1.11 & 7.80 & 0.21 & 227 & 0.60 & 0.00 & 0 & 0.00 & 0.22 & 0.51 \vst \\
A0539 & 71 & 126.08 & 1.56 & 0.71 & 4.40 & 0.83 & 30 & 0.53 & 0.18 & 223 & 0.75 & 0.09 & 0.32 \vst \\
S540 & 72 & 157.55 & 1.53 & 0.56 & 4.20 & 0.47 & 92 & 0.64 & 0.00 & 0 & 0.00 & 0.05 & 0.25 \vst \\
A0548w & 73 & 187.52 & 1.19 & 0.36 & 2.00 & 0.12 & 141 & 0.67 & 0.00 & 0 & 0.00 & 0.05 & 0.17 \vst \\
A0548e & 74 & 181.14 & 1.33 & 0.42 & 2.70 & 0.33 & 84 & 0.48 & 0.00 & 0 & 0.00 & 0.09 & 0.19 \vst \\
A3395n & 75 & 221.45 & 2.12 & 0.55 & 10.90 & 0.12 & 480 & 0.98 & 0.00 & 0 & 0.00 & 0.09 & 0.25 \vst \\
UGC03957 & 76 & 149.43 & 1.65 & 0.63 & 5.20 & 0.57 & 101 & 0.74 & 0.00 & 0 & 0.00 & 0.04 & 0.29 \vst \\
PKS0745 & 77 & 474.79 & 2.04 & 0.25 & 9.80 & 6.71 & 51 & 0.70 & 0.70 & 167 & 0.65 & 0.01 & 0.11 \vst \\
A0644 & 78 & 317.77 & 2.14 & 0.39 & 11.20 & 0.92 & 144 & 0.70 & 0.00 & 0 & 0.00 & 0.03 & 0.18 \vst \\
ANTLIA & 79 & 50.13 & 1.35 & 1.54 & 2.80 & 0.08 & 245 & 0.75 & 0.00 & 0 & 0.00 & 0.28 & 0.71 \vst \\
A1413 & 80 & 677.27 & 2.23 & 0.19 & 12.70 & 1.44 & 110 & 0.80 & 0.26 & 399 & 0.91 & 0.01 & 0.09 \vst \\
M49 & 81 & 18.91 & 1.07 & 3.24 & 1.40 & 1.57 & 7 & 0.59 & 0.00 & 0 & 0.00 & 0.04 & 1.48 \vst \\
A3528n & 82 & 240.86 & 1.80 & 0.43 & 6.70 & 0.40 & 126 & 0.62 & 0.00 & 0 & 0.00 & 0.05 & 0.20 \vst \\
A3528s & 83 & 245.97 & 1.57 & 0.37 & 4.50 & 0.57 & 71 & 0.46 & 0.00 & 0 & 0.00 & 0.08 & 0.17 \vst \\
A3530 & 84 & 242.72 & 1.78 & 0.42 & 6.50 & 0.15 & 300 & 0.77 & 0.00 & 0 & 0.00 & 0.07 & 0.19 \vst \\
A3532 & 85 & 240.40 & 2.00 & 0.48 & 9.20 & 0.34 & 137 & 0.74 & 0.11 & 543 & 1.09 & 0.06 & 0.22 \vst \\
A1689 & 86 & 897.29 & 2.51 & 0.16 & 18.20 & 2.46 & 108 & 0.88 & 0.46 & 336 & 0.91 & 0.01 & 0.07 \vst \\
A3560 & 87 & 220.07 & 1.58 & 0.41 & 4.50 & 0.20 & 182 & 0.57 & 0.00 & 0 & 0.00 & 0.09 & 0.19 \vst \\
A1775 & 88 & 343.00 & 1.76 & 0.29 & 6.30 & 0.35 & 385 & 2.05 & 0.07 & 1030 & 1.70 & 0.04 & 0.13 \vst \\
A1800 & 89 & 338.70 & 1.94 & 0.33 & 8.40 & 0.21 & 279 & 0.77 & 0.00 & 0 & 0.00 & 0.05 & 0.15 \vst \\
A1914 & 90 & 827.98 & 2.35 & 0.16 & 14.90 & 1.32 & 164 & 0.75 & 0.00 & 0 & 0.00 & 0.01 & 0.07 \vst \\
NGC5813 & 91 & 27.55 & 0.95 & 1.98 & 1.00 & 1.06 & 17 & 0.77 & 0.00 & 0 & 0.00 & 0.03 & 0.90 \vst \\
NGC5846 & 92 & 26.25 & 0.76 & 1.66 & 0.50 & 4.15 & 2 & 0.51 & 0.70 & 39 & 4.78 & 0.02 & 0.76 \vst \\
A2151w & 93 & 162.53 & 1.36 & 0.48 & 2.90 & 0.97 & 48 & 0.56 & 0.00 & 0 & 0.00 & 0.04 & 0.22 \vst \\
NORMA & 94 & 70.69 & 1.84 & 1.49 & 7.20 & 0.22 & 213 & 0.56 & 0.00 & 0 & 0.00 & 0.33 & 0.68 \vst \\
TRIANGUL & 95 & 226.98 & 2.50 & 0.63 & 18.00 & 0.63 & 177 & 0.71 & 0.17 & 500 & 0.80 & 0.08 & 0.29 \vst \\
OPHIUCHU & 96 & 122.51 & 3.28 & 1.53 & 40.50 & 0.80 & 234 & 1.04 & 0.12 & 850 & 1.40 & 0.10 & 0.70 \vst \\
ZwCl1742 & 97 & 343.00 & 2.25 & 0.38 & 13.10 & 0.71 & 165 & 0.72 & 0.00 & 0 & 0.00 & 0.03 & 0.17 \vst \\
A2319 & 98 & 252.01 & 2.44 & 0.55 & 16.70 & 0.58 & 273 & 1.06 & 0.16 & 624 & 0.82 & 0.08 & 0.25 \vst \\
A3695 & 99 & 407.09 & 2.03 & 0.29 & 9.70 & 0.24 & 284 & 0.64 & 0.00 & 0 & 0.00 & 0.06 & 0.13 \vst \\
IIZw108 & 100 & 219.60 & 1.72 & 0.45 & 5.90 & 0.17 & 260 & 0.66 & 0.00 & 0 & 0.00 & 0.08 & 0.21 \vst \\
A3822 & 101 & 344.43 & 1.82 & 0.30 & 6.90 & 0.25 & 250 & 0.64 & 0.00 & 0 & 0.00 & 0.06 & 0.14 \vst \\
A3827 & 102 & 451.10 & 2.53 & 0.32 & 18.70 & 0.31 & 423 & 0.99 & 0.00 & 0 & 0.00 & 0.04 & 0.15 \vst \\
A3888 & 103 & 720.64 & 3.04 & 0.24 & 32.40 & 0.61 & 285 & 0.93 & 0.00 & 0 & 0.00 & 0.02 & 0.11 \vst \\
A3921 & 104 & 429.52 & 2.00 & 0.27 & 9.20 & 0.40 & 234 & 0.76 & 0.00 & 0 & 0.00 & 0.03 & 0.12 \vst \\
HCG94 & 105 & 184.32 & 1.49 & 0.46 & 3.80 & 0.80 & 42 & 0.53 & 0.19 & 142 & 0.58 & 0.06 & 0.21 \vst \\
RXJ2344 & 106 & 356.88 & 2.17 & 0.35 & 11.80 & 0.51 & 91 & 0.72 & 0.32 & 285 & 0.92 & 0.03 & 0.16 \vst \\
\hline
\hline
\end{tabular}
\begin{quote}
  Notes:
  (1) The HIFLUGCS cluster ID and luminosity distance
  ($D_\rmn{lum}$) are taken from \cite{2002ApJ...567..716R}.
  (2) The virial mass ($\mvir$) for each cluster is derived from the $M_{500}$
  mass in \cite{2007A&A...466..805C}. We solve for $\mvir$ using
  $\mvir=M_{500}\,200/500\,(c_{200}(\mvir)/c_{500}(\mvir))^3$ \cite{2005RvMP...77..207V}.
  The virial radius ($\rvir$) is derived from $\mvir$.
  (3) The electron number density profile of each cluster follows either a single- or double-beta profile;
  $n_\rmn{e}(r) = \sqrt{\sum_i n_\rmn{e,i}(0)^2(1+r^2/r_\rmn{c,i}^2)^{3\beta_i}}$
  where the central electron density, $n_\rmn{e}(0)$, core radius, $r_\rmn{c}$, and slope, $\beta$, are derived from
  \cite{2007A&A...466..805C}.
  (4) Half light radius ($r_\rmn{hlr,CR}$) of the flux from CR-$\pi^0$.
  (5) Half light radius ($r_\rmn{hlr,DM}$) of the flux from the BM-$\Kp$
  continuum and LP-IC-CMB emission. The flux has been
  boosted by substructures and includes the Sommerfeld enhancement for the LP model.
  \end{quote}
\end{minipage}
\end{table}

\end{document}